\documentclass[onecolumn,showpacs,preprintnumbers,nofootinbib,prd,superscriptaddress]{revtex4-1}
\usepackage{graphicx, epsfig, amssymb} 
\usepackage{amsmath, amsfonts}
\usepackage{bm} 

\usepackage[linktocpage]{hyperref}
\usepackage[caption=false]{subfig}
\usepackage[usenames]{color}

\def\Lie{\mathcal{L}}
\def\A{\mathcal{A}}

\def\al{\alpha}
\def\be{\beta}

\def\eps{\epsilon}
\def\ga{\gamma}

\def\si{\sigma}

\def\De{\Delta}

\def\Si{\Sigma}
\def\p{\partial}
\def\na{\nabla}
\def\non{\nonumber}

\def\non{\nonumber}

\newcommand{\ben}{\begin{enumerate}}
\newcommand{\een}{\end{enumerate}}

\begin{document}

\title{\large Superradiant instabilities in astrophysical systems}

\author{Helvi Witek} 
\email{h.witek@damtp.cam.ac.uk}
\affiliation{CENTRA, Departamento de F\'{\i}sica, Instituto Superior T\'ecnico, Universidade T\'ecnica de Lisboa - UTL,
Avenida Rovisco Pais 1, 1049 Lisboa, Portugal.}
\affiliation{Department of Applied Mathematics and Theoretical Physics,
Centre for Mathematical Sciences, University of Cambridge,
Wilberforce Road, Cambridge CB3 0WA, UK}

\author{Vitor Cardoso} 
\affiliation{CENTRA, Departamento de F\'{\i}sica, Instituto Superior T\'ecnico, Universidade T\'ecnica de Lisboa - UTL,
Avenida Rovisco Pais 1, 1049 Lisboa, Portugal.}
\affiliation{Department of Physics and Astronomy, The University of Mississippi, University, MS 38677, USA.}

\author{Akihiro Ishibashi}
\affiliation{Department of Physics, Kinki University, Higashi-Osaka 577-8502, Japan}

\author{Ulrich Sperhake} 
\affiliation{CENTRA, Departamento de F\'{\i}sica, Instituto Superior T\'ecnico, Universidade T\'ecnica de Lisboa - UTL,
Avenida Rovisco Pais 1, 1049 Lisboa, Portugal.}
\affiliation{Department of Applied Mathematics and Theoretical Physics,
Centre for Mathematical Sciences, University of Cambridge,
Wilberforce Road, Cambridge CB3 0WA, UK}
\affiliation{Institute of Space Sciences, CSIC-IEEC, 08193 Bellaterra, Spain}
\affiliation{California Institute of Technology, Pasadena, CA 91125, USA}

\date{\today} 

\begin{abstract} Light bosonic degrees of freedom have become a
serious candidate for dark matter, which seems to pervade our entire
universe.  The evolution of these fields around curved spacetimes
is poorly understood but is expected to display interesting effects.
In particular, the interaction of light bosonic fields with supermassive
black holes, key players in most galaxies, could provide colourful
examples of superradiance and nonlinear bosenova-like collapse. In turn,
the observation of spinning black holes is expected to impose stringent
bounds on the mass of putative massive bosonic fields in our universe.

Our purpose here is to present a comprehensive study of the evolution
of linearized massive scalar and vector fields in the vicinities of
rotating black holes. The evolution of generic initial data has a very
rich structure, depending on the mass of the field and of the black hole.
Quasi-normal ringdown or exponential decay followed by a power-law tail at
very late times is a generic feature of massless fields at intermediate
times. Massive fields generically show a transition to power-law tails
early on. For a certain boson field mass range, the field can become
trapped in a potential barrier outside the horizon and transition to
a bound state. Because there are a number of such quasi-bound states,
the generic outcome is an amplitude modulated sinusoidal,
or beating,
signal, whose envelope is well described by the two lowest
overtones. We believe that the appearance of such beatings has gone
unnoticed in the past, and in fact mistaken for exponential growth. The
amplitude modulation of the signal depends strongly on the relative
excitation of the overtones, which in turn is strongly tied to the
bound-state geography.

A fine tuning of the initial data allows one to see the evolution of
the nearly pure bound state mode which turns unstable for sufficiently
large black hole rotation. For the first time we explore massive vector
fields in generic BH background which are hard, if not impossible, to
separate in the Kerr background. Our results show that spinning BHs are
generically strongly unstable against massive vector fields.
\end{abstract}

\pacs{
98.80.Es,
11.25.Wx,
14.80.Va,
04.70.-s
}
\maketitle
\tableofcontents

\section{Introduction}
One of the most exciting outcomes of General Relativity (GR)
are black holes (BHs), the physics of which has grown into a
mature and fully developed branch of GR and extensions thereof
\cite{Ruffini,Frolov:1998wf}. Observations of, e.g., $X$-ray binaries
indicate that solar mass ($3-30M_{\odot}$) BHs mark the endpoint of
the life of massive stars and are anticipated to be a significant
component of the galaxies' population. Supermassive BHs (SMBHs) with
masses $10^6-10^9M_{\odot}$ or higher are conjectured to be hosted in the
center of most galaxies, controlling galaxy growth and evolution, stellar
birth and powering active galactic nuclei and other powerful phenomena.

Tremendous progress has been made in actually observing some of the
fascinating general relativistic effects. From $X$-ray spectra on the
inner edge of accretion disks, which probe the innermost stable circular
orbit of the geometry, to gravitational wave (GW) physics, ``precision BH
physics'' is a new and rapidly developing field \cite{Arvanitaki:2009fg,
Arvanitaki:2010sy,McClintock:2009as}. The future holds the promise to
observe some of these effects accurately by monitoring the supermassive
BH at the center of our own galaxy.

One of the fundamental reasons why precision BH physics is
possible at all, are the no-hair and uniqueness theorems: BHs in
$4$-dimensional, asymptotically flat spacetimes must belong to
the Kerr-Newman family and are, thus, fully specified by three
parameters only: their mass, angular momentum and electric charge
(see e.g. Ref.~\cite{Heusler:1998ua,Chrusciel:2012jk}, or Carter's
contribution to Ref.~\cite{Hawking:1979ig}).  In more colloquial terms,
this is commonly expressed by saying that BHs have no hair or, rather,
have three hairs only.  This simple yet powerful result has far reaching
consequences: Given some arbitrary perturbations with the same conserved
charges,
they must all decay to the same final state, namely one BH with those
charges.  By now, there are a plethora of studies, at the perturbative
and fully non-linear level, investigating how this unique final state
is approached (see, e.g., Ref.~\cite{Berti:2009kk,Konoplya:2011qq} for
recent overviews). In the following we briefly summarize these studies.

\vspace{0.5cm}
\noindent{\bf{\em Generic response of a BH spacetime to external perturbations.}}
%
\begin{figure}
\begin{center}
\includegraphics[width=0.50\textwidth]{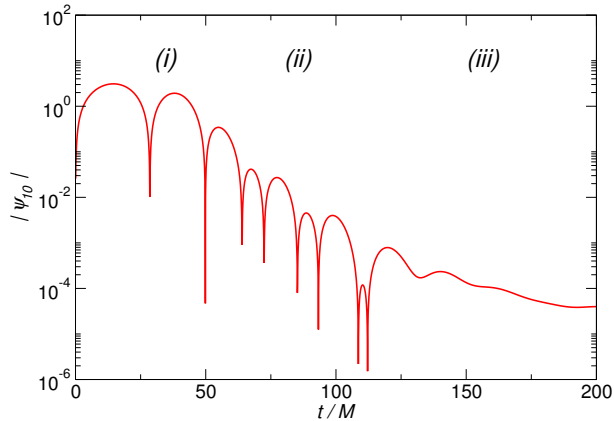}
\end{center}
\caption{\label{fig:MFresponse}
Time evolution of a dipole ($l=1,m=0$) scalar Gaussian wave packet in
Schwarzschild background.  We clearly observe the main features of
such a field: (i) a prompt response at early times followed by (ii)
the quasi-normal mode ringdown and (iii) a late-time tail.
}
\end{figure}
The generic behavior of massless fields around a BH is illustrated in
Fig.~\ref{fig:MFresponse}, where we plot the evolution of a Gaussian
wave packet $\Psi=e^{-(r-10)^2/10^2}$ around a Schwarzschild BH.
The particular initial data
refers to a scalar field, but the qualitative results are universal and
independent of the initial conditions.
The generic behavior of massless fields around a BH can be divided into
three parts (c.f. Fig.~\ref{fig:MFresponse}):

\noindent(i) An initial data-dependent prompt response at early times,
which is the counterpart to light-cone propagation in flat space;

\noindent(ii) An exponentially decaying ``ringdown'' phase at intermediate
times, where the BH is ringing with its characteristic quasi-normal
modes (QNMs).  This stage typically dominates the signal, and its
properties, such as vibration frequency and decay timescale, depend
solely on the parameters of the {\it final} BH \cite{Berti:2009kk}.
Because of the no-hair theorem, the detection of QNMs allows one to
uniquely determine the BH mass and spin and provides tests of GR
\cite{Berti:2009kk,Berti:2005ys,Kamaretsos:2011um};

\noindent (iii) At late times, the signal is dominated
by a power-law fall-off, known as ``late-time tail''
\cite{Price:1971fb,Leaver:1986gd,Ching:1995tj}.  Tails are caused by
backscattering off spacetime curvature and more generically by a failure
of Huygen's principle. As such, tails also appear in other situations
where light propagation is not {\it on} the light cone such as in massive
field propagation in Minkowski spacetime \cite{feshbach}, or massless
field propagation in odd-dimensional spacetimes \cite{Cardoso:2003jf}.

\noindent{\bf{\em Superradiant effects.}}
The long-lasting oscillation of the lowest QNMs is the most important
stage in the life of any field around a BH.  Its lifetime, or quality
factor, depends solely on the BH spin \cite{Berti:2009kk}.  Specifically,
the lifetime tends to increase with growing spin and the decay timescale
approaches zero 
for nearly extremal BHs.
This behaviour is tightly connected to {\textit{superradiance}}
\cite{zeldovich1,zeldovich2,Bekenstein:1973mi}:
In a scattering experiment of low-frequency waves off a BH the scattered
wave is amplified if the real part of its frequency $\omega_R$ satisfies the superradiant
condition
\begin{align}
  \label{eq:MFSRcond}
  \omega_R < m\Omega_H\,,
\end{align}
where $m$ is the azimuthal ``quantum'' number and $\Omega_H$ is the
angular velocity of the BH horizon.  
We refer the reader to App.~\ref{app:Fluxformula} for a derivation of this condition for both, scalar and vector fields.
The excess energy is withdrawn
from the object's rotational energy \cite{zeldovich1,zeldovich2} and,
in a dynamical scenario, the BH would presumably spin down.  The effect
can be attributed to the existence of negative-energy states in the
ergo-region, and dissipation at the event horizon.

Superradiance is the chief cause of a number of exciting phenomena in
BH physics:

\noindent{(i)}
Generic perturbations are damped away to infinity and across the event
horizon.  Because rotating BHs amplify waves that fulfill the superradiant
condition, Eq.~\eqref{eq:MFSRcond}, the amplification factors as well
as the quality factor of these superradiant modes increase with rotation.

\noindent{(ii)} Satellites around BHs typically spiral inwards as time
goes by, due to gravitational wave emission and energy conservation.
Emission of radiation to infinity
results in a larger binding energy of
the particle.  Because superradiance implies the extraction of the BH's
rotational energy, it is possible that the energy deficit comes entirely
from the BH kinetic energy.  In this way, satellites around rapidly
spinning BHs can {\it in principle} orbit at a nearly fixed radius --
on so-called floating orbits -- for a much longer time, tapping the
BH's kinetic energy.  In BH binaries, this effect can dominate in the
presence of resonances \cite{Cardoso:2011xi,Yunes:2011aa,Cardoso:2012zn}.
This phenomenon is analogous to tidal acceleration, e.g., in case of
the Earth-moon system ~\cite{Hut:1981,Verbunt}.

\noindent{(iii)}
A further interesting effect can be triggered by enclosing the spinning
BH inside a perfectly reflecting cavity. As was recognized already
by Zel'dovich \cite{zeldovich1,zeldovich2}, any initial perturbation
will get successively amplified near the BH and reflected back at the
mirror, thus creating an instability, which was termed the ``BH bomb''
\cite{Press:1972zz,Cardoso:2004nk}.  Whereas the setup appears physically
artificial at first glance, the role of the mirror can actually be
realized naturally in many ways, including an anti-de Sitter spatial
infinity. In this case, the BH bomb translates into a real, physical
instability of (small) rotating BHs in asymptotically AdS spacetimes
\cite{Cardoso:2004hs,Cardoso:2006wa,Kodama:2007sf,Uchikata:2009zz}.

\noindent{(iv)} Finally, of direct interest for the present
study is the fact that massive fields around
Kerr BHs are also prone to a BH bomb-like instability,
because the mass term effectively confines the field
\cite{Damour:1976kh,Zouros:1979iw,Detweiler:1980uk,Furuhashi:2004jk,
Cardoso:2005vk,Dolan:2007mj,Hod2012,Hod:2012px}.

Consider a scalar field surrounding a black hole with mass $M$ and angular momentum $J=a M$.
The instability is regulated by the dimensionless parameter $\mu_S M$
(from now on we set $G = c = 1$), where $m_s = \mu_S \hbar$ is the
scalar field mass, and is described by the time dependence of the field,
$\Psi\sim e^{-i\omega t}$ with complex frequency $\omega=\omega_R +
\imath \omega_I$.  For small coupling $M\mu_S\ll 1$ the characteristic
(unstable mode) frequency giving rise to the instability is
\cite{Pani:2012vp,Pani:2012bp} \footnote{Notice the difference of a
factor $2$ to the original result \cite{Detweiler:1980uk}.}
\begin{align}
  \label{eq:MFtsScaSmallMu}
  M\omega_I = & \frac{a}{48M}(M\mu_S)^9\,.
\end{align}
In the opposite limit, i.e., for very large mass couplings $M\mu_S>>1$,
the characteristic inverse time is \cite{Zouros:1979iw}
\begin{align}
\label{eq:MFtsScaLargeMu}
M\omega_I = & 10^{-7}\exp(-1.84 M\mu_S)\,.
\end{align}
The instability timescales are typically large. The scalar field growth
rate has a global maximum of $\tau \equiv 1/\omega_I \sim 10^{7}M$ for
the dipole with mass coupling $M\mu_S=0.42$ in the background of
a Kerr BH with $a/M=0.99$ \cite{Cardoso:2005vk,Dolan:2007mj}.

The above results refer to massive scalar fields in the background of Kerr BHs.
It was widely believed that massive vectors would be subject
to a similar instability. Unfortunately, the non-separability of the
field equations renders this is a non-trivial problem.  Recently,
significant progress has been made, with a thorough study of massive
vector fields around Schwarzschild BHs \cite{Rosa:2011my}
and slowly rotating Kerr BH backgrounds \cite{Pani:2012bp,Pani:2012vp}.
Pani {\it et al} use a slow-rotation expansion of the Fourier transformed
field equations, accurate to second order in rotation, to prove that the
Kerr spacetime is indeed unstable against massive vector fields
\cite{Pani:2012bp,Pani:2012vp}. The massive vector field instability
can be orders of magnitude stronger (i.e., shorter timescales) than its
scalar counterpart.

All these calculations have been performed in the linear regime, thus
neglecting backreaction effects such as the BH spin-down or effects
due to non-linear self-interaction of the scalar field.  Therefore,
the final state of the superradiant instability in the fully dynamical
regime is not known, partly because it requires the non-linear
evolution of Einstein's equations for a timescale of order $10^6M$.
A plausible evolution scenario consists of an exponentially growing
scalar condensate outside the BH, extracting energy and angular
momentum from the BH until the superradiant extraction stops, i.e.,
until the condition~\eqref{eq:MFSRcond} is no longer satisfied.
Further interesting new phenomena arise when we consider non-linear
interaction terms, such as bosenova-type collapse presented in
Refs.~\cite{Kodama:2011zc,Yoshino:2012kn,Mocanu:2012fd}, or higher
dimensional background spacetimes, such as the boosted black string
recently reported in~\cite{Rosa:2012uz}
or the Schwarzschild-Tangherlini solution discussed in \cite{Herdeiro:2011uu,Wang:2012tk}.

\noindent{\bf{\em {Superradiant instability in astrophysical systems.}}}
Massive fields in the vicinity of BHs are subject to a BH bomb-like,
superradiant instability, and they grow exponentially with time.
However, the effect is very weak for known standard model particles
in astrophysical environments: For example, the mass coupling for the
lightest known elementary scalar particle, the pion, around a solar mass
BH is $M\mu_S\sim10^{18}$, resulting in an instability timescale much
larger than the age of the universe.
Nevertheless,  the superradiant instability might become significant 
if we consider standard model particles around primordial BHs 
(see, e.g., \cite{Khlopov:1985jw,Khlopov:2008qy})
or if there exist fields with small, but non-vanishing mass.
One exciting possibility for these fields is provided by axions,
ultralight bosonic states emerging from string-theory compactifications,
which have not been ruled out by current experiments.  In the ``axiverse''
scenario an entire landscape of ultra-light pseudo-scalar fields covering
a mass range from $10^{-33}eV\le\mu_S\le10^{-8}eV$ has been proposed
(see ~\cite{Arvanitaki:2009fg, Arvanitaki:2010sy,Kodama:2011zc} for
recent overviews).  
In fact, the existence of ultra-light axions leads
to a plethora of possible observational implications and signatures,
such as modifications of the cosmic-microwave background polarization
(for $10^{-33}eV\le\mu_S\le10^{-28}eV$).  They are also anticipated to
make up a fraction of dark matter if $10^{-28}eV\le\mu_S\le10^{-22}eV$
\cite{Arvanitaki:2009fg,Burt:2011pv,Barranco:2012qs}.  Of
particular interest in the context of BH physics are axions in the
mass range $10^{-22}eV\le\mu_S\le10^{-10}eV$ \cite{Arvanitaki:2009fg,
Arvanitaki:2010sy,Kodama:2011zc}.
Then, the time scales for the superradiant instability become
astrophysically significant, giving rise to a number of interesting
effects:

\noindent{(i)}
A bosonic cloud bounded in the vicinity of a Kerr BH might create a
``gravitational atom'', which can be de-excited by the emission of
gravitons, thus carrying away BH angular momentum;

\noindent{(ii)}
If the accretion of bosons from this cloud is efficient enough, the
rotation of the BH can be sustained and it might be turned into a
GW pulsar;

\noindent{(iii)}
If, on the other hand, the accretion from the axionic cloud is not
efficient enough, the BH will eventually spin down, thus yielding
gaps in the Regge plane (the phase-space spanned by mass and spin
parameter of the BH).  Further possible effects have been discussed in
Refs.~\cite{Arvanitaki:2009fg,Arvanitaki:2010sy,Cardoso:2011xi,Yunes:2011aa,
Alsing:2011er,Kodama:2011zc,Yoshino:2012kn,Mocanu:2012fd}.

Similar superradiant instabilities are expected to
occur for massive hidden $U(1)$ vector fields, which are
also a generic feature of extensions of the standard model
\cite{Goodsell:2009xc,Jaeckel:2010ni,Camara:2011jg,Goldhaber:2008xy}.
As already stated, while superradiant
instabilities have been widely studied for massive scalar
fields~\cite{Press:1972zz,Damour:1976kh,Cardoso:2004nk,Cardoso:2005vk,
Dolan:2007mj,Rosa:2009ei,Cardoso:2011xi,Hod:2011zzd},
the case of massive vector fields is still in its infancy,
though significant work along these lines was recently reported
\cite{Rosa:2011my,Pani:2012bp,Pani:2012vp,Herdeiro:2011uu,Wang:2012tk}.

So far most studies on the massive boson instability have been performed
in Fourier space.  
An early attempt at studying the massive scalar field instability in the time domain,
with generic initial conditions was presented by Strafuss and Khanna
\cite{Strafuss:2004qc}.  
We believe that, while the technical study may
be correct, some of its conclusions are not; specifically, the authors
reported an instability growth rate of $M\omega_I\sim2\cdot10^{-5}$, which
is two orders of magnitude larger than previous results in the frequency
domain~\cite{Cardoso:2005vk,Dolan:2007mj} and more recent numerical
studies in the time domain~\cite{Yoshino:2012kn,Dolan:2012yt}.  We will attempt a
correct explanation for these puzzling results in the body of this work.

The purpose of the present study is to investigate the time evolution of
generic linearized massive scalar and vector fields in the vicinities
of spinning BHs. Surprisingly, not much seems to have been done on
this problem.  Our ``generic'' initial data consists of Gaussian wave
packets, but we will also study the evolution of bound state modes. The
exploration of nonlinear gravitational dynamics or self-interactions
will be presented elsewhere.

This work is organized as follows: In Sec.~\ref{sec:MFFramework} we
present the numerical framework, describing the formulation as a Cauchy
problem, the setup of initial configurations and the background spacetime.
Sec.~\ref{sec:SFresults} is devoted to the numerical results of massive
scalar field evolution.  In particular, we present a number of benchmark
tests to verify our implementation before studying more generic setups.
We will show that the evolution of a massive scalar has a non-trivial
pattern, which can be explained in terms of multi-mode excitation.  We
believe that this pattern also describes the results reported by Strafuss
and Khanna \cite{Strafuss:2004qc}.  In Sec.~\ref{sec:Procaresults} we
discuss our investigations of the massive vector (also known as Proca \cite{Goldhaber:2008xy})
field in generic Kerr
BH backgrounds, where we show, for the first time in a time evolution
of rapidly spinning BHs, that Kerr BHs are strongly unstable against
these fields.  Finally, we summarize our results and present concluding
remarks in Sec.~\ref{sec:MFconclusion}.

\section{Setup: action, equations of motion and background metric}
\label{sec:MFFramework}

\subsection{Action and equations of motion}\label{ssec:MFEoM}

We consider a generic action \cite{Arvanitaki:2009fg,Balakin:2009rg}
involving one complex, massive scalar $\Psi$ and a massive vector
field $A_{\mu}$ with mass $m_S = \mu_S \hbar$ and $m_V=\mu_V\hbar$,
respectively,
\begin{align}
\label{eq:MFaction}
S = & \int d^4x \sqrt{-g} 
      \left( \frac{R}{k} - \frac{1}{4}F^{\mu\nu}F_{\mu\nu} - \frac{\mu_V^2}{2}A_{\nu}A^{\nu}
            -\frac{k_{\rm axion}}{2}\Psi \,^{\ast}F^{\mu\nu}F_{\mu\nu} 
            -\frac{1}{2}g^{\mu\nu}\Psi^{\ast}_{,\mu}\Psi^{}_{,\nu} - \frac{\mu_S^2}{2}\Psi^{\ast}\Psi-V(\Psi) \right)\,.
\end{align}
Here, the potential $V(\Psi)$ is of cubic or higher order in the scalar field.
The scalar and vector fields are allowed to interact through the
axion-like coupling constant $k_{\rm axion}$.  $F_{\mu\nu} \equiv
\na_{\mu}A_{\nu} - \na_{\nu} A_{\mu}$ is the Maxwell tensor and
$\,^{\ast}F^{\mu\nu} \equiv \frac{1}{2}\eps^{\mu\nu\rho\si}F_{\rho\si}$
is its dual. Here, $\epsilon^{\mu\nu\rho\si}\equiv
\frac{1}{\sqrt{-g}}E^{\mu\nu\rho\si}$ and $E^{\mu\nu\rho\si}$ is
the totally anti-symmetric Levi-Civita symbol with $E^{0123}=1$.
\footnote{The identity $\nabla_{\nu} \,{^*}F^{\mu\nu}=0$ is useful to
derive the equations of motion for the Chern-Simons term.}
The resulting equations of motion are
\begin{subequations}
\label{eq:MFEoMgen}
\begin{eqnarray}
  \label{eq:MFEoMScalar}
  \left(\nabla_{\nu}\nabla^{\nu}-\mu_S^2\right)\Psi &=&
      \frac{k_{\rm axion}}{2}\,{^*}F^{\mu\nu}F_{\mu\nu}+V'(\Psi) \,,\\
  \label{eq:MFEoMVector}
  \nabla_{\nu} F^{\mu\nu}+\mu_V^2A^\mu &=&
      -2k_{\rm axion}\,{^*}F^{\mu\nu}\partial_{\nu}\Psi \,,\\
  \label{eq:MFEoMTensor}
  \frac{1}{k} \left(R^{\mu \nu} - \frac{1}{2}g^{\mu\nu}R\right) &=&
      - \frac{1}{8}F^{\alpha\beta}F_{\alpha\beta}g^{\mu\nu}
      +\frac{1}{2}F^{\mu}_{\,\,\alpha}F^{\nu\alpha}
      - \frac{1}{4}\mu_V^2A_{\alpha}A^{\alpha}g^{\mu\nu}
      +\frac{\mu_V^2}{2}A^{\mu}A^{\nu}
      \nonumber\\
   && -\frac{1}{2}g^{\mu\nu}\left(\frac{1}{2} \Psi^{\ast}_{,\alpha}\Psi^{,\alpha}
      +\frac{\mu_S^2}{2}\Psi^{\ast}\Psi + V(\Psi)\right)
      +\frac{1}{4}\left( \Psi^{\ast,\mu}\Psi^{,\nu}
      + \Psi^{,\mu}\Psi^{\ast,\nu} \right) \,.
\end{eqnarray}
\end{subequations}
We note that these equations describe the fully non-linear evolution of
the system. Also, we have written the equations such that all terms quadratic
or of higher order in the vector or scalar fields appear on the right hand
side. In the remainder of this work, we will restrict ourselves to the
case of scalar and vector fields with small amplitudes and will ignore
the higher-order contributions on the right-hand sides of
(\ref{eq:MFEoMScalar})-(\ref{eq:MFEoMTensor}).

Under this assumption Eq.~\eqref{eq:MFEoMTensor} is equivalent
to Einstein's equations
in vacuum and a solution to this equation is the Kerr metric
which in Boyer-Lindquist coordinates is given by
\begin{align}
  \label{eq:MFKerrBL}
  ds^2 = & - \left(1-\frac{2Mr_{\rm BL}}{\Sigma}\right) dt^2
           + \left(1+\frac{2Mr_{\rm BL}}{\Sigma}\right) dr_{\rm BL}^2
           + \Sigma d\theta^2 + \sin^2\theta\left(r_{\rm BL}^2
           +a^2+\frac{2Ma^2r_{\rm BL}\sin^2\theta}{\Sigma}\right) d\phi^2
           \nonumber \\
         & + \left(\frac{4Mr_{\rm BL}}{\Sigma}\right) dtdr_{\rm BL}
           - \left(\frac{4Mr_{\rm BL}a\sin^2\theta}{\Sigma}\right) dtd\phi
           - 2a\sin^2\theta\left(1+\frac{2Mr_{\rm BL}}{\Sigma}\right)
           dr_{\rm BL}d\phi \,,
\end{align}
with
\begin{align}
  \Sigma = & r_{\rm BL}^2+a^2\cos^2\theta, \quad
  \Delta = r_{\rm BL}^2-2Mr_{\rm BL}+a^2 \,.
\end{align}
This geometry describes a rotating BH with mass $M$ and angular
momentum $J=aM$.  Note that in order to ensure the regularity of 
the spacetime, i.e. the existence of an event horizon, the BH spin is
constrained by the Kerr bound $a/M\leq 1$.

A second consequence of our assumptions is
that the axionic coupling can be neglected.
This means that we effectively study
minimally coupled massive scalar and vector fields separately and our
results will describe small linearized fields around the Kerr background.
Any potential instability we find is consistent with the above assumptions
for timescales small enough such that the fields are small. Over long
timescales, the fields may grow to large amplitudes where our
assumption no longer remains valid and a non-linear study becomes
necessary. We postpone such non-linear evolutions to a future
investigation.

In our approximation, the scalar and vector field dynamics are governed
by the linearized version of Eqs.~\eqref{eq:MFEoMScalar}
and~\eqref{eq:MFEoMVector}
\begin{subequations}
\begin{align}
  \label{eq:MFEoMScalar1}
  & \left(\nabla_{\nu}\nabla^{\nu}-\mu_S^2\right)\Psi=0\,,\\
  \label{eq:MFEoMVector1}
  & \nabla_{\nu} F^{\mu\nu}+\mu_V^2A^\mu= 0\,,
\end{align}
\end{subequations}
while the Kerr metric (\ref{eq:MFKerrBL}) satisfies $G_{\mu \nu}=0$,
i.~e.~Eq.~(\ref{eq:MFEoMTensor}) linearized in $\Psi$ and $A_{\mu}$.

\vspace{0.2cm}

\noindent{\bf{\em {Evolution equations for scalar fields.}}}
Because we intend to solve the equations of motion
~\eqref{eq:MFEoMScalar1} and~\eqref{eq:MFEoMVector1} numerically,
it is convenient to reformulate them as time evolution problem.
For this purpose we employ the $3+1$-decomposition of the spacetime
(see e.g. \cite{Alcubierre:2008}) and consider the
background spacetime in generic $3+1$-form
\begin{align}
  \label{eq:MFlineelement}
  ds^2= & -\alpha^2 dt^2 + \gamma_{ij} (dx^i + \beta^i dt) (dx^j + \beta^j dt)
\,.
\end{align}
Here, $\gamma_{ij}$ is the spatial metric and $\alpha$ and $\beta^i$ are
the lapse function and shift vector which represent the coordinate or
gauge freedom of general relativity.
We introduce the conjugated momenta
\begin{align}
\label{eq:MFdefPi}
\Pi_R = & -\frac{1}{\alpha}(\p_t - \Lie_{\beta}) \Psi_R
\,,\quad
\Pi_I = -\frac{1}{\alpha}(\p_t - \Lie_{\beta}) \Psi_I
\,,
\end{align}
where $X_R := \Re(X)$ and $X_I := \Im(X)$ denote the real and imaginary
parts, respectively.
Definition~\eqref{eq:MFdefPi} provides evolution equations for the scalar field $\Psi$
\begin{align}
\label{eq:MFevolPsi}
\p_t \Psi_R = & \Lie_{\beta}\Psi_R - \alpha \Pi_R
\,,\quad 
\p_t \Psi_I= \Lie_{\beta}\Psi_I - \alpha \Pi_I
\,,
\end{align}
where $\Lie_{\beta}\Psi_{R,I} = \beta^k \p_k\Psi_{R,I}$. 
By applying the $3+1$-split to the Klein-Gordon equation
(\ref{eq:MFEoMScalar1}), we obtain the evolution equations
for the momentum
\begin{subequations}
\label{eq:MFevolPi}
\begin{align}
  \p_t \Pi_R = & \Lie_{\beta}\Pi_R - D^i\alpha D_i \Psi_R
                 + \alpha ( - D^i D_i \Psi_R + K \Pi_R + \mu_S^2 \Psi_R) \,\\
  \p_t \Pi_I = & \Lie_{\beta}\Pi_I - D^i\alpha D_i \Psi_I
                 + \alpha ( - D^i D_i \Psi_I + K \Pi_I + \mu_S^2 \Psi_I) \,,
\end{align}
\end{subequations}
where $\Lie_{\beta}\Pi_{R,I} = \beta^k \p_k \Pi_{R,I}$.  $D_i$ is the
covariant derivative associated with the $3$-metric $\gamma_{ij}$ and $K$
is the trace of the extrinsic curvature.

\vspace{0.2cm}
\noindent{\bf{\em {Evolution equations for vector fields.}}} 
%
We next apply the 3+1 decomposition to the evolution equation
(\ref{eq:MFEoMVector1}) and obtain
\begin{align}
  \label{eq:MFEoMA1}
  \nabla^{\nu}\nabla_{\mu} A_{\nu} - \nabla^{\nu}\nabla_{\nu} A_{\mu}
      + \mu_V^2 A_{\mu} = &
      - \left[  \nabla^{\nu}\nabla_{\nu} A_{\mu}
      - \nabla_{\mu}(\nabla^{\nu} A_{\nu})
      - R_{\mu}{}^{\nu} A_{\nu} - \mu_V^2 A_{\mu} \right]
      = 0
\,.
\end{align}
By operating with $\nabla^{\mu}$ on Eq.~\eqref{eq:MFEoMA1} it is
straight-forward to show that the Lorenz gauge
\begin{align}
  \label{eq:MFLG}
  \nabla^{\mu} A_{\mu} = & 0
\,
\end{align}
{\it needs} to be satisfied. For a vacuum background spacetime as considered
in our work, we also have $R_{\mu \nu}=0$ and Eq.~(\ref{eq:MFEoMA1})
simplifies to
\begin{equation}
  \nabla^{\nu} \nabla_{\nu} A_{\mu} - \mu_V^2 A_{\mu} = 0\,.
  \label{eq:DDA}
\end{equation}
Note, that in case of a non-vanishing cosmological constant $\Lambda$,
$R_{\mu \nu} = \Lambda g_{\mu \nu}$ and in place of Eq.~(\ref{eq:DDA})
we would obtain
\begin{equation}
  \nabla^{\nu} \nabla_{\nu} A_{\mu} - (\Lambda + \mu_V^2) A_{\mu} = 0\,,
\end{equation}
i.~e.~the cosmological constant enters as an additional ``mass''-like term. In
particular, it changes the evolution equation of a massless vector field
to that of a massive one. 
Note, however, that the Maxwell equations in Kerr-(anti-)de Sitter background
are known to be separable and can be written in the form of a Teukolsky
type equation \cite{Chambers:1994ap,Giammatteo:2005vu,Cardoso:2006wa}.
Therefore, one might expect that the equations of motion for a massive vector field
in a vacuum Kerr spacetime should also be separable. We emphasize, however,
that this analogy between a massless field with cosmological constant and
a massive one without needs to be taken with care:
a massless vector field has only 2 dynamical degrees of freedom
irrespective of the value of the cosmological constant,
whereas a massive vector field has 3. 
In fact, up to date, a separation of the equations of motion for a massive vector field
has not been accomplished.
It is not immediately obvious, for this reason, whether there exists
a well-defined correspondence between the two cases. In this work we
focus on $\Lambda=0$ and therefore leave a detailed investigation of
this question for future work.

We now apply the 3+1 decomposition to the vector field and split
$A_{\mu}$ into its spatial part and normal component
\begin{align}
\A_{\mu} =  \ga^{\nu}{}_{\mu} A_{\nu}\,,&\quad\text{and}\quad
\varphi =  - n^{\mu} A_{\mu}\,,
\end{align}
where $n^{\mu}$ is the vector normal to the spatial hypersurface $\Si$.
The vector field can be reconstructed from its projections
according to $A_{\mu} = \A_{\mu} + n_{\mu} \varphi$. Furthermore,
the projection of the Maxwell tensor along the normal vector $n^{\mu}$ 
yields the electric field
\begin{align}
  \label{eq:MFdefE}
  E_{\mu} = & F_{\mu\nu} n^{\nu} \,,
\end{align}
which is a purely spatial quantity, i.e., $E_{\mu} n^{\mu} = 0$.

With all the necessary ingredients at hand we now proceed by
performing the $3+1$-split Eqs.~\eqref{eq:MFLG} and \eqref{eq:DDA}.
In terms of the dynamical variables $\{\varphi,\A_i,E_i\}$
this procedure results in the constraint
\begin{align}
  \label{eq:MFCE}
  C_E = & D^i E_i + \mu^2_V \varphi = 0 \,.
\end{align}
and in the evolution equations
\begin{subequations}
  \label{eq:EvolProca}
  \begin{align}
    \label{eq:MFdtphi}
    (\p_t - \Lie_{\beta} ) \varphi = &
        - \alpha\left( D^i \A_i - K \varphi \right)-\A_i D^i\alpha \,, \\
    \label{eq:MFdtA}
    (\p_t - \Lie_{\beta})\A_i = &
        - \alpha \left( E_i + D_i \varphi\right) - \varphi D_i \alpha \,, \\
    \label{eq:MFdtE}
    (\p_t - \Lie_{\beta}) E_i = &
    \alpha\left(  \mu^2_V \A_i + K E_i - 2 E^j K_{ij}
        + D^j ( D_i \A_j - D_j \A_i) \right)
        + D^j\alpha ( D_i\A_j - D_j \A_i )    \,,
  \end{align}
\end{subequations}
where $\Lie_{\be}\varphi = \be^k\p_k\varphi$, $\Lie_{\be}\A_i
= \be^k\p_k\A_i + \A_k\p_i\be^k$ and
$\Lie_{\be}E_i = \be^k\p_k E_i + E_k\p_i\be^k$. 
%

\subsection{Background in horizon penetrating coordinates}
\label{ssec:MFbackground2}
%
In practice, it is convenient to employ horizon penetrating coordinates
and consider the Kerr spacetime
in Cartesian Kerr-Schild coordinates $(t,x,y,z)$. 
Without loss of generality, we assume the
angular momentum to point in the $z$ direction. Then, the Kerr-Schild form
of the lapse function $\al$, shift vector $\be^i$, $3$-metric $\ga_{ij}$
and extrinsic curvature $K_{ij}$ is given by
\begin{subequations}
  \label{eq:MFKSlapse}
  \begin{align}
    \alpha = & (1 + 2 H l^t l^t)^{-1/2} \,,\quad
    \beta^i = -\frac{2 H l^t l^i}{1 + 2 H l^t l^t} \,,\quad
    \gamma_{ij}=\delta_{ij} + 2 H l_i l_j \,,\\
    K_{ij} = & - \frac{1}{\alpha} (l_i l_j \p_t H + 2 H l_{(i}\p_t l_{j)} )
             - 2 \alpha
             \left( \p_{(i} ( l_{j)} H l^t )+ 2 H^2 l^t l^k l_{(i}
             \p_{|k|}l_{j)}+ H l^t l_i l_j l^k \p_k H \right) \,,
  \end{align}
\end{subequations}
where
\begin{align}
  \label{eq:MFHfunc}
  H = & \frac{M r_{\rm BL}^3}{r_{\rm BL}^4 + a^2 z^2}
  \,,\quad
  l_{\mu} = \left( 1, \frac{r_{\rm BL} x + a y}{r_{\rm BL}^2 + a^2},
            \frac{r_{\rm BL} y - a x}{r_{\rm BL}^2 + a^2}, \frac{z}{r_{\rm BL}} \right)
            \,.
\end{align}
and the Boyer-Lindquist radial coordinate $r_{\rm BL}$
is related to the Cartesian
Kerr-Schild coordinates by
\begin{align}
  \label{eq:MFdefrBL}
  \frac{x^2 + y^2}{r_{\rm BL}^2 + a^2} + \frac{z^2}{r_{\rm BL}^2} = & 1 \,.
\end{align}
%

\subsection{Initial data}\label{ssec:MFinitdata}
In this work we consider two types of initial configurations: 
(i) generic pulses of Gaussian shape and 
(ii) bound states which are particularly suitable for identifying putative instabilities.
We describe each of these initial data in detail.

\noindent{\bf{\em {Gaussian initial data.}}}
We specify Gaussian wave packets of the form
\begin{subequations}
  \begin{align}
    \label{eq:MFinitdataQNM}
    \Psi(t=0) = & 0 
\,,\quad
    \Pi(t=0)=\exp{\left(-\frac{(r-r_0)^2}{w^2}\right)}\,\,
        {_0}\Sigma(\theta, \phi) 
\,,\\
    \label{eq:MFID_A}
    \varphi(t=0) = & 0 
\,,\quad
    \A_i(t=0)=\exp{\left(-\frac{(r-r_0)^2}{w^2}\right)}\,\,
        {_{-1}}\Sigma(\theta,\phi) \,,\quad 
    E_i(t=0)=0 \,,
\quad i=1,\,2,\,3 \,,\quad
  \end{align}
\end{subequations}
where $r=\sqrt{x^2 + y^2 + z^2}$ is the Kerr-Schild radial coordinate. 
$r_0$ and $w$ are the center and width of the Gaussian,
while
${_0}\Sigma(\theta,\phi)$ and
${_{-1}}\Sigma(\theta,\phi)$
represent superpositions of spherical harmonics
${}_sY_{l m}(\theta,\phi)$ of spin weight $s=0$ and $s=-1$, respectively.
Expressed in Cartesian coordinates $(x,y,z)$
\begin{align}
  \label{eq:CartCoords}
  & x = r \sin\theta \cos\phi \,,\quad
    y = r \sin\theta \sin\phi \,,\quad
    z = r \cos\theta \,,
\end{align}
the spin-weighted spherical harmonics up to $l=2$ are given by
Eqs.~\eqref{eq:s0Y00}-\eqref{eq:sm1Y} in 
Appendix~\ref{app:Ylmm}.

\noindent{\bf{\em {Bound state initial data.}}}
%
Our second type of initial data is given by
the quasi-bound states of massive scalar fields around BHs
which represent long-lived modes of massive scalar field perturbations
around Schwarzschild or Kerr BHs and have been studied extensively
in the literature in the frequency domain
\cite{Cardoso:2005vk,Konoplya:2006br,Dolan:2007mj,Berti:2009kk,
Pani:2012vp,Pani:2012bp,Detweiler:1980uk}.
Our particular interest in these modes arises from their pure nature;
they represent potentially superradiant, single-frequency states.
By specifying single-mode states of this type, we are able to suppress 
interference or beating effects of the kind discussed below for Gaussian
initial data. Evolutions of such modes additionally serve as a useful
test for our code \cite{Yoshino:2012kn}.

There exist powerful and simple methods
to construct the bound states for massive scalars, either by direct
numerical integration or a continued fraction approach
\cite{Leaver:1986gd,Cardoso:2005vk,Dolan:2007mj,Berti:2009kk}.
Here, we adopt Leaver's continued fraction method
and obtain in Boyer-Lindquist coordinates
\begin{align}
  \Psi_{lm} = & e^{-\imath \omega t_{\rm BL}}
      e^{-\imath m \phi_{\rm BL}} S_{lm}(\theta_{\rm BL}) R_{lm}(r_{\rm BL}) \,.
\end{align}
Here, $S_{lm}(\theta_{\rm BL})$ are spheroidal harmonics \cite{Berti:2005gp} and
the radial dependence is given by
\begin{align}
  R_{lm}(r_{\rm BL}) = & (r_{\rm BL}-r_{\rm BL,+})^{-\imath \sigma}
      (r_{\rm BL} - r_{\rm BL,-})^{\imath \sigma + \chi -1} e^{r_{\rm BL} q}
      \sum_{n=0}^{\infty} a_n
      \left(\frac{r_{\rm BL}-r_{\rm BL,+}}{r_{\rm BL}-r_{\rm BL,-}}\right)^n
      \,,
\end{align}
with
\begin{align}
\sigma = & \frac{2r_{BL,+} (\omega - \omega_c)}{r_{BL,+} - r_{BL,-}} 
\,,\quad 
q = \pm \sqrt{\mu_S^2 - \omega^2} 
\,,\quad 
\chi=\frac{\mu_S^2 - 2\omega^2}{q}
\,.
\end{align}
$r_{\rm BL,\pm} = M \pm \sqrt{M^2-a^2}$ are the radii of the inner and outer horizon
and $\omega_c = m\Omega_H = m\frac{a}{2 M r_{\rm BL,+}}$ is  the critical frequency for superradiance. 
All remaining terms in this expression are known in closed form and
the characteristic frequency $\omega$ can be obtained by solving a
three-term recurrence relation
for the coefficients $a_n$ given by, e.g., Eqs.~(35)-(48) of \cite{Dolan:2007mj}.
For our purposes, we still need to transform these results from
Boyer-Lindquist to Kerr-Schild coordinates 
denoted in this discussion for clarity by a subscript ``KS'',
\begin{align}
  dt_{\rm KS} = & dt_{\rm BL} + \frac{2M r_{\rm BL}}{\Delta} dr_{\rm BL} \,,\quad
  dr_{\rm KS} = dr_{\rm BL} \,,\quad
  d\theta_{\rm KS} = d\theta_{\rm BL} \,,\quad
  d\phi_{\rm KS} = d\phi_{\rm BL} + \frac{a}{\Delta} dr_{\rm BL} \,.
\end{align}
Then, the bound state scalar field is given by
\begin{align}
  \label{eq:MFinitialBS}
  \Psi_{lm} = & e^{-\imath \omega t_{\rm KS}} (r_{\rm KS}-r_{KS,+})^A
      (r_{\rm KS}-r_{\rm KS,-})^B
      \left(\frac{r_{\rm KS}-r_{\rm KS,+}}{r_{\rm KS}-r_{\rm KS,-}}\right)^C
      Y_{lm}(\theta_{\rm KS},\phi_{\rm KS}) R_{lm} \,,
\end{align}
where $A = -\frac{2\imath \omega M r_{\rm KS,+}}{r_{\rm KS,-}-r_{\rm KS,+}}$, 
$B= \frac{2\imath \omega M r_{\rm KS,-}}{r_{\rm KS,-}-r_{\rm KS,+}}$, 
$C = \frac{\imath m a}{r_{\rm KS,-}-r_{\rm KS,+}}$.
Henceforth we will drop the subscript ``KS'' and denote the
Kerr-Schild radius and angles by $(r,\,\theta,\,\phi)$.

The corresponding conjugated momenta $\Pi_{lm}$ are computed from their
definition, Eq.~\eqref{eq:MFdefPi}. Unfortunately, a simple
construction of these modes exists only for scalar fields,
while for vector fields a fully numerical procedure is required
\cite{Cardoso:2005vk,Dolan:2007mj,Pani:2012vp,Pani:2012bp}. Partly for
this reason and partly because Gaussian initial data turn out to be adequate
for the identification of superradiant instabilities we will not consider
vector bound states in the remainder of this work.

\subsection{Wave extraction and output}\label{sec:waveoutput}

The main diagnostic quantities extracted from our simulations are the
radiated scalar and vector waves. The scalar multipoles are directly
obtained from interpolating the fields $\psi$ and $\Pi$ onto spheres
of constant coordinate radius $r=r_{\rm ex}$ and projecting onto
$s=0$ spherical harmonics according to
\begin{align}
  \label{eq:MFmodedecompScalar}
  \Psi_{lm}(t) = \int d\Omega \Psi(t,\theta,\phi) Y^{\ast}_{lm}(\theta,\phi)
  \,, &\quad
  \Pi_{lm}(t)  = \int d\Omega \Pi(t,\theta,\phi)  Y^{\ast}_{lm}(\theta,\phi)
  \,.
\end{align}
For vector fields, we construct the gauge invariant Newman Penrose scalar
\cite{Newman:1961qr} 
(see also, e.g., \cite{Palenzuela:2009hx,Mosta:2009rr,Zilhao:2012gp} for 
recent applications in numerical simulations).
\begin{align} \label{eq:MFNP2}
  \Phi_2 = & F_{\mu\nu} \ell^{\mu} \bar{m}^{\nu} \,,
\end{align}
where $\ell^{\mu} = \frac{1}{\sqrt{2}}(n^{\mu} - u^{\mu})$ and
$\bar{m}^{\mu} = \frac{1}{\sqrt{2}}(v^{\mu} - \imath w^{\mu})$ are
vectors of a null tetrad. In practice, the vectors of the null tetrad are
constructed from a Cartesian orthonormal basis $\{u^i, v^i, w^i\}$ on the
spatial hypersurface and the timelike orthonormal vector $n^{\mu}$.
Together with the reconstruction of the Maxwell tensor from
\begin{align}
  F_{\mu\nu} = & n_{\mu} E_{\nu} - n_{\nu} E_{\mu}
      + D_{\mu} \A_{\nu} - D_{\nu} \A_{\mu}
\,
\end{align}
we can straightforwardly derive the Newman-Penrose scalar $\Phi_2$
whose real and imaginary components are given by
\begin{align}
  \Phi^R_2 = & - \frac{1}{2} 
             \left[  E^R_i v^i + u^i v^j (D_i\A^R_j - D_j\A^R_i)
             + E^I_i w^i + u^i w^j (D_i\A^I_j - D_j\A^I_i) \right]
  \,,\\
  \Phi^I_2 = & \frac{1}{2} 
               \left[  E^R_i w^i + u^i w^j (D_i\A^R_j - D_j\A^R_i)
                     - E^I_i v^i - u^i v^j (D_i\A^I_j - D_j\A^I_i) \right]
  \,.
\end{align}
Finally, we obtain the multipoles by projecting
$\Phi_2$ onto $s=-1$ spin-weighted spherical harmonics
\begin{align}
  \Phi^R_{2_{lm}}(t) = & \int d\Omega \left[
        \Phi^R_2(t,\theta,\phi) _{-1}Y^R_{lm}(\theta,\phi)
        + \Phi^I_2(t,\theta,\phi) _{-1}Y^I_{lm}(\theta,\phi) \right]
  \,,\\
  \Phi^I_{2_{lm}}(t) = & \int d\Omega \left[
        \Phi^I_2(t,\theta,\phi) _{-1}Y^R_{lm}(\theta,\phi)
        - \Phi^R_2(t,\theta,\phi) _{-1}Y^I_{lm}(\theta,\phi) \right]
  \,.
\end{align}
%

\subsection{Numerical implementation}\label{ssec:MFImplementation}
We have implemented this framework as a module {\textsc{Lin-Lean}}
in the {\textsc{Lean}} code \cite{Sperhake:2006cy},
which is based on the {\textsc{Cactus}} computational toolkit
\cite{Goodale02a, cactus} and the {\textsc{Carpet}} mesh refinement
package \cite{Schnetter:2003rb, carpet}.
The evolution equations are integrated in time using the
method of lines with a fourth-order Runge-Kutta scheme
and fourth-order spatial discretization on all refinement levels
except the innermost where we excise the black-hole singularity
and use second-order accurate stencils instead.
The excision is implemented by removing a ``legosphere'' of radius
$r_{\rm ex} \le 1M$ centered on the singularity
\cite{Shoemaker:2003td,Sperhake:2005uf} and therefore guaranteed to be
inside the event horizon for all values of $a$.
Inside the excision region, we set all evolution variables
to their flat spacetime values
$\alpha = 1$, $\beta^i=0$, $\gamma_{ij} = \delta_{ij}$ and $K_{ij} = 0$,
and extrapolate values onto the excision boundary from
the exterior regular grid; see \cite{Shoemaker:2003td,Sperhake:2005uf}
for more details of this procedure.

\section{Scalar field evolutions}\label{sec:SFresults}
%
\begin{table}
\begin{center}
\begin{tabular}{lcccc}
\hline
Run             &$M \mu_S$ & $a/M$  & $\,_0\Sigma(\theta,\phi)$ & Grid Setup   \\ 
\hline
sS\_m000        &0.00      &0.00    &$Y_{00}+Y_{10}+Y_{1-1}-Y_{11}$      & $\{(384,192,96,48,24,12,6,3,1.5),~h=M/100\}$ \\
\hline
sK\_m000        &0.00      &0.99    &$Y_{00}+Y_{10}+Y_{1-1}-Y_{11}$      & $\{(384,192,96,48,24,12,6,3,1.5),~h=M/108\}$ \\
\hline
sS\_m001        &0.10      &0.00    &$Y_{11}$ & $\{(1024,512,256,128,64,32,8,4,2),~h=M/40\}$ \\
sS\_m042        &0.42      &0.00    &$Y_{10}+Y_{11}+Y_{20}+Y_{22}$ 
& $\{(1536,384,192,96,48,24,12,6,3,1.5),~h=M/60\}$ \\
sS\_m100        &1.00      &0.00     &$Y_{11}$ & $\{(1024,512,256,128,64,32,8,4,2), 1/40\}$ \\
\hline
sK\_m042$_c$ &0.42      &0.99    &$Y_{00}+Y_{10}+Y_{1-1}-Y_{11}$ & $\{(1536,384,192,96,48,24,12,6,3,1.5),~h=M/60\}$ \\
sK\_m042$_m$ &0.42      &0.99    &$Y_{11,1-1}$ & $\{(1536,384,192,96,48,24,12,6,3,1.5),~h=M/72\}$ \\
sK\_m042$_f$ &0.42      &0.99    &$Y_{11,1-1}$ & $\{(1536,384,192,96,48,24,12,6,3,1.5),~h=M/84\}$ \\
\hline
\end{tabular}
\end{center}
\caption{\label{tab:MFScaMassiveSetup} 
Initial setup for simulations of a scalar field with Gaussian initial data
located at $r_0=12~M$ and with width $w= 2M$ in a Schwarzschild or Kerr
background.  We denote the mass parameter $M\mu_S$, the dimensionless spin
parameter $a/M$, the modes of the initial pulse $\,_0\Sigma(\theta,\phi)$
and the specific grid setup, measured in units of the BH mass $M$,
following the notation of Sec.~II E in \cite{Sperhake:2006cy}.
}
\end{table}

In this section we report our results obtained for the evolution of
scalar fields in BH background spacetimes. For this purpose,
we have evolved Gaussian wave pulses of width $w=2~M$ centered
at $r_0=12~M$ around either a non-rotating Schwarzschild BH or
a rapidly spinning Kerr BH with $a/M=0.99$. The set of our simulations
is summarized in Table~\ref{tab:MFScaMassiveSetup}.

As mentioned in the introduction, the resulting signal is
typically composed of three stages, a transient, an
exponential ringdown phase and late-time tails.
Whereas we identify these stages clearly in evolutions of massless
scalar fields, the massive case exhibits a richer phenomenology
which we will discuss in detail further below. First, however, we summarize
results from the literature and present tests of our numerical
infrastructure.

\subsection{A summary of results in the literature}
%
\begin{table}
\begin{center}
\begin{tabular}{ccllc}
\hline
$M \mu_S$& $a/M$   & $M \omega_{11}$ ($n=0$)     & $M \omega_{11}$ ($n=1$) & $M\delta_{10}$   \\ 
\hline
$0.00$ & $0.00$  & $0.2929-i0.09766$           & $0.2645-i0.3063$               & $0.0284$\\ 
$0.00$ & $0.99$  & $0.4934-i0.03671$           & $0.4837-i0.09880$              & $0.0097$\\ 
\hline
$0.42$ & $0.00$  & $0.4075-i0.001026$          & $0.4147-i0.0004053$            & $0.0072$\\ 
$0.42$ & $0.99$  & $0.4088+i1.504\cdot10^{-7}$ & $0.4151+i5.364\cdot10^{-8}$    & $0.0063$\\ 
\hline
$1.00$ & $0.00$  & $0.9222-i0.11842$           & $0.9570-i0.04792$             & $0.0348$\\ 
$1.00$ &  $0.99$ & $0.8765-i0.18994$          &  $0.9515-i0.06292$             & $0.075$
\\
\hline
\end{tabular}
\end{center}
\caption{\label{tab:MFScaAnaModes} Frequencies for the fundamental
 ($n=0$) and first overtone ($n=1$) modes of 
(i) the quasi-normal modes for $M\mu_S=0$ and 
(ii) the bound states for $M\mu_S=0.42, 1.0$
for a set of BH spins $a/M$
obtained with the continued fraction method.
The beating frequency $\delta_{10}$ is defined as
the difference of the real parts of the frequencies for $n=0$ and
$n=1$.
}
\end{table}

In the following discussion,
we will make frequent use of the characteristic frequencies of
the two lowest (fundamental and first overtone) modes of the dipole
as obtained from a linearized analysis in the frequency domain
\cite{Berti:2005ys,Berti:2009kk}. These values are summarized for
a set of chosen mass parameters $\mu_S$ of the scalar field and
our two choices of the rotation parameter $a/M$ in
Table~\ref{tab:MFScaAnaModes}.

In the case of massless perturbations the classification of oscillation
modes is relatively straightforward; there exists one family of
oscillation modes, the quasi-normal modes, 
characterized by an integer number and the ``lowest'' or fundamental mode is defined
as having the largest damping time. In fact,
massless perturbations are generally short-lived and decay fast (on timescales
of order of the BH mass)
unless the BH is rotating at nearly extremal rate.
In contrast, massive perturbations generally decay on larger timescales
and their class of oscillation modes contains a second family
with profiles concentrated near the effective potential well.
These long-lived modes, referred to as ``bound states''
\cite{Dolan:2007mj}, are conventionally ordered
by {\em decreasing} absolute value of the imaginary part, i.~e.~the fundamental oscillation
mode is defined as that with the shortest damping time. This is in contrast
to the usual quasinormal modes, which are conventionally ordered by increasing imaginary part.
The values for the fundamental ($n=0$) and the first
overtone ($n=1$) bound state mode obtained from the continued fraction method 
are listed in Table~\ref{tab:MFScaAnaModes}
together with the difference $\delta_{10}\equiv \omega^{n=1}_R-\omega^{n=0}_R$
which will play an important role in our interpretation of the
results further below.

In the regime of small mass couplings $M\mu_S$,
one finds an analytic approximation for the oscillation frequencies
of these bound states, resembling a hydrogen-like spectrum ~\cite{Pani:2012vp,Pani:2012bp}, 
\begin{subequations}
  \label{eq:boundstates}
  \begin{align}
  M\omega_R = & M\mu_S - \frac{(M\mu_S)^3}{2(l+n+1)^2}
  \,,\\
  %
  M\omega_I = & -\imath \delta\nu \left(\frac{M\mu_S}{l+n+1}\right)^3
\,,
  \end{align}
\end{subequations}
where $\delta\nu$ is given by Eq.~(C8) of \cite{Pani:2012bp}.

At late times, after the relaxation of the BH, power-law tails arise \cite{Price:1971fb,Leaver:1986gd,Ching:1995tj,Zenginoglu:2012mc}.
Power-law tails exist both in flat and in curved spacetime and are generically caused by a failure of Huygen's principle.
In the case of flat spacetime, they can arise in massive interactions \cite{feshbach} or even with massless fields in odd-dimensional spacetimes
due to the peculiar nature of the Green's function \cite{Cardoso:2003jf}.
Curved backgrounds give generically rise to power-law tails which are a consequence of the continued
scattering of the scalar field off the background curvature. These tails have the form
\begin{align}
  \label{eq:massivetails}
  \Psi \sim t^{p}\sin(\mu_S t)\,.
\end{align} 
For massless scalar fields, one finds one exponent $p$ given by
\begin{align}
  \label{eq:MFScaPLmassless}
  p=-(2l + 3)\,.
\end{align} 
Tails of massive perturbations, however, exhibit a more complex behaviour;
details depend on the mass parameter $\mu_S$
\cite{Hod:1998ra,Koyama:2001ee,Koyama:2001qw,Burko:2004jn}, but their intermediate
and late-time behaviour is characterized by
\begin{subequations}
  \label{eq:MFScaPLmassive}
  \begin{align}
  \label{eq:MFScaPLmassiveA}
  p= -(l+3/2)\,,  & \quad {\textrm{at intermediate times}}\,,\\
  \label{eq:MFScaPLmassiveB}
  p= -5/6\,,      & \quad {\textrm{at very late times}}\,.
  \end{align}
\end{subequations}
These exponents are
loosely associated with the two timescales $M,\,\mu_S$ of the problem;
in particular, intermediate times refer to the window (we assume small $M\,\mu_S$) $M\mu_S \ll t\mu_S \ll (M\mu_S)^{-2}$.

Note, that the intermediate-time behavior is identical to the power-law
behavior of massive fields in {\it flat} spacetime
\cite{feshbach} whereas the late-time behavior is also affected by
scattering off the spacetime curvature. 

\subsection{Space dependent mass coupling and massless scalars}
\label{ssec:MFScaMuVar}
\begin{figure}
\begin{center}
\includegraphics[width=0.50\textwidth]{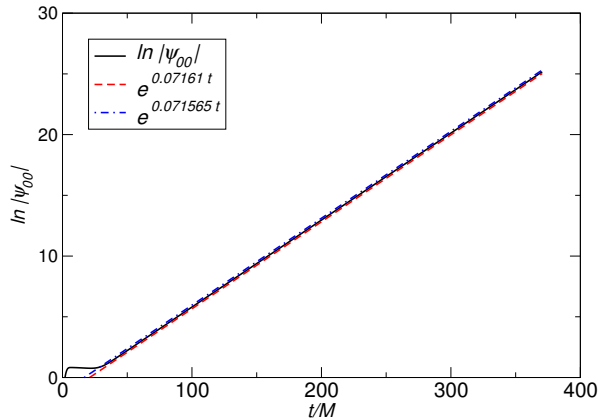}
\end{center}
\caption{\label{fig:MFSca_a0muVar}
Amplitude of the scalar field extracted at $r_{\rm ex}=10~M$ as function of
time obtained from the evolution of a scalar field initialized as
spherical shell
with space dependent mass $\mu_S^2 = -10M^2/r^4$,
around a non-rotating BH.}
\end{figure}

%
We next employ the results summarized in the previous section to test
our numerical framework.
For the first test, we consider the unphysical scenario of
a scalar field with space-dependent mass term $\mu_S^2=-10M^2/r^4$
in a Schwarzschild background. This choice is motivated by the
strong instability it generates and provides a unique and fast
setup to test the code in a particularly violent regime.
A mode analysis of the Klein-Gordon equation for this choice
of $\mu_S$ is straightforward and demonstrates the
existence of at least one unstable mode with time dependence
$\Psi\sim e^{0.071565 t}$.
The results obtained for the evolution of
a spherical shell described by $\Psi_{t=0}=\frac{1}{\sqrt{4\pi}}\exp\left(-\frac{(r-12)^2}{4}\right)$
are shown in Fig.~\ref{fig:MFSca_a0muVar} which shows the amplitude of the
scalar field extracted at $r_{\rm ex}=10~M$ as a function of time.
Our numerical results are consistent with an exponential growth,
$\Psi \sim e^{0.07161 t}$ in excellent agreement with the frequency
calculation.

\begin{figure}
\begin{center}
\begin{tabular}{ccc}
\includegraphics[width=0.5\textwidth]{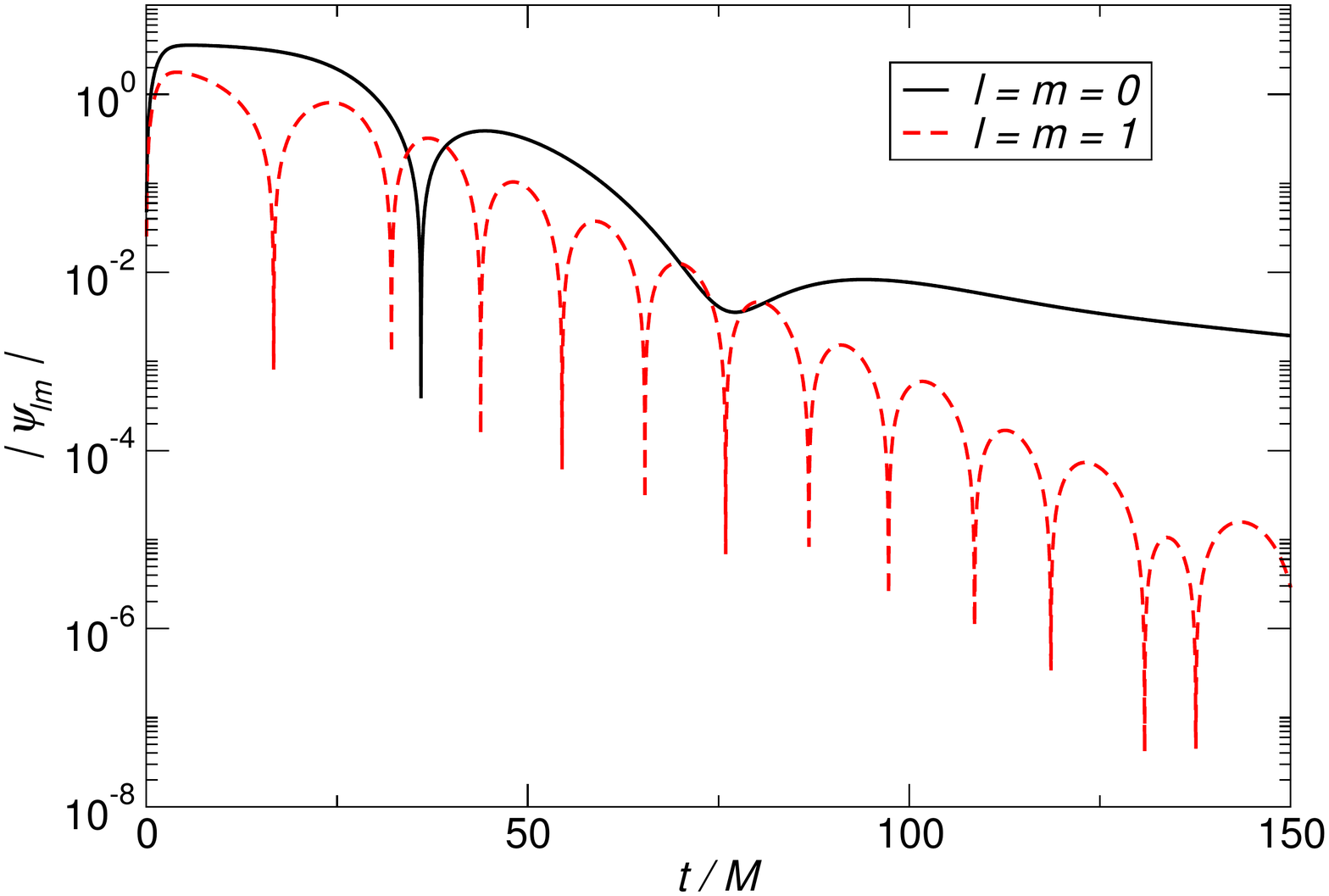} &
\includegraphics[width=0.5\textwidth]{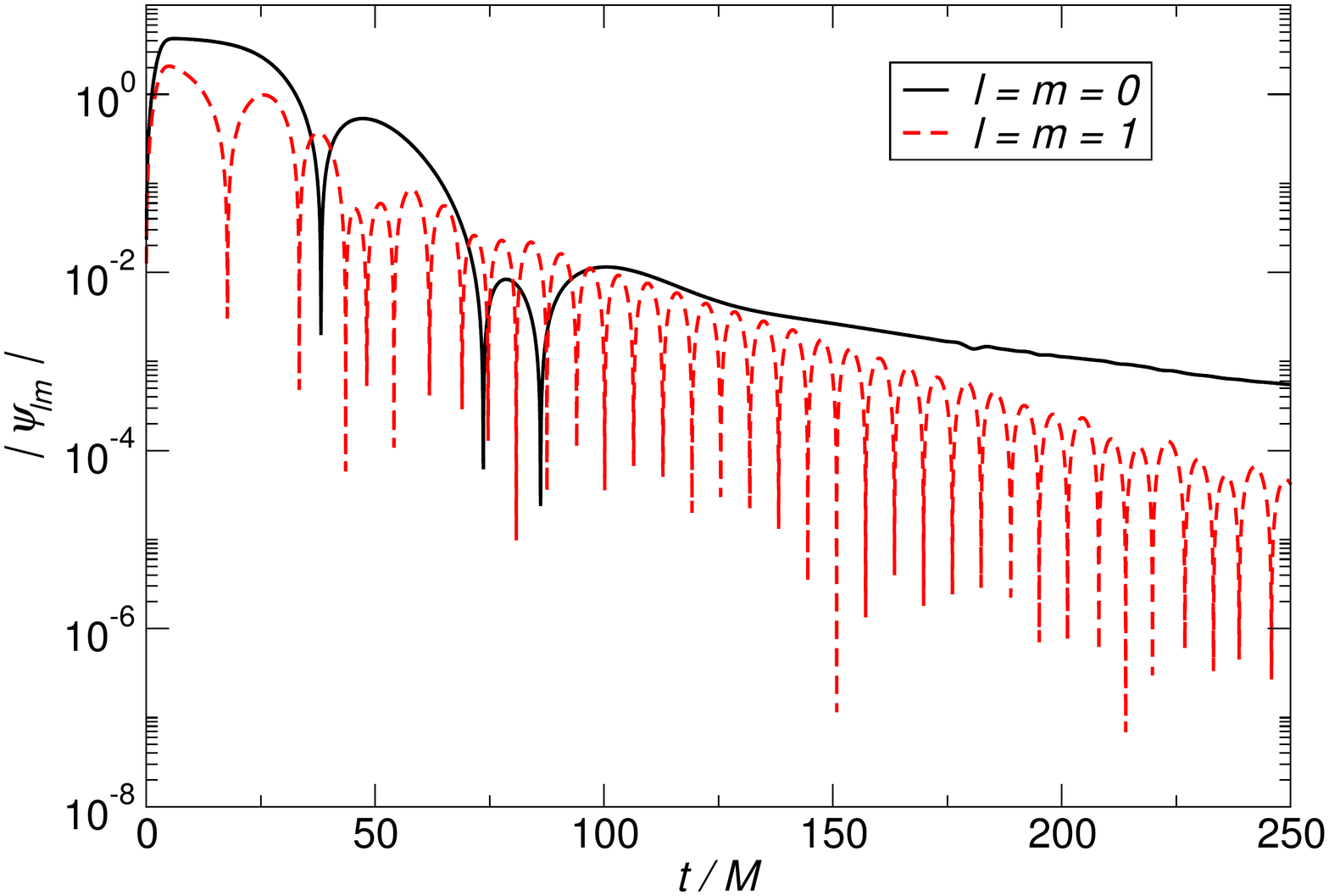} 
\end{tabular}
\end{center}
\caption{\label{fig:MFScaMLwaveform} 
Evolution of a Gaussian profile of a massless scalar field with width
$w=2~M$ centered at $r_0=12~M$ around a Schwarzschild BH (left panel)
and around a Kerr BH with $a/M = 0.99$ (right panel).
We depict the $l=0$ (solid black line) and $l=m=1$ (red dashed line) multipoles.
}
\end{figure}

A second test for our code is provided by the evolution of massless
scalar fields around BHs, a case
well studied in the literature \cite{Berti:2009kk,Konoplya:2011qq}.
For this purpose we have initialized the field by a Gaussian
with $r_0=12~M$ and $w=2~M$ and extracted the monopole and dipole
at $r_{\rm ex}=10$. The results are shown in Fig.~\ref{fig:MFScaMLwaveform} for a BH
background with $a/M=0$ and $a/M=0.99$, respectively.
The waveform displays the familiar features; an early transient
followed by an exponentially decaying sinusoid and a power-law tail
at late times.
A fit to the ringdown phase of the dipole
yields numerical QNM frequencies
within less than $2\%$ of the values in Table \ref{tab:MFScaAnaModes}.
Likewise, we obtain a numerical power-law tail of the form $t^{p}$ for the monopole,
with $p=-3.08$ for $a/M=0$ and $p=-3.07$ for $a/M=0.99$ in good agreement
with the prediction $p=-3$ obtained from the low-frequency expansion
of the wave equation underlying Eq.~\eqref{eq:MFScaPLmassless}.

Additionally, we have performed a convergence analysis for the more
challenging of these two cases, that of a highly rotating BH
background with $a/M=0.99$. 
This analysis provides the estimate of the numerical 
uncertainties $\De\Psi_{11}/\Psi_{11}\le8\%$ for the $l=m=1$ mode at
late times of the evolution and $\De\Psi_{00}/\Psi_{00}\le3\%$
for $l=m=0$.

%

\subsection{Massive scalar fields}

Having tested the numerical framework, we next explore the dynamics of
generic, massive scalar fields.  A mass term introduces a new scale
to the problem and new features appear in the evolution which depend
on the particular details of the initial configuration. Roughly, these
configurations can be classified into three groups.
\begin{itemize}
\item Bound state configurations. These configurations are characterized
      by unusually long-lived modes \cite{Detweiler:1980uk,Zouros:1979iw,
      Damour:1976,Furuhashi:2004jk,Cardoso:2005vk,
      Dolan:2007mj,Cardoso:2011xi,Yoshino:2012kn,Barranco:2012qs},
      and exist for any $m,\mu_S\neq 0$. These modes are described
      by Eq.~\eqref{eq:boundstates} for small $M\mu_S$.
      If they, furthermore, satisfy the condition $\omega_R<\mu_S \le m\Omega_H$
      they are subject to the superradiant instability.
\item Rapidly damped configurations.
      Generic initial profiles of massless or massive scalar fields
      typically decay on short time scales via
      quasi-normal ringdown followed by a power-law tail.
\item Beating regime. 
      Additionally, massive fields may exhibit long-lived,
      strongly modulated oscillations which
      result from the interplay between different overtones of the
      same multipole. 
\end{itemize}
In the following we will discuss numerical evolutions of initial configurations
for each of these classes in more detail.

\subsubsection{Bound states}\label{ssec:MFScaBS}
%
\begin{figure}[htpb!]
\begin{center}
\begin{tabular}{ccc}
\includegraphics[width=0.33\textwidth]{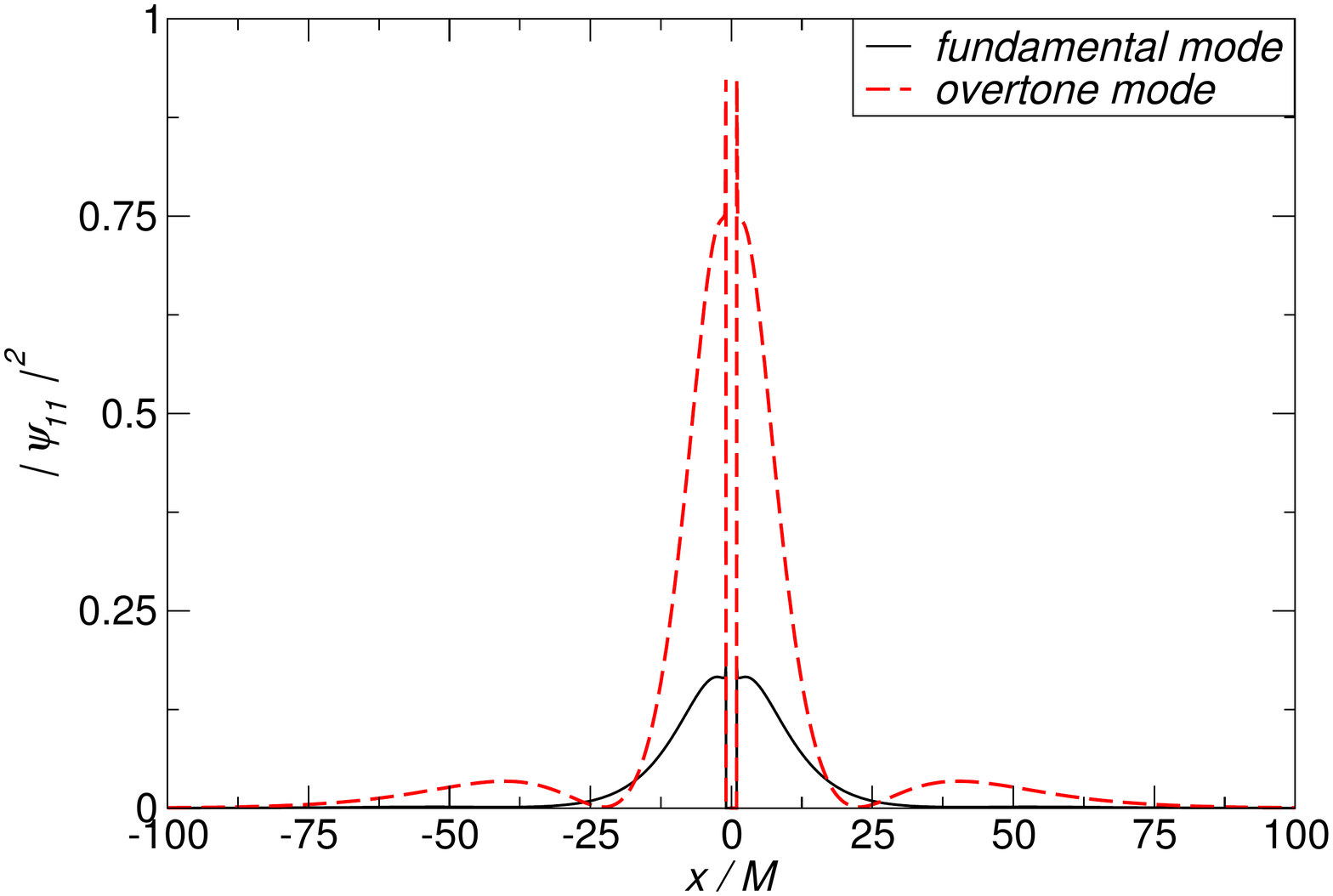} &
\includegraphics[width=0.33\textwidth]{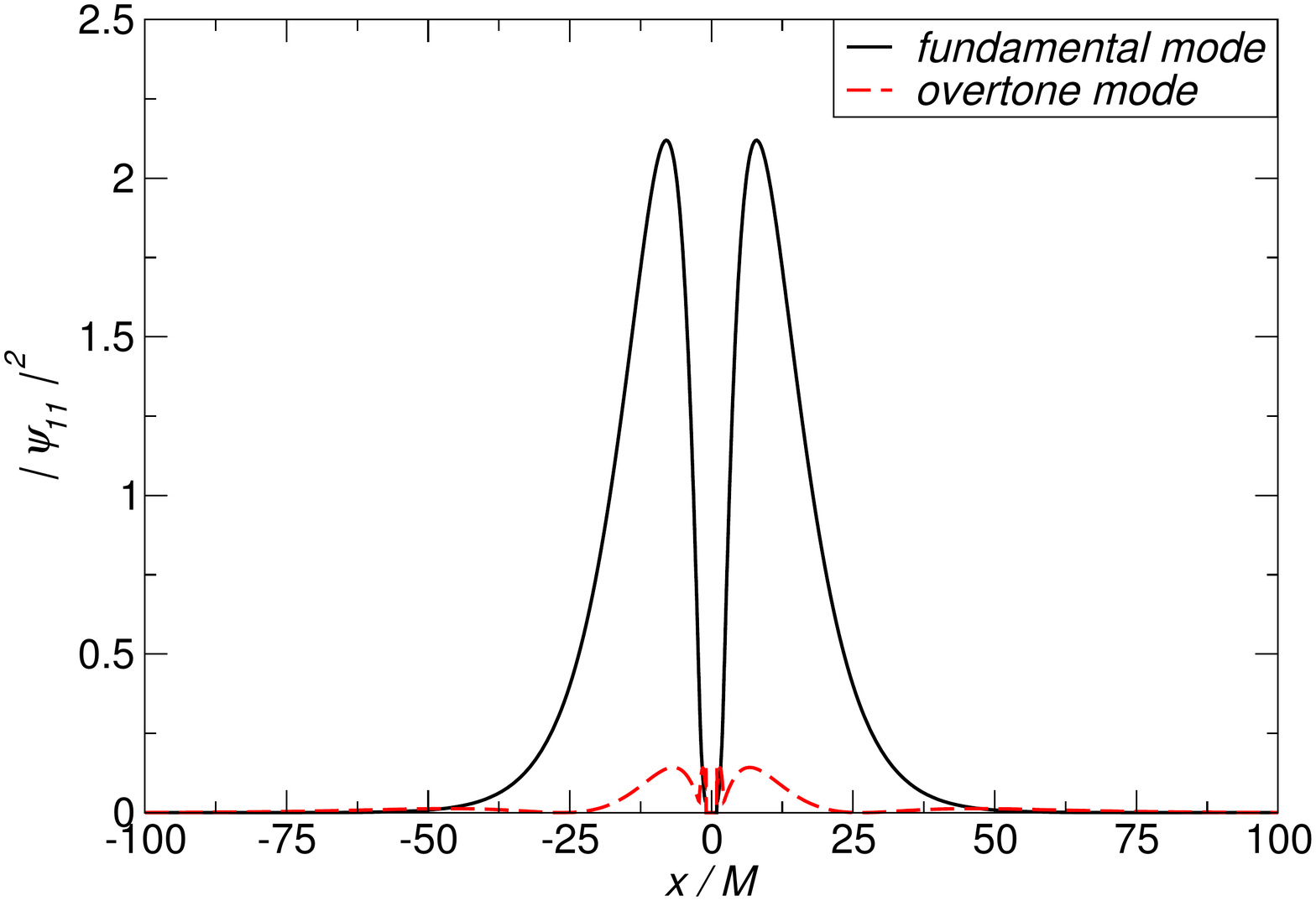} &
\includegraphics[width=0.33\textwidth]{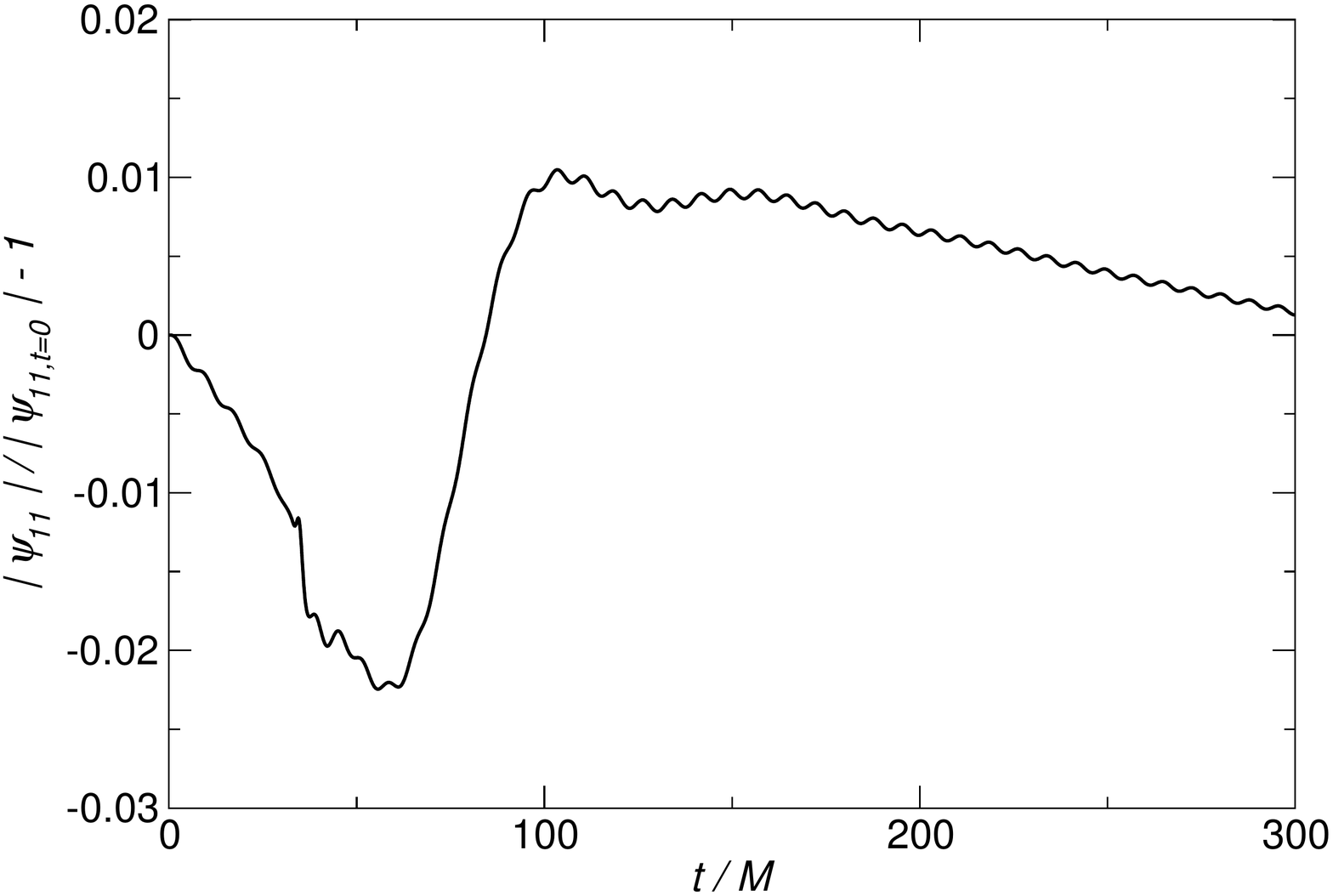} 
\end{tabular}
\end{center}
\caption{\label{fig:MFScaBSID}
Left and center: 
Initial profile of $|\Psi_{11}|^2$ for a bound state $m=1$ dipole configuration
with $M\mu_S=0.42$ in a Schwarzschild (left panel) and $a/M=0.99$
Kerr background (mid panel).  Black solid lines represent the fundamental
mode and red dashed lines the first overtone.
Right:
Relative change of the modulus of the $(1,1)$ mode extracted
at $r_{\rm ex}=20~M$.
}
\end{figure}

We have constructed bound-state initial configurations
for fields with a mass coupling $M\mu_S=0.42$, cf. Eq.~\eqref{eq:MFinitialBS},
following \cite{Dolan:2007mj,Cardoso:2005vk,Berti:2009kk},
and evolved them in a Schwarzschild or a
Kerr background with $a/M=0.99$.
In the left and center panels of Fig.~\ref{fig:MFScaBSID} we show the modulus of
the fundamental mode and first overtone of the $m=1$ dipole bound state
along the $x$-axis; cf.~Table~\ref{tab:MFScaAnaModes}. 
For non-rotating BHs the fundamental bound state is localized near the origin,
whereas its maxima are shifted to larger radii as the rotation
parameter $a/M$ is increased.
For a highly spinning BH with $a/M=0.99$, the fundamental mode is peaked at
around $r\sim 12~M$. For the first overtone we observe a node
at $r_{\rm node}\sim 22.5~M$ and at $r_{\rm node}\sim26.5~M$ for $a/M=0$ and
$a/M=0.99$, respectively. The specific structure of the bound-state
profiles will play an important role in our analysis of beating effects
below.

Throughout the time evolution, we expect the bound state scalar field
to remain localized in the vicinity of the BH.  Thus, by construction,
its absolute value $\Psi_{11}\Psi_{11}^{\ast} \sim \exp(-\imath \omega
t)\exp(\imath \omega t) \sim const.$ should remain almost constant
in time, with a small growth rate of $M\omega_I\sim1.5\cdot 10^{-7}$
\cite{Dolan:2007mj,Cardoso:2005vk}.  This behavior is confirmed in
the animations generated from our numerical data and
made available online \cite{webpageAnimation}.
Here we have tested these properties numerically by extracting the
dipole mode as a function of time at $r_{\rm ex}=20~M$.
The result is shown in the right panel of Fig.~\ref{fig:MFScaBSID}
for a Kerr background with $a/M=0.99$.
The scalar field varies by less than $\sim 2\%$
until $t\sim 200~M$ and by less than $\sim 1\%$ at late times,
which is within the numerical uncertainties.

We note that bound states are unstable states, but have a long instability
time scale of $\sim10^{7}M$ (see Table~\ref{tab:MFScaAnaModes}),
about three orders of magnitude larger than the evolution times
feasible within our framework. Over the time range covered
in the figure the instability has not yet generated a visible growth
in amplitude.

\subsubsection{Damped states: ringdown and tails}
\label{ssec:MFScaMassiveSW}

%
\begin{figure}
\begin{center}
\begin{tabular}{cc}
\includegraphics[width=0.50\textwidth]{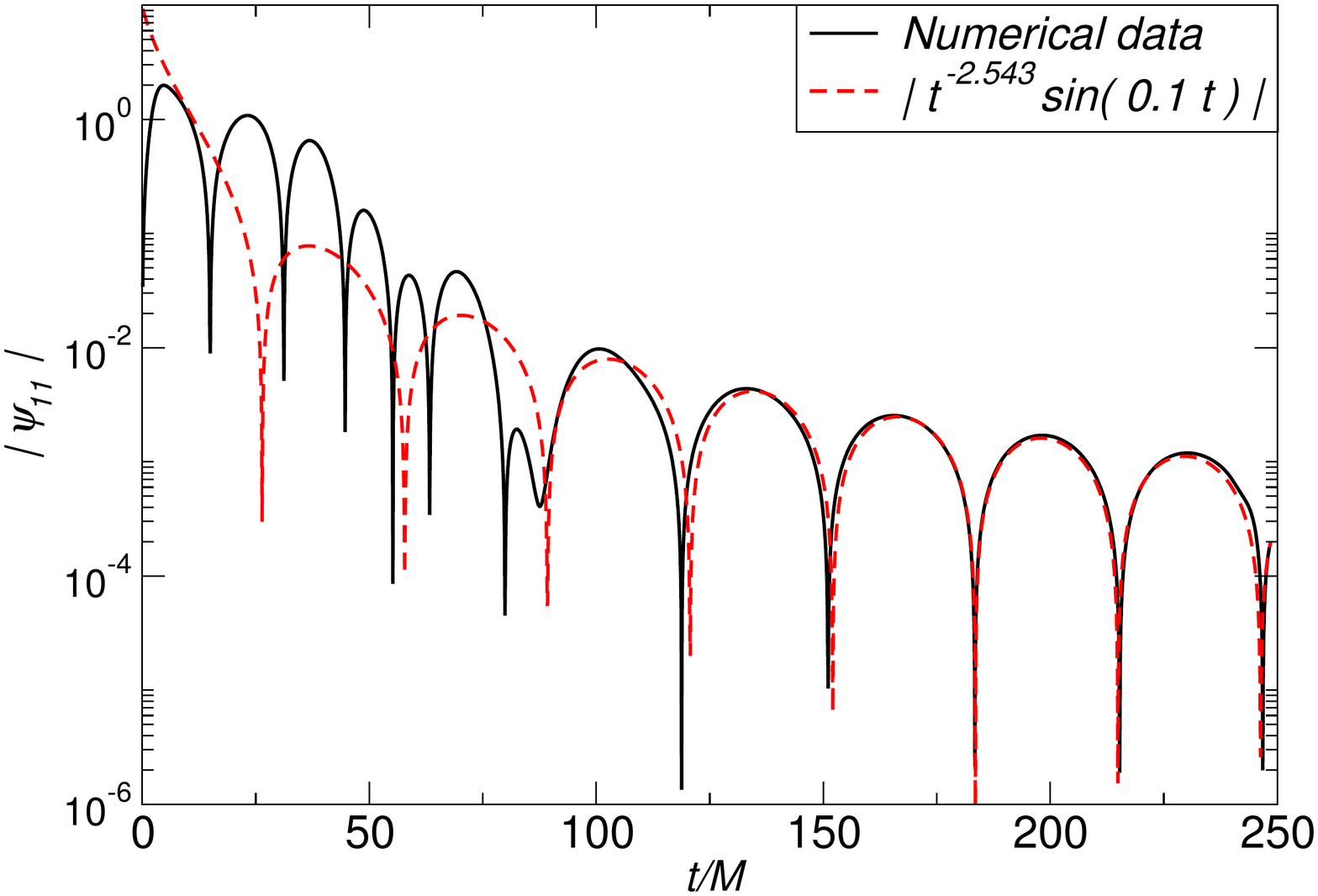} & 
\includegraphics[width=0.50\textwidth]{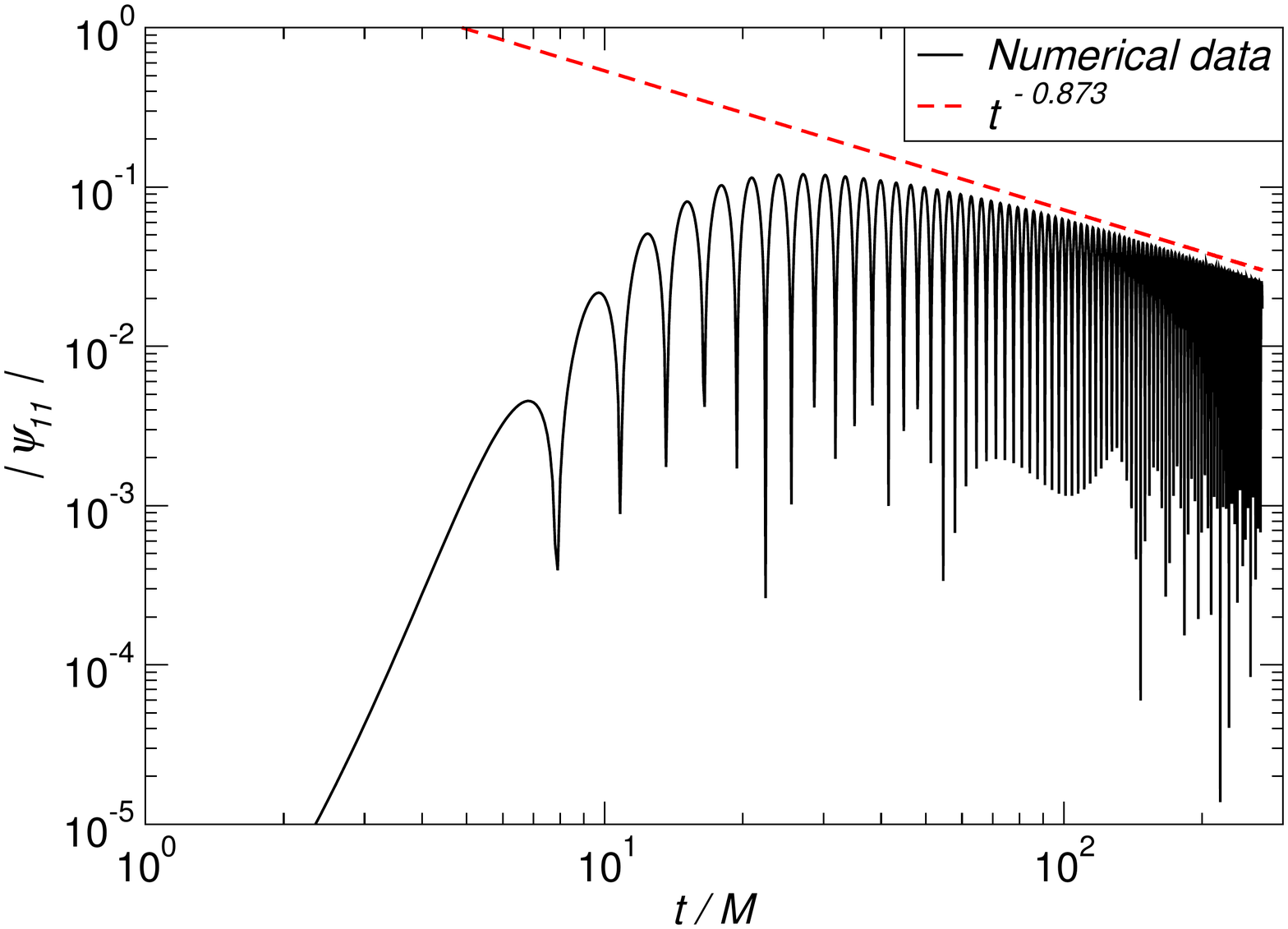} \\ 
\includegraphics[width=0.50\textwidth]{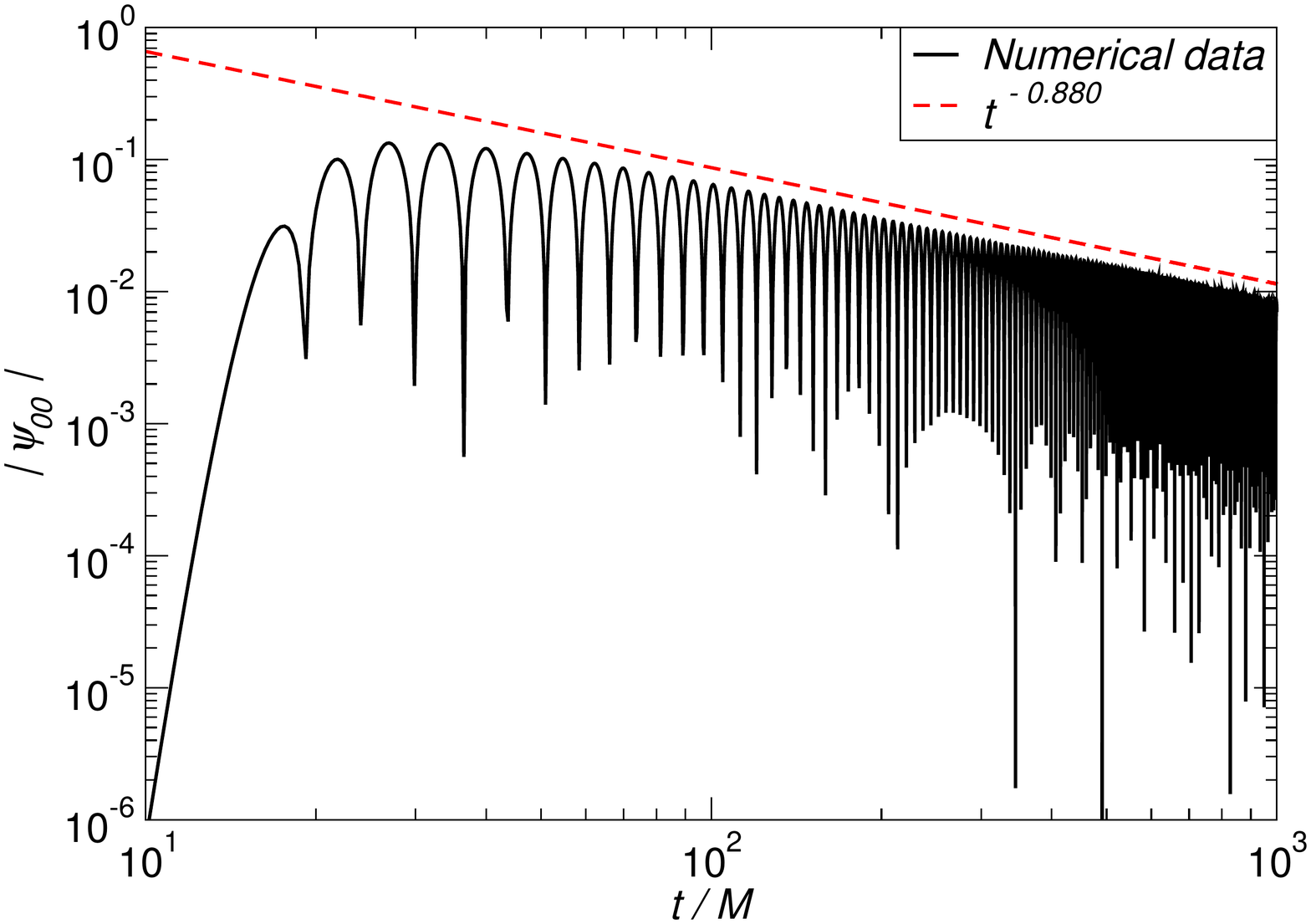} & 
\includegraphics[width=0.50\textwidth]{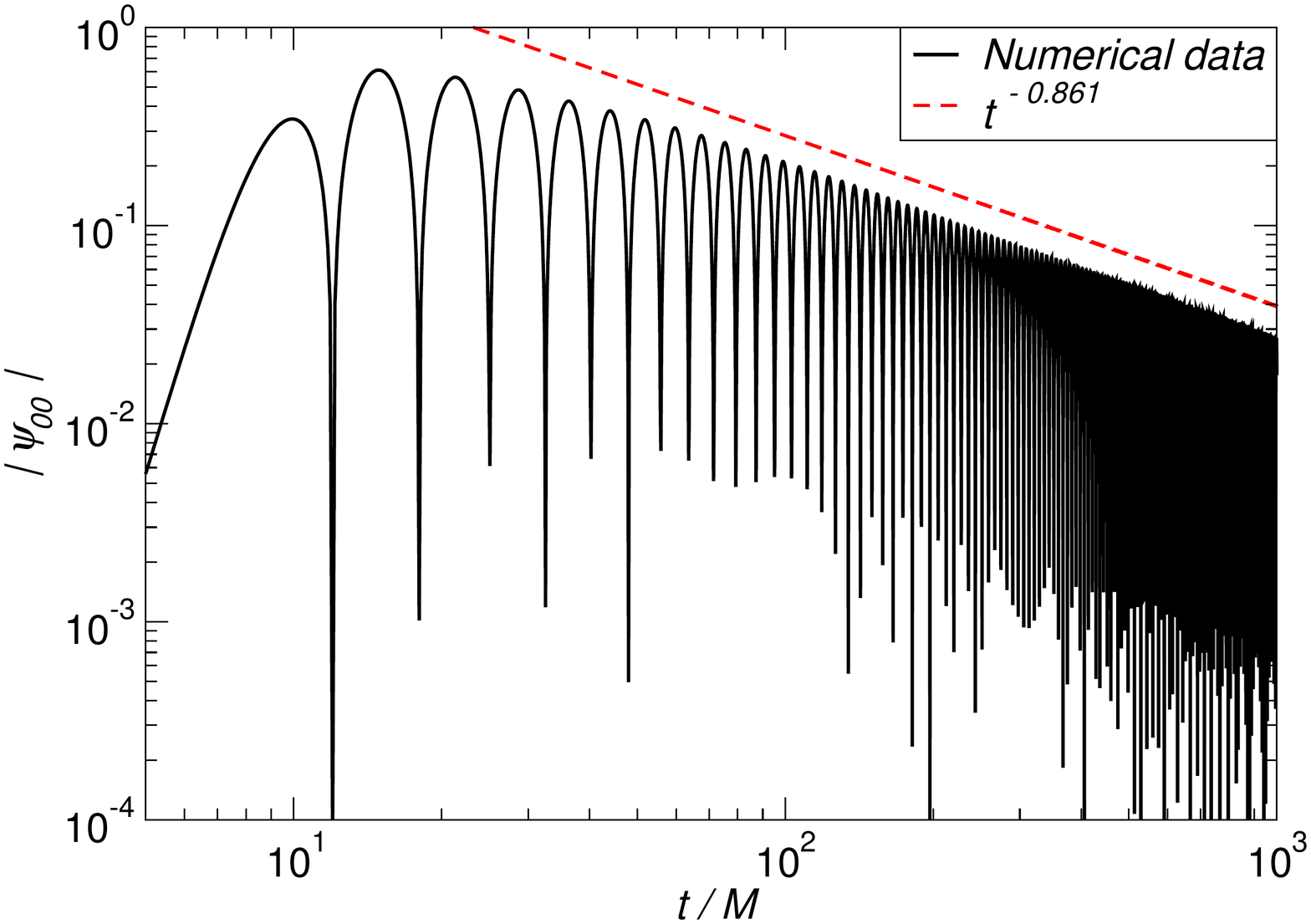} \\ 
\end{tabular}
\end{center}
\caption{\label{fig:waveformsSSm01}
Upper left:
$l=m=1$ multipole of run sS\_m001 extracted at $10M$.
The solid (black) line refers to numerical data, the dashed (red) to an
oscillatory tail fit.  Upper right: Same for sS\_m100, extracted at
$r_{\rm ex}=20~M$ (black solid line) together with the tail fit (red dashed
line, we omit the oscillatory term for clarity).
Lower Panels:
$l=m=0$ multipole of runs sS\_m042 (left, extracted at $20M$)
and sK\_m042 (right, extracted at $25M$). The solid black lines refer to
the numerical data while the red dashed lines denote the the fit to the
envelope of the oscillatory late-time tail.
}
\end{figure}
In order to study the behaviour of rapidly damped configurations, we
initialize the field by a Gaussian wave pulse with $r_0=12~M$ and $w=2~M$
according to Eq.~\eqref{eq:MFinitdataQNM}. The specific choices of the
mass coupling $M \mu_S$, rotation rate $a/M$
and the initial mode contributions for our
set of simulations are summarized in Table~\ref{tab:MFScaMassiveSetup}.
The resulting dipole and monopole amplitudes for a subset of our simulations
are shown in Fig.~\ref{fig:waveformsSSm01}. For all simulations we observe
the expected pattern of an early transient followed by quasi-normal
ringdown and a late-time tail which is
dominated by an oscillatory behaviour characterized by the mass term
according to Eq.~\eqref{eq:massivetails}
and governed by scattering off spacetime curvature according to
Eq.~\eqref{eq:MFScaPLmassiveB} at late times
\cite{Koyama:2001ee,Koyama:2001qw,Burko:2004jn}.

At intermediate times we find
for the case of an $m=1$ dipole with $M\mu_S=0.1$ (left upper panel
of Fig.~\ref{fig:waveformsSSm01}), an oscillatory decay of the field as
$\Psi_{11} \sim t^{-2.543} \sin(0.1 t)$, within $~2\%$
of the expected value for tails at intermediate times
\cite{Hod:1998ra,Koyama:2001ee,Koyama:2001qw}; cf. Eqs.~\eqref{eq:massivetails} and~\eqref{eq:MFScaPLmassiveA}.
At {\it very late times}, we expect a tail of the form~\eqref{eq:massivetails}
with $p=-5/6$ to dominate,
but simulation of sufficient duration are computationally too expensive with
our present computational framework.

Furthermore, we consider larger mass couplings $M\mu_S=0.42$ and $M\mu_S=1$. 
We present the $l=m=1$ dipole mode of sS\_m100 in the top right panel of Fig.~\ref{fig:waveformsSSm01}
and the monopole modes for both runs sS\_m042 and sK\_m042 in the bottom panels
of Fig.~\ref{fig:waveformsSSm01}.
Here, the intermediately-late time tail with $p=-(l+3/2)$ in Eq.~\eqref{eq:massivetails} 
appears to be suppressed. 
Instead, the decay with $p=-5/6$ expected at very late times 
clearly dominates the signal.
Focusing on the case with mass coupling $M\mu_S=0.42$ we observe that the
tail is present for both spinning and non-spinning BH backgrounds; cf. bottom panel 
of Fig.~\ref{fig:waveformsSSm01}.
For $a/M=0$ and $a/M=0.99$, we find the exponents
$p=-0.880$ and $p=-0.861$, respectively, which agree within $~5\%$
with the theoretically expected late-time behaviour
\cite{Koyama:2001ee,Koyama:2001qw,Burko:2004jn}.
We note that these results are independent of the extraction
radii $r_{\rm ex}$ at which the field is observed.

\subsubsection{Massive scalar fields: mode excitation and beating
\label{ssec:excitationcoefficients}}
%
\begin{figure}
\begin{center}
\begin{tabular}{ccc}
\includegraphics[width=0.33\textwidth]{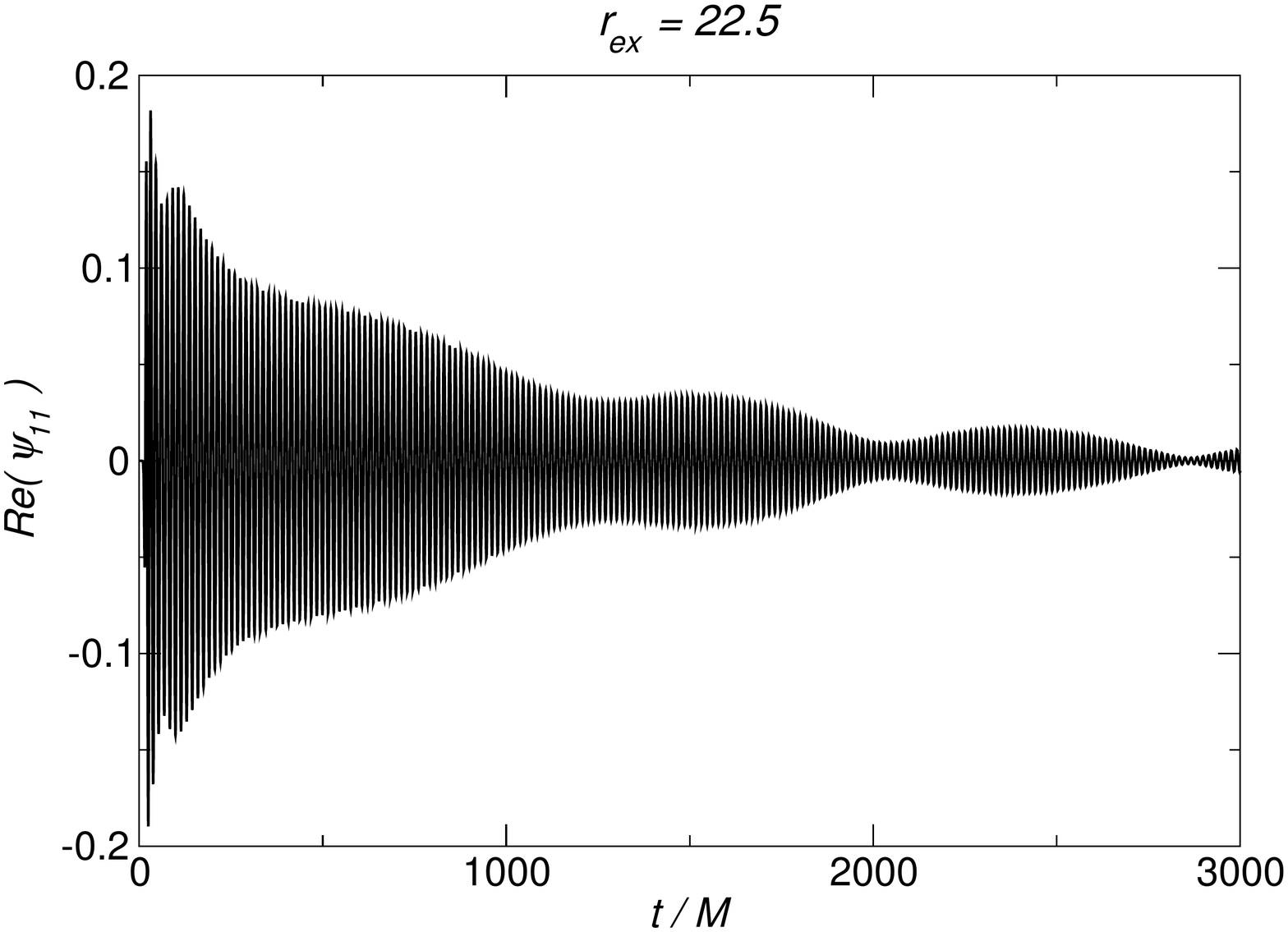} & 
\includegraphics[width=0.33\textwidth]{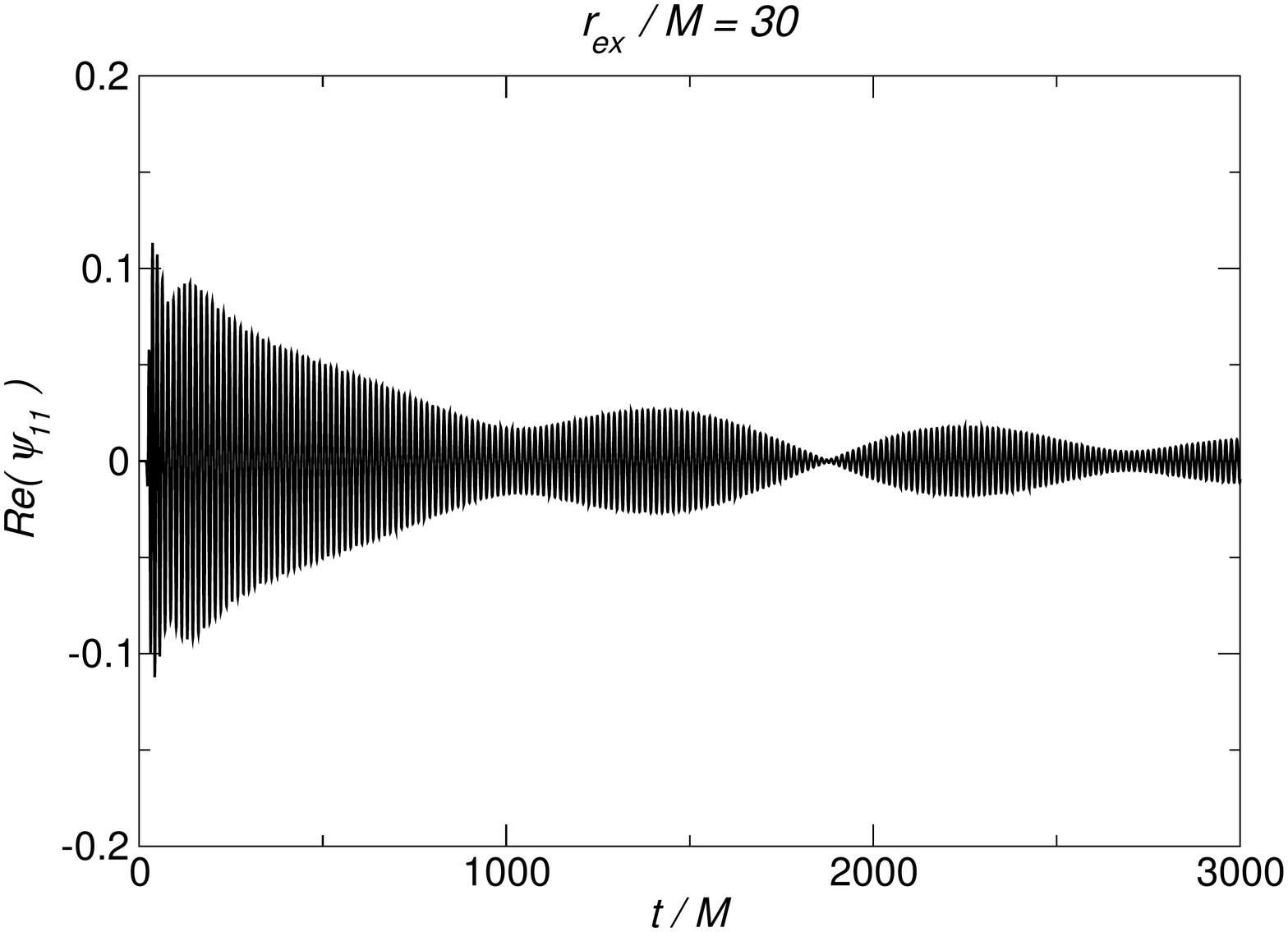} & 
\includegraphics[width=0.33\textwidth]{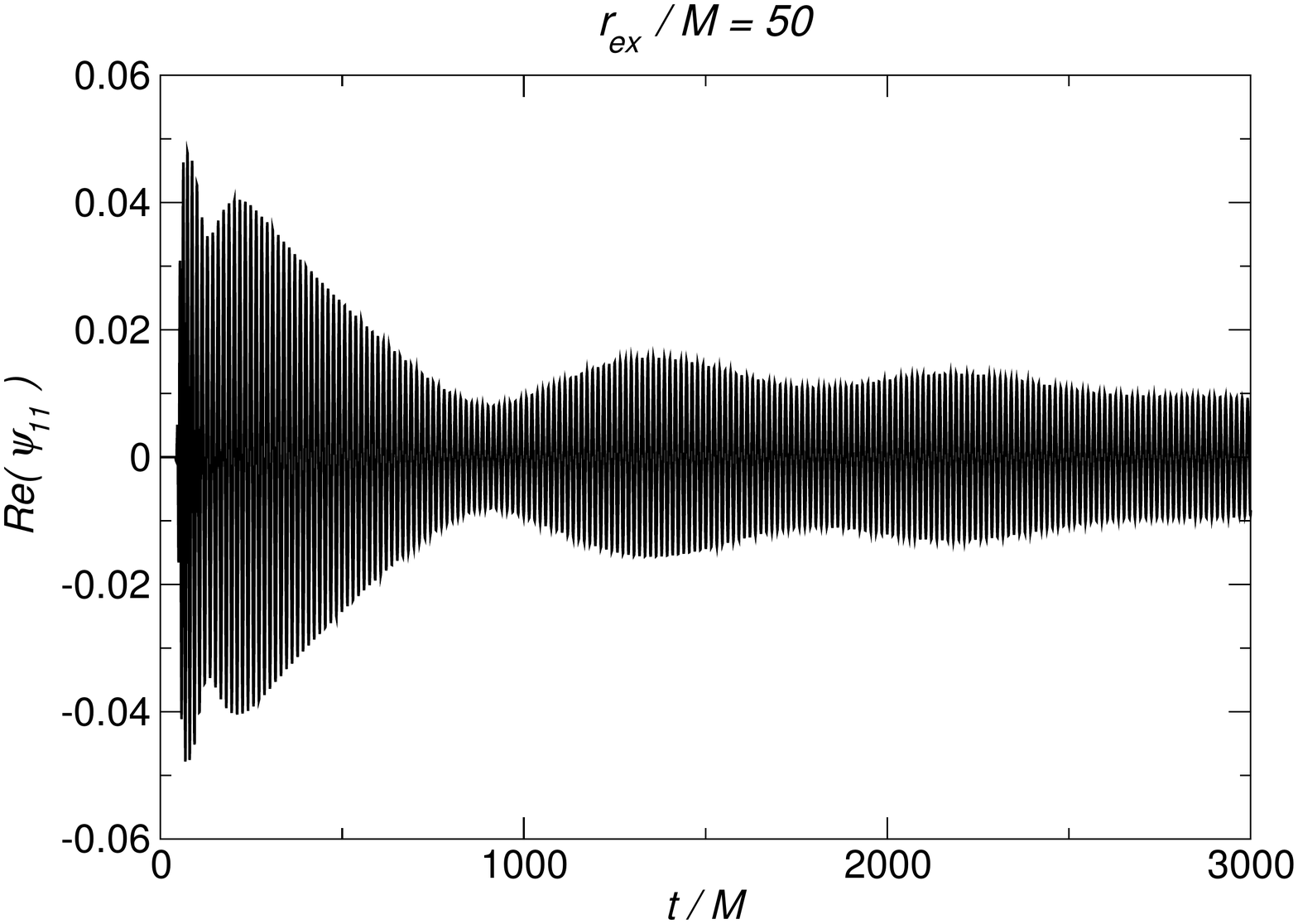} \\ 
\includegraphics[width=0.33\textwidth]{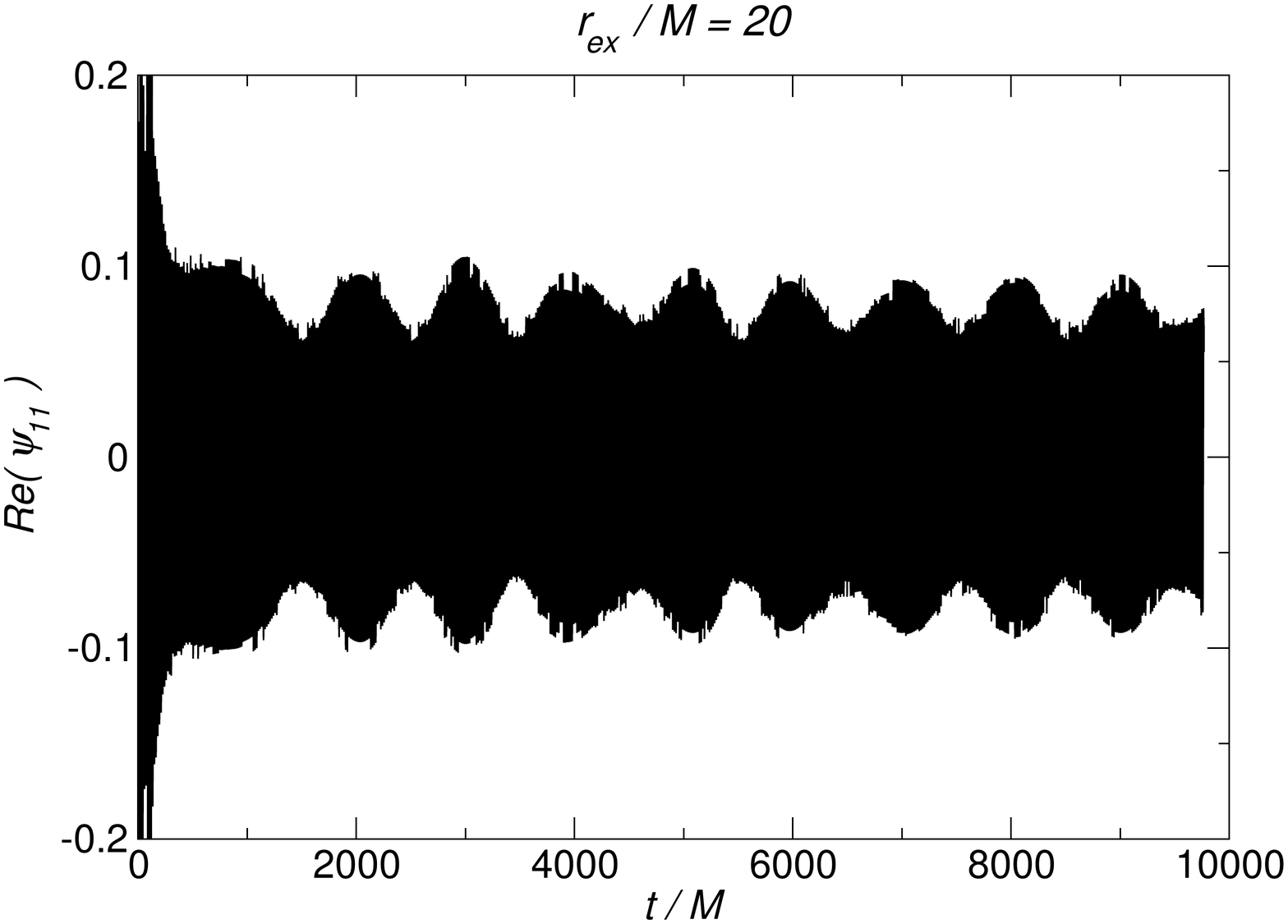} &
\includegraphics[width=0.33\textwidth]{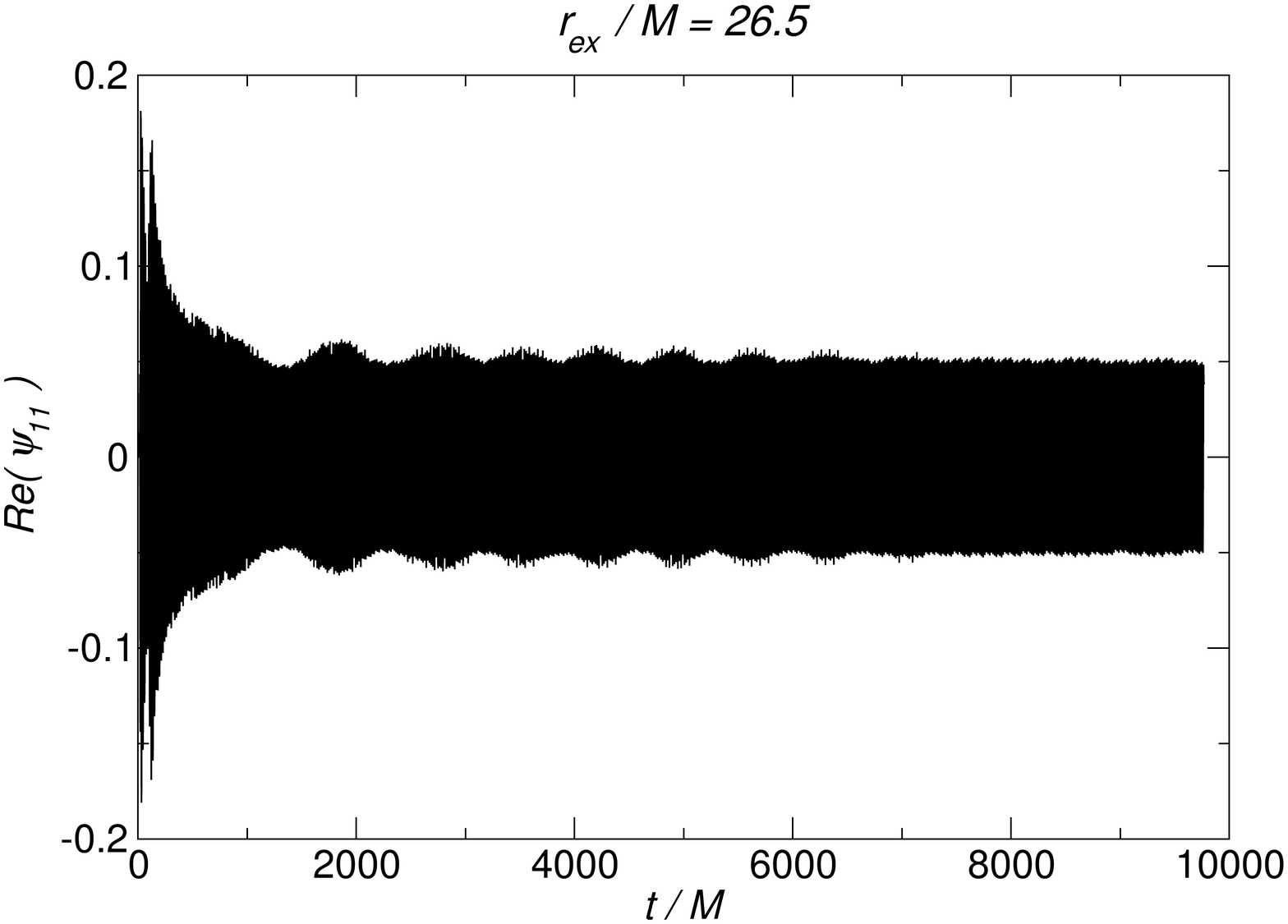} &
\includegraphics[width=0.33\textwidth]{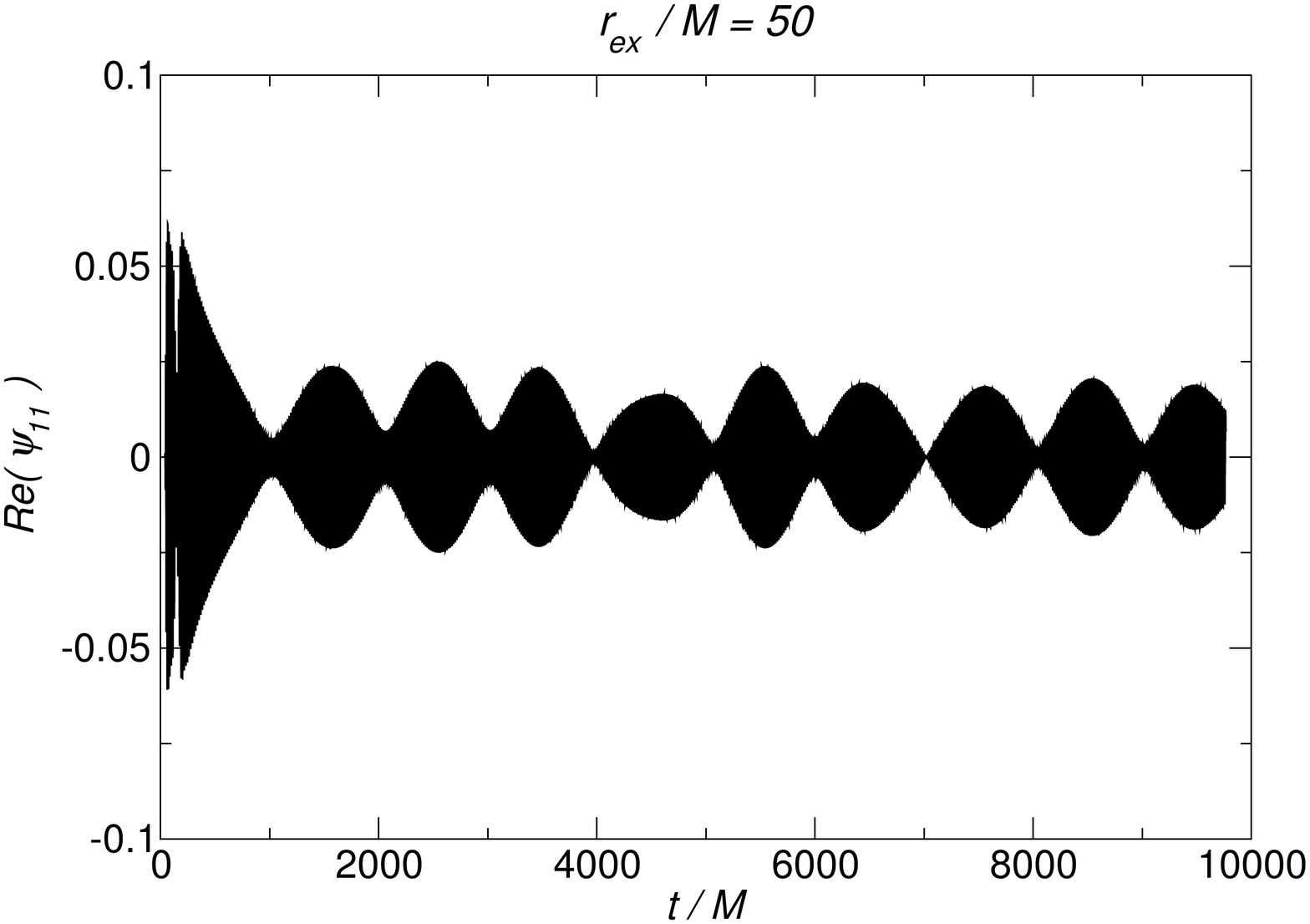} \\
\end{tabular}
\end{center}
\caption{\label{fig:waveformsSSm042_11}
Real part of the $m=1$ dipole obtained at selected extraction
radii $r_{\rm ex}$ for a massive scalar field with $M\mu_S = 0.42$ in
a Schwarzschild (upper panels) and Kerr background with $a=0.99M$
(bottom panels). The location of the node of the first overtone is
$r_{\rm ex}=22.5~M$ for the Schwarzschild case (upper left most panel)
and $r_{\rm ex}=26.5~M$ for the Kerr case (bottom center panel).
$r_{\rm ex}\sim50~M$ corresponds to the overtone's local maximum in the Kerr case.
}
\end{figure}

In addition to the well-known ringdown and decay, the evolution
of massive Gaussian wave packets around BHs can exhibit more
complex patterns. 
In particular, we expect massive scalar fields in Kerr backgrounds
to eventually show exponential growth if they satisfy the superradiance
condition Eq.~\eqref{eq:MFSRcond} as well as $\omega_R \le \mu_S$.
The timescale for this instability, however, is of the order
$\gtrsim 10^7~M$ which is computationally too expensive
to be realized within our numerical framework. In the
following, we therefore focus on the complex signals
observed at earlier times up to $\sim 10^4~M$ in the evolution
of massive scalar fields around Schwarzschild or Kerr black holes.

In particular, the time evolution of the $l=m=1$ mode of the 
scalar field with mass coupling $M\mu_S=0.42$
exhibits an oscillatory pattern with significant modulation of the amplitude.
The quantitative behaviour of the oscillations, however,
depends sensitively not only on the initial data but also on the radius where the modes are measured.
This is illustrated in Fig.~\ref{fig:waveformsSSm042_11},
where we show the $m=1$ dipole extracted at different
radii for $M\mu_S=0.42$ considering a Schwarzschild or a Kerr BH
background with $a/M=0.99$. For each configuration shown in the
figure, we have chosen three extraction radii, 
including the location $r_{\rm node}$ of the node, $22.5~M$ and $26.5~M$ respectively,
for the Schwarzschild and Kerr case.

This phenomenon can be explained in terms of a beating modulation
between two or more long-lived modes described by
Eq.~\eqref{eq:boundstates}. This beating modulation depends on the
relative strength of the different overtones which, in turn, depends
on the extraction radius; for the case of a guitar string, for example,
a specific mode cannot be excited at the location of its nodes.
Our choice of initial parameters for this example did not involve
any finetuning and we expect these features to be present
in the time evolution of generic massive fields around BHs
provided only that at least two long-lived modes are excited.

\begin{figure}
\begin{center}
\begin{tabular}{cc}
\includegraphics[width=0.50\textwidth]{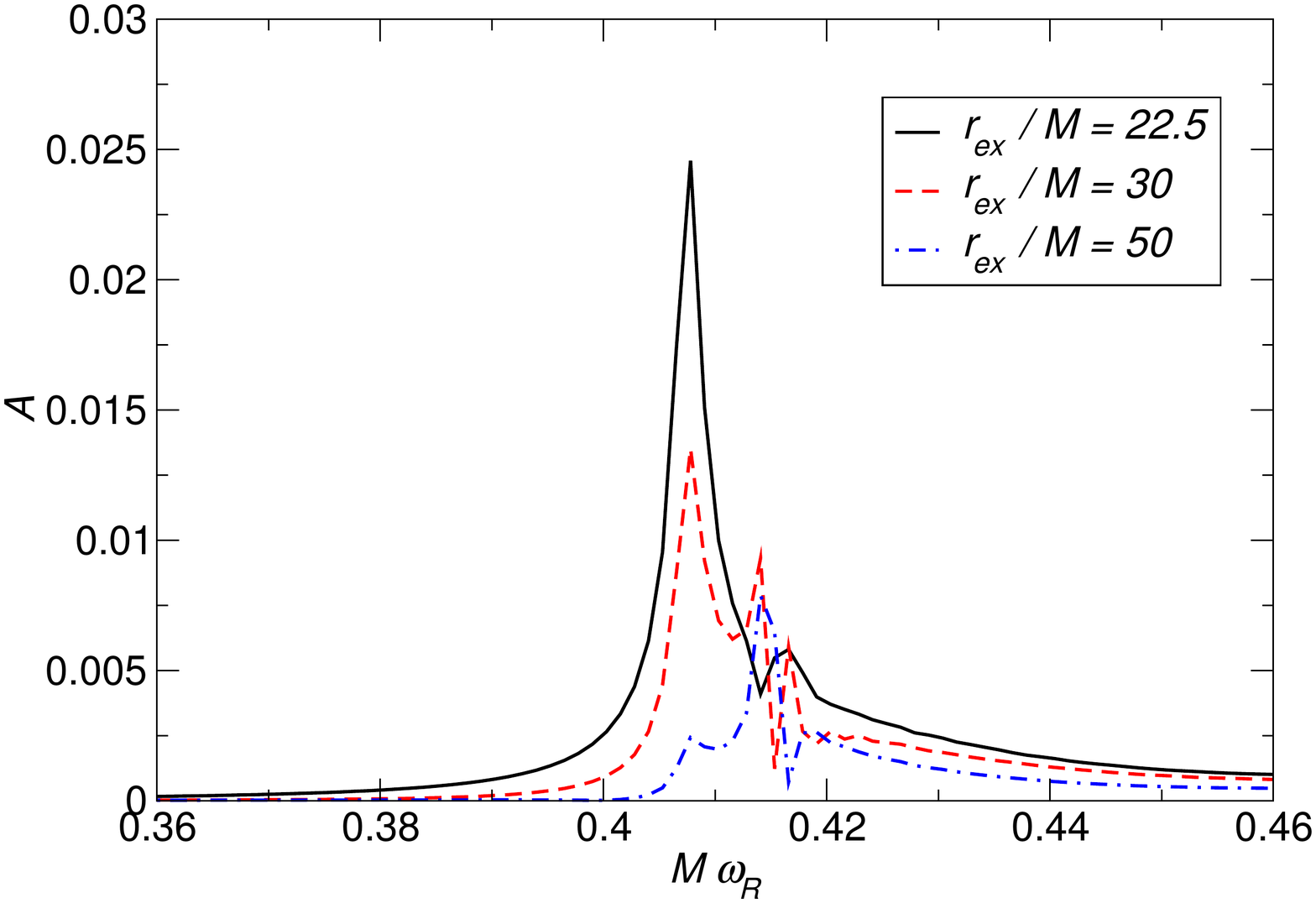} &
\includegraphics[width=0.50\textwidth]{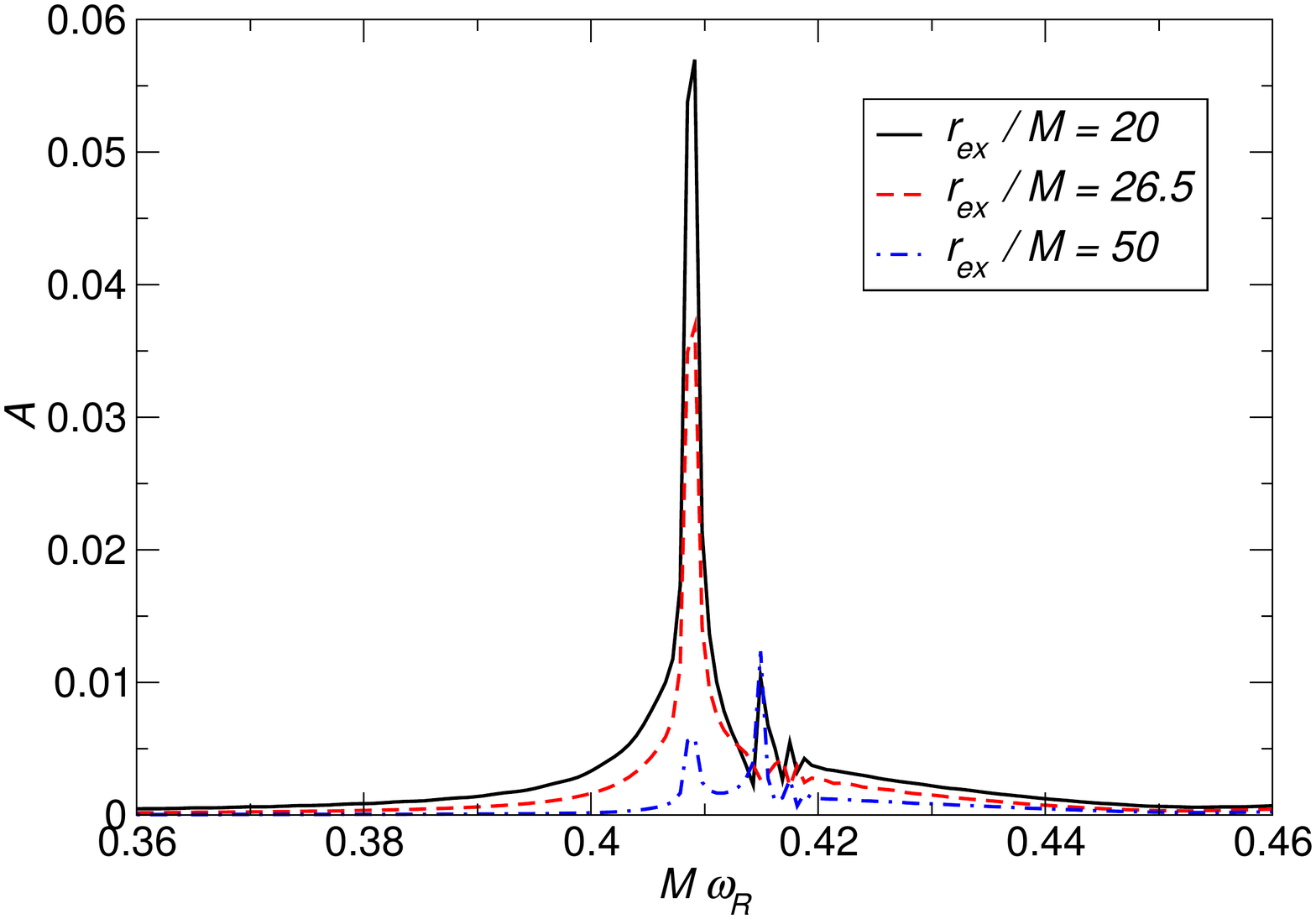} 
\end{tabular}
\end{center}
\caption{\label{fig:SpecKerrmu042_l1}
Spectra of the $l=m=1$ mode of the massive scalar field with $M\mu_S=0.42$
evolved in the background of a Schwarzschild (left panel) or Kerr BH with $a/M=0.99$ (right panel).
The lines correspond to the waveforms measured at different extraction radii. 
In particular, $r_{ex}=22.5~M$ (left) and $r_{ex}=26.5~M$ (right) correspond to the nodes
of the first overtone.
}
\end{figure}

Our interpretation is confirmed by
the Fourier spectra of the $m=1$ dipole obtained at different
radii which are shown in Fig.~\ref{fig:SpecKerrmu042_l1} for the
two background spacetimes. For the Kerr case with $a/M=0.99$ (right panel),
we see that the signal is dominated by the fundamental mode at
$r_{\rm ex}=20~M<r_{\rm node}$ and has two overtones of smaller amplitude.
This is in agreement with the weakly modulated high frequency
signal in the time evolution in the bottom left panel
of Fig.~\ref{fig:waveformsSSm042_11}. As expected, the amplitude
modulation is particularly weak for $r_{\rm ex}=r_{\rm node}=26.5~M$ which
coincides with the node of the first
overtone. In fact, the small amount of modulation visible
in the bottom center panel of Fig.~\ref{fig:waveformsSSm042_11} at early times
is mostly due to the {\em second} overtone which, however, damps out
at later times. Then, the envelope is almost constant
with a small growth rate $M\omega_{I}\sim10^{-7}$, in order-of-magnitude agreement
with previous studies \cite{Dolan:2007mj,Yoshino:2012kn}.
The situation is markedly different at $r_{\rm ex}=50~M$, where
the first overtone has a local maximum and the comparable strength
of fundamental mode and overtone result in the strong beating
modulation displayed in the bottom right panel of
Fig.~\ref{fig:waveformsSSm042_11}. 
This analysis can be repeated for the non-rotating case, with similar conclusions.
In contrast to the Kerr case, however, all modes decay resulting in overall damped signals.

\begin{figure}
\begin{center}
\begin{tabular}{cc}
\includegraphics[width=0.50\textwidth]{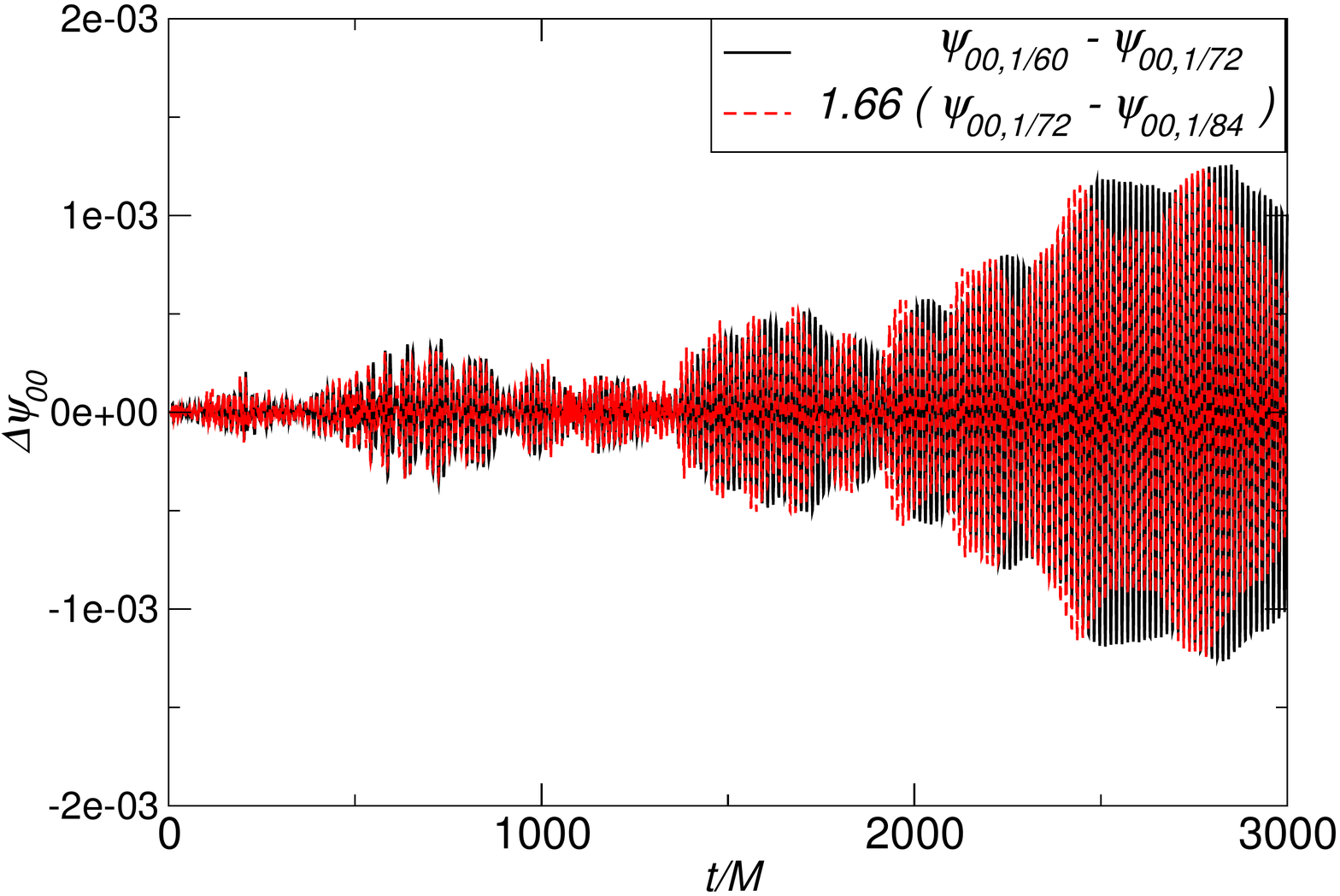} &
\includegraphics[width=0.50\textwidth]{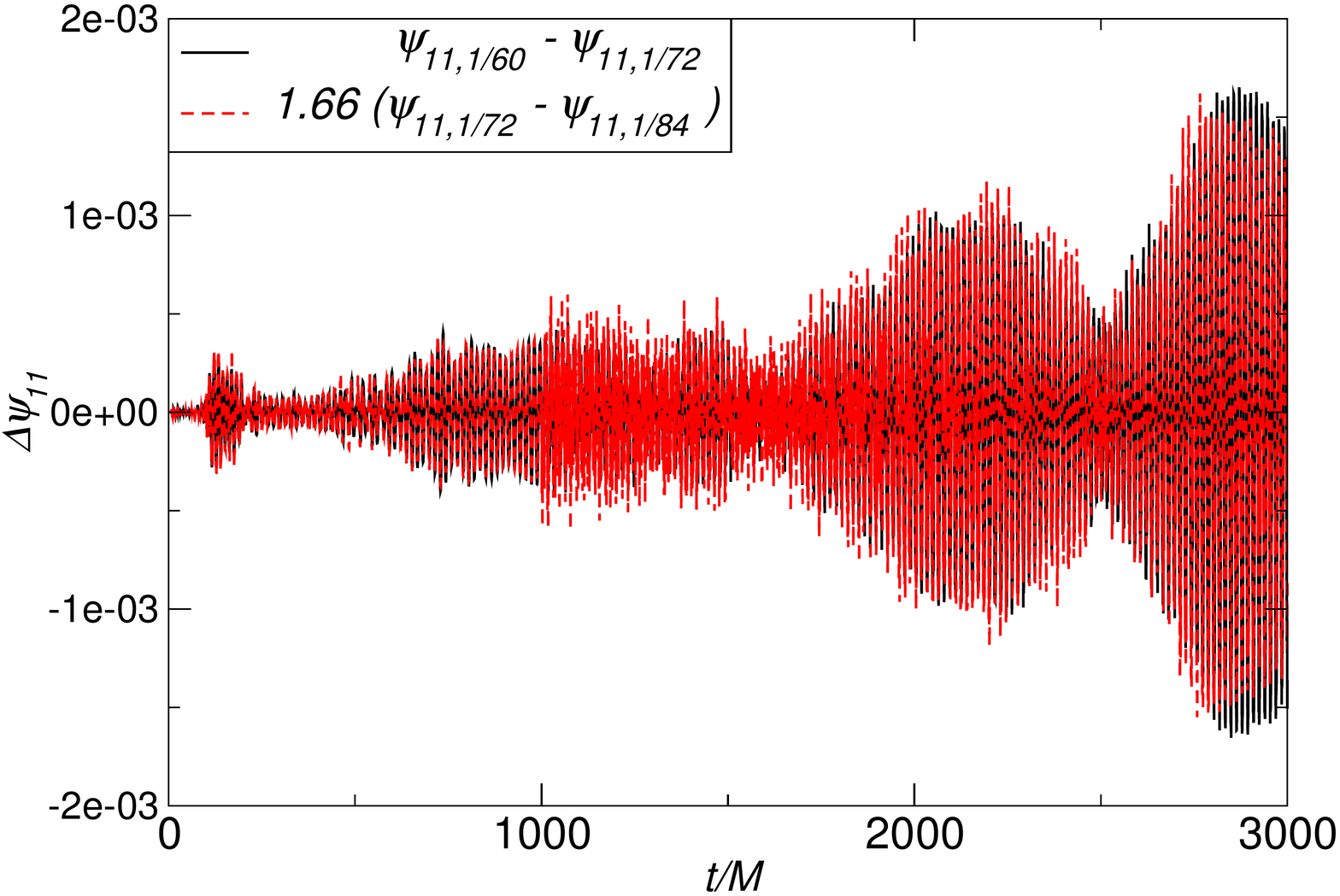} 
\end{tabular}
\end{center}
\caption{\label{fig:convergenceKerr042}
Lower panels: Convergence plot of the $l=m=0$ (left panel) and $l=m=1$
(right panel) modes of simulation sK\_042.
We present the differences between the coarse-medium and
medium-high resolution runs, where the latter is amplified by $Q_2=1.66$
indicating second order convergence.
}
\end{figure}

Before proceeding with a detailed analysis of the beating modulation and
mode excitation, we perform a convergence analysis of this setup which
represents the most demanding and longest of our simulations.
For this purpose,
we have evolved the setup sK\_m042 using three different resolutions
$h_c = M/60$, $h_m = M/72$ and $h_h = M/84$.
We present the convergence plot, i.e. the differences between the coarse-medium and medium-high
resolution runs in Fig.~\ref{fig:convergenceKerr042} for the monopole (left
panel) and dipole (right panel).
Our numerical results show second order
convergence throughout the simulation.
We estimate the discretization
error to be about $\De\psi_{11}/\psi_{11} \le 1.1\%$
for the entire interval and $\De\psi_{00}/\psi_{00} \le 1\%$ 
$l=m=0$ mode at early time which increases to $\De\psi_{00}/\psi_{00}
\le 7\%$ at late times.
%

\noindent{\bf{\em Beating.}}
In order to better understand the beating pattern quantitatively, let us for simplicity
consider the presence of only two modes with similar frequencies:
the long-lived fundamental mode with
frequency $\omega_0 = \omega_{R,0} + \imath \omega_{I,0}\sim \omega_{R,0}$ and amplitude $A_0$
and the first overtone with frequency $\omega_1 = \omega_{R,0} + \delta_{10}$,
where $\delta_{10}\ll 1$, and amplitude $A_1$. 
The generalization to a larger number of modes is straightforward. 
For illustration, we explicitly list $\delta_{10}$
for a selected subset of configurations in Table \ref{tab:MFScaAnaModes}.
Because all these modes are long-lived, they are well approximated
by pure sinusoids over short periods of time. 
The waveform $\Psi$ is then described by the superposition
\begin{align}
  \label{eq:FitBeating}
  \Psi \sim & \left(A_0-A_1\right)\sin (\omega_{R,0}t)
            +A_1\sin(\omega_{R,0}t) \cos(\delta_{10}t / 2)  
\,.
\end{align}
The modulation in amplitude is governed by the low-frequency signal
$\cos\delta_{10}t$ whereas the beating amplitude depends on the
relative strength of the modes $A_0$, $A_1$.
For equal amplitudes $A_0=A_1$, for instance,
the total signal is given by a sinusoid modulated by a cosine.
The beating frequency $\delta_{10}$ can be estimated by fitting
the envelope of the total signal. 
For the two cases displayed in Fig.~\ref{fig:waveformsSSm042_11}
we obtain
$\delta_{10}\sim 0.0074$ (for sS\_m042) and $\delta_{n0}\sim 0.0063$ (for sK\_m042),
in excellent agreement with the corresponding predictions
listed in the third and fourth row of Table~\ref{tab:MFScaAnaModes}.

This picture is confirmed by calculating the mode spectra
from a Fourier transformation of the time series.
The spectra thus obtained for the two configurations and
different extraction radii are
shown in Fig.~\ref{fig:SpecKerrmu042_l1} and each
exhibit three pronounced peaks which correspond to
the real parts of the fundamental and first two overtone QN frequencies.
A spectral analysis applied to the Kerr case $a/M=0.99$
reveals  $\delta_{10}=0.0063$ and $\delta_{20}=0.0087$. Thus, the evolution of
Gaussian initial data excites a third long-lived mode
which is not listed
in Table~\ref{tab:MFScaAnaModes}.
We have verified that this mode indeed exists, via a continued fraction scheme in the frequency domain,
corresponding to $M\omega=0.41730 + i2.265\times 10^{-8}$.

\vspace{0.5cm}
\noindent{\bf{\em Mode excitation.}}
%
The mode excitation can be put on a more rigorous framework:
Leaver's seminal work, in particular,
has established important results in this context and
for further details we refer the
reader to the original work \cite{Leaver:1986gd} 
as well as
comprehensive follow-up studies
\cite{Andersson:1995zk,Berti:2006wq,Berti:2009kk}.
The upshot is that each quasi-normal mode, which corresponds to a pole in the complex-frequency plane, 
is excited to a different degree depending on the initial data and on the mode in question. 
The QNM contribution can be isolated from other features of the signal, such
as the late-time tail, using the Green's function technique
\cite{Leaver:1986gd,Andersson:1995zk,Berti:2006wq}. 
In this formalism, the scalar field amplitude at intermediate times is given by a sum over
quasinormal modes as
\begin{equation}
\Psi=\sum C_n e^{-i\omega_n t}\psi_n(\omega_n,r)\,,
\end{equation}
where $\psi_n(\omega_n,r)$ is the quasinormal mode eigenfunction and $\omega_n$ its frequency, both quantities can be computed 
via the Fourier-domain ordinary differential equation that governs massive scalars in the Kerr background \cite{Leaver:1986gd,Andersson:1995zk,Berti:2006wq}.
The numbers $C_n$, called excitation coefficients characterize the amplitude to which each mode is excited.
Two quantities are crucial to determine the excitation coefficients \cite{Berti:2006wq}:
the behavior of $\psi_n$ close to the eigenfrequency $\omega_n$, and
the convolution of the eigenfunction $\psi_n$ with the initial data.
Thus, for instance, the relative amplitude between different modes depends strongly 
on the point where this amplitude is evaluated: if it is close to a node of one of the modes, 
the mode in question will have a very small amplitude: 
by definition a mode is not excited at its node. 
Likewise, localized initial data close to the node of the mode do
not excite the mode in question, a well-known result for closed systems \cite{Berti:2006wq}.

We have not attempted a complete quantitative understanding of mode excitation for this work, 
a preliminary analysis indicates that the excitation coefficients are indeed of comparable magnitude
away from the nodes of the eigenmodes, but vary substantially close to the nodes.

\section{Proca field evolutions}\label{sec:Procaresults}
%
\begin{table}
\begin{center}
\begin{tabular}{cccccc}
\hline
Run             &$a/M$ &$M\mu_V$ & $\,_{-1}\Sigma(\theta,\phi)$                 &$w/M$ & Grid Setup \\ 
\hline
v1S\_m000       &0.00  &0.00     & Sup1 &2.0   &$\{(192,96,48,24,12,6,3,1.5),~h=M/60\}$ 
\\
v2S\_m000       &0.00  &0.00     & Sup1 &30.0  &$\{(192,96,48,24,12,6,3,1.5),~h=M/60\}$ 
\\
v1S\_m010       &0.00  &0.10     & Sup1 &2.0   &$\{(192,96,48,24,12,6,3,1.5),~h=M/60\}$ 
\\
v1S\_m020       &0.00  &0.20     & Sup1 &2.0   &$\{(192,96,48,24,12,6,3,1.5),~h=M/60\}$ 
\\
\hline
v2K1\_m040      &0.50  &0.40     & Sup2 &30.0  &$\{(1536,384,192,96,48,24,12,6,3,1.5),~h=M/60\}$
\\
\hline
v1K2\_m000      &0.99 &0.00      & Sup1 &2.0   & $\{(192,96,48,24,12,6,3,1.5),~h=M/96\}$ 
\\
v2K2\_m040      &0.99 &0.40      & Sup1 &30.0  & $\{(1536,384,192,96,48,24,12,6,3,1.5),~h=M/64\}$ 
\\
v2K2\_m042      &0.99 &0.42      & Sup1 &30.0   & $\{(1536,384,192,96,48,24,12,6,3,1.5),~h=M/64\}$ 
\\
v2K2\_m044      &0.99 &0.44      & Sup1 &30.0    & $\{(1536,384,192,96,48,24,12,6,3,1.5),~h=M/64\}$ 
\\
v1K2\_m100      &0.99 &1.00      & Sup1 & 2.0    & $\{(1536,384,192,96,48,24,12,6,3,1.5),~h=M/60\}$  
\\
\hline
\end{tabular}
\end{center}
\caption{\label{tab:ProcaSetup} 
Initial setup for simulations of Proca fields with mass coupling $M\mu_V$ in
BH background with dimensionless spin parameter $a/M$.
The initial Gaussian pulse with width $w/M$ is located at $r_0=12~M$
and consists of a superposition of $s=-1$ spin-weighted spherical harmonics
given by Eq.~\eqref{eq:VecSup1} (``Sup1'') or Eq.~\eqref{eq:VecSup2} (``Sup2'').
We further denote the grid setup, in units of the BH mass $M$
following the notation of Sec.~II E in \cite{Sperhake:2006cy}.
}
\end{table}
The evolution of vector fields, as for example the photon,
are governed by Eq.~(\ref{eq:MFEoMVector1})
which, after substitution of the definition $F_{\mu \nu}=\nabla_{\mu}
A_{\nu}-\nabla_{\nu}A_{\mu}$
and the Lorenz condition, Eq.~\eqref{eq:MFLG}, 
becomes rather similar to its scalar
counterpart (\ref{eq:MFEoMScalar1}). For this reason, it has
been believed for a long time that massive vector fields should
also be prone to a ``BH bomb''-like superradiant instability.
In fact, Rosa and Dolan \cite{Rosa:2011my}
conjectured that instabilities of vector fields should not
only be present but can be much stronger than those of scalar fields.
This has recently been verified explicitly
by Pani {\it et al} \cite{Pani:2012vp,Pani:2012bp}
who derived the instability growth rates of
massive vector fields in the slow-rotation approximation 
and found them to be several orders of magnitude
larger than their scalar field counterparts.
Their results, however, have been obtained
only in the regime of slowly rotating black holes. Even though their
results are conjectured to hold for arbitrary spins, a definitive answer
to this question calls for the modelling of vector fields in generic
Kerr geometries. In the frequency domain, such modelling represents
a formidable challenge because the equations of motion appear to be
non-separable. Here we therefore address this question in the
framework of numerical evolutions in the time domain, where the
non-separability of the equations does not represent a serious
obstacle.

For this purpose, we prescribe initial data in the form of a
superposition of Gaussian pulses centered around $r_0=12~M$ and composed of
several multipoles with angular dependence given by
$s=-1$ spin-weighted spherical harmonics ${}_{-1}Y_{lm}$. Specifically,
we have chosen linear combinations
\begin{subequations}
\begin{align}
\label{eq:VecSup1}
\,_{-1}\Sigma(\theta,\phi) = &
-\frac{1}{\sqrt{3}} \left( \,_{-1}Y_{1-1} + \,_{-1}Y_{11} \right)
-\frac{1}{\sqrt{5}} \left( \,_{-1}Y_{2-1} - \,_{-1}Y_{21} \right) 
-\frac{1}{\sqrt{6}} \,_{-1}Y_{10} - \frac{1}{\sqrt{30}} \,_{-1}Y_{20}
\non\\ &
-\frac{1}{3 \sqrt{5}} \left( \,_{-1}Y_{2-2} - \,_{-1}Y_{22} \right) 
-\frac{\sqrt{2}}{3\sqrt{35}} \left( \,_{-1}Y_{3-2} + \,_{-1}Y_{32} \right) 
\,,\\
\label{eq:VecSup2}
\,_{-1}\Sigma(\theta,\phi) = &
- \frac{1}{\sqrt{3}} \left( \,_{-1}Y_{1-1} + \,_{-1}Y_{11} \right) 
- \frac{1}{\sqrt{5}} \left( \,_{-1}Y_{2-1} - \,_{-1}Y_{21} \right)
\,,
\end{align}
\end{subequations}
such that the imaginary parts of the fields vanish. 
By virtue of the evolution
equations (\ref{eq:MFdtphi})-(\ref{eq:MFdtE}), purely real initial
data remain real throughout the evolution so that our choice
reduces the computational requirements to evolving only seven
-- instead of $14$ -- independent variables. We have summarized the
initial configurations of our set of simulations together with
the grid setup in Table~\ref{tab:ProcaSetup}.
In order to analyse the time evolutions of the vector field
$A_{\mu}$, we decompose its time component $\varphi$
and the Newman-Penrose scalar $\Phi_2$, constructed from the spatial
components according to Eq.~\eqref{eq:MFNP2}, into multipoles by
projecting them onto spherical harmonics with spin weight $s=0$
and $s=-1$, respectively.

\subsection{Massless vector fields}\label{ssec:VecMassless}
\begin{figure}[htpb!]
\begin{center}
\includegraphics[width=0.50\textwidth]{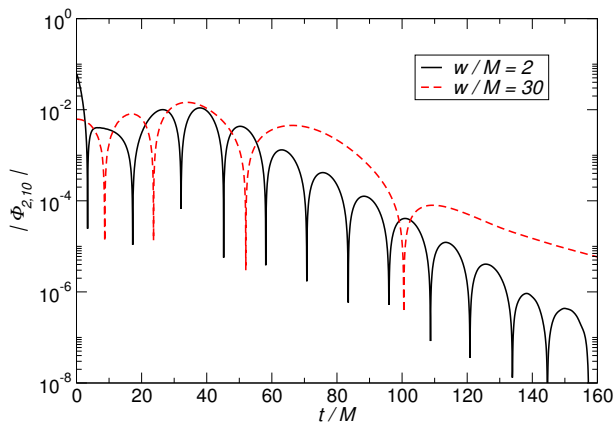}
\end{center}
\caption{\label{fig:VecSchWidth}
The $l=1, m=0$ multipole of $\Phi_2$, extracted at $r_{\rm ex}=10~M$
from the time evolution of a Gaussian pulse of
width $w=2~M$ (black solid line) and $w=30~M$ (red dashed line)
around a Schwarzschild BH. 
}
\end{figure}

\begin{figure}[htpb!]
\begin{center}
\begin{tabular}{c}
\includegraphics[width=0.50\textwidth]{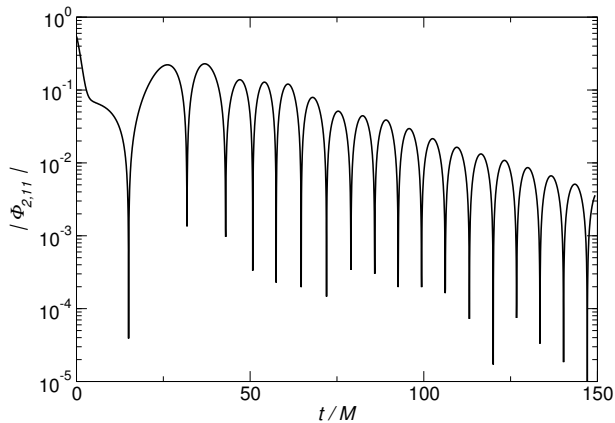} 
\end{tabular}
\end{center}
\caption{\label{fig:VecMasslessConv}
Time evolution of the $l=m=1$ mode of $\Phi_2$ of run v1K2\_m000, i.e., 
a massless vector field in a Kerr background with $a/M = 0.99$ extracted at $r_{\rm ex}=10~M$.
}
\end{figure}

\begin{table}
\begin{center}
\begin{tabular}{cccc}
\hline
$a/M$   & ($lm$) & $M \omega^{\rm fd}_{lm}$     & $M\omega^{\rm num}_{lm}$      
\\ \hline
$0.00$  & ($10$) & $0.2483 - \imath 0.0925$ & $0.248-\imath0.092$       
\\
$0.00$  & ($20$) & $0.4576 - \imath 0.0950$ & $0.455-\imath0.093$       
\\ \hline
$0.99$  & ($10$) & $0.2743 - \imath 0.0759$ & $0.274-\imath0.075$       
\\
$0.99$  & ($11$) & $0.4634 - \imath 0.0313$ & $0.464-\imath0.033$       
\\
$0.99$  & ($20$) & $0.4999 - \imath 0.0800$ & $0.498-\imath0.079$       
\\
$0.99$  & ($22$) & $0.9099 - \imath 0.0301$ & $0.887-\imath0.037$       
\\
\hline
\end{tabular}
\end{center}
\caption{\label{tab:VecMasslessModes}
Quasi-normal mode frequencies of massless vector perturbations in a
Schwarzschild or Kerr BH background with $a/M=0.99$.
The values for $M\omega^{\rm num}_{lm}$ have
been obtained from fits to our numerical evolution of the field, whereas
those for $M\omega^{\rm fd}_{lm}$
have been computed with the continued fraction method
\cite{Berti:2009kk,Berti:2005ys,QNMwebpage}.
}
\end{table}
We first study the behaviour of massless vector field perturbations
with $M\mu_V=0$. This case has been studied extensively in
the literature \cite{Berti:2009kk} and therefore also enables us to
compare our findings with previous investigations.
Our results for massless Proca fields are summarized in Table~\ref{tab:VecMasslessModes} and
Figs.~\ref{fig:VecSchWidth} and~\ref{fig:VecMasslessConv}.
Let us first consider the simplest case of a massless vector field in
Schwarzschild background. The time evolutions of the $l=m=1$ multipole
obtained for initial Gaussian pulses of width
$w=2~M$ and $w=30~M$ are shown in Fig.~\ref{fig:VecSchWidth}.
For the narrow pulse (solid curve in the figure) we clearly identify
the pattern familiar from our scalar field evolutions in
Sec.~\ref{ssec:MFScaMuVar}:
an early transient whose details depend on the
initial configuration is followed
by a quasi-normal ringing characterized entirely by the
BH parameters and a late-time tail. For an initially broad pulse (red dashed curve),
however, the initial transient is directly followed by a power-law tail
with no visible intermediate ringdown stage. This
feature has been reported for scalar fields in
Refs.~\cite{Andersson:1995zk,Vishveshwara:1970zz} and is a
consequence of the negligible excitation of the
long-lived fundamental mode and low overtones by broad pulses;
the high overtones which are excited significantly by this type
of initial data rapidly decay before the transient gives way
for a clear QN ringdown pattern to emerge. 

For a quantitative
comparison with calculations performed in the frequency
domain,
we have fitted the ringdown part of our numerically extracted multipoles
for the case of a narrow initial Gaussian
with exponentially damped sinusoids. The resulting complex
frequencies $\omega_{lm}^{\rm num}$ are listed in
Table \ref{tab:VecMasslessModes} and agree well with
the values $\omega_{lm}^{\rm fd}$ obtained from 
frequency-domain calculations
\cite{Leaver:1986gd,Berti:2009kk,Berti:2005ys,QNMwebpage}. 
The time evolution of a massless vector field initialized as a narrow
Gaussian of width $w=2~M$ around a rapidly spinning
Kerr BH with $a/M=0.99$ is displayed in Fig.~\ref{fig:VecMasslessConv}
and qualitatively agrees with the corresponding simulation around
a Schwarzschild background. Note, however that the quasi-normal
ringdown is significantly slower for the spinning case which is
also reflected by the relatively small imaginary quasi-normal mode
frequencies listed for this case in Table~\ref{tab:VecMasslessModes}.
Accurately measuring such small imaginary components in numerical
simulations represents a considerable challenge which is why we have
chosen a higher numerical resolution for this particular simulation;
cf.~Table~\ref{tab:ProcaSetup}.
We thus obtain agreement of
a few \% with frequency-domain predictions for the imaginary
part. In contrast the real part of the frequency is much easier
to extract numerically and shows excellent agreement around $1~\%$
or less with the values calculated in the frequency domain.
%

\subsection{Proca field in Schwarzschild backgrounds and in the slow-rotating
limit}
\label{ssec:ProcaSchw}
%
\begin{table}
\begin{center}
\begin{tabular}{ccccccccc}
\hline
$l$ & $M\omega^{\rm fd}_{l0}$ (ES) & $M\omega^{\rm fd}_{l0}$ (O)& $M\omega^{\rm fd}_{l0}$ (EV) & $M\omega^{\rm num}_{l0}$ ($\varphi_{l}$) & $M\omega^{\rm num}_{l0}$ ($\Phi_{2,l}$) & $p$ ($\varphi_{l0}$)   & $p$ ($\Phi_{2,l0}$) \\ \hline
0   & $0.1216 - \imath 0.0791$     &                            &                              & $0.125-\imath0.080$                      &                                         & $1.51$                 & \\ 
1   & $0.3054 - \imath 0.0914$     & $0.2435 - \imath 0.0944$   & $0.2539 - \imath 0.0887$     & $0.245-\imath0.098$                      & $0.252-\imath0.087$                     & $0.49$                 & $0.49$\\ 
2   & $0.4921 - \imath 0.0943$     & $0.4552 - \imath 0.0955$   & $0.4610 - \imath 0.0938$     & $0.462-\imath0.091$                      & $0.452-\imath0.096$                     & $0.48$                 & $0.49$\\ \hline
\end{tabular}
\end{center}
\caption{\label{tab:ProcaSchw}
QNM frequencies and tail exponents for Proca field perturbations in the
Schwarzschild background, for $M\mu_V=0.1$.  Modes $\omega^{\rm fd}$ have
been computed with the continued fraction and forward-integration method
in the frequency domain \cite{Rosa:2011my,Pani:2012vp,Pani:2012bp}.
These modes are divided in odd (O), even scalar (ES)
and even vector (EV) parity; see text for details.
The QNM frequencies extracted from our numerical simulations are shown as
$\omega^{\rm num}$, as extracted from $\varphi_{l0}$ and $\Phi_{2,l0}$.
Finally, we also show the decay exponent $p$ of the oscillatory tail;
c.~f.~Eq.~\eqref{eq:ProcaTail}.
}
\end{table}
We now consider time evolutions of massive vector, or Proca, fields
in a Schwarzschild background. For this purpose, we
choose the mass parameters $M\mu_V=0.1$ and $M\mu_V=0.2$
which allow for a direct comparison with
recent QNM computations by Rosa \& Dolan \cite{Rosa:2011my}
and Pani {\it et al} \cite{Pani:2012vp,Pani:2012bp}.
In contrast to the massive scalar field, the Proca field has
three degrees of freedom and the resulting radiation
multipoles can be 
classified into three groups:
axial modes with spin $s=0$ and two polar modes with
$s=\pm1$. Following Refs.~\cite{Rosa:2011my,Pani:2012vp,Pani:2012bp},
we denote $s=0$ multipoles as {\em even scalars} (ES) and
the $s=+1,\,-1$ as {\em odd} (O) and {\em even vector} (EV), respectively.
We emphasize that these three different degrees of freedom have
different spectra except for the massless case 
where the spectra of the vector modes are degenerate while the scalar mode reduces to a gauge field
\cite{Berti:2009kk,QNMwebpage,Rosa:2011my}. 
The quasi-normal ringdown frequencies
$\omega_{lm}^{\rm fd}$ obtained from frequency-domain
calculations for the three types of multipoles
are shown in Table~\ref{tab:ProcaSchw}
for $M\mu_V=0.1$ considering a Schwarzschild background.

\begin{figure}[htpb!]
\begin{center}
\includegraphics[width=0.50\textwidth]{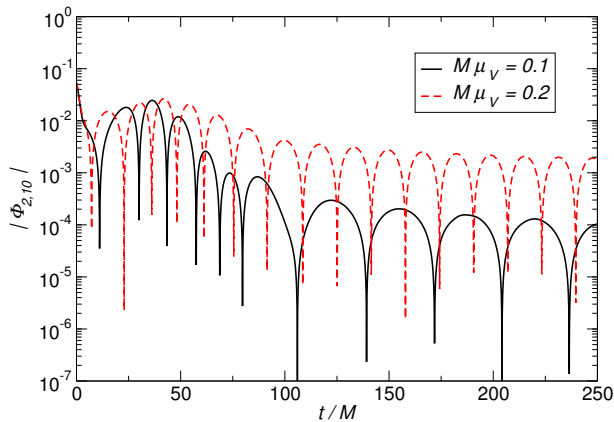}
\end{center}
\caption{\label{fig:ProcaSchw01}
Time evolution of the Proca field with mass coupling $M\mu_V=0.1$
(black solid line) and $M\mu_V=0.2$ (red dashed line) in a Schwarzschild
background.  We show the $l=1,m=0$ multipole 
of the Newman-Penrose scalar $\Phi_{2}$, extracted at $r_{\rm ex}=10~M$.
}
\end{figure}

In order to obtain numerical estimates for the frequencies, we have
evolved Gaussian initial data of width $w=2~M$ centered around $r_0=12~M$
for our two choices of $M\mu_V=0.1$ and $0.2$. The resulting dipoles
of $\Phi_2$
extracted at $r_{\rm ex}=10~M$ are shown in Fig.~\ref{fig:ProcaSchw01}
and reveal the by now familiar pattern of early transient, ringdown
and tail. Fitting an exponentially damped sinusoid to the ringdown
part of the dipoles of $\Phi_2$ (solid curve in the figure) as well as
the scalar component $\varphi$ for the case $M\mu=0.1$
yields numerical estimates for the frequencies listed in
Table~\ref{tab:ProcaSchw} as $\omega_{lm}^{\rm num}$.
These estimates agree with the frequency-domain predictions within
a few percent or less. Note, however, that the frequencies for
some types of modes are very similar, so that we cannot unambiguously
identify which modes are excited by our particular choice of initial data.
For $l=2$, for instance, our numerical results are compatible with both,
an odd or even vector mode.

In Fig.~\ref{fig:ProcaSchw01} we see that from time $t\sim100~M$ onwards,
the signal is dominated by the tails whose functional form
is given by a decaying sinusoid of the form \cite{Konoplya:2006gq}
\begin{align}
  \label{eq:ProcaTail} \Phi\sim t^{-(l+p)} \sin (\mu_V t)\,,
\end{align}
at intermediate times,
where $p=3/2+s$ depends on the spin, or parity, of the mode.
At late times, on the other hand,
the signal is expected to follow the universal behaviour
\begin{equation}
  \Phi\sim t^{-5/6} \sin (\mu_V t)\,,
\end{equation}
independent of the spin $s$ of the field
\cite{Koyama:2001ee,Koyama:2001qw,Konoplya:2006gq}.
Numerical estimates for the exponent $p$ extracted from our numerical
results for $\Phi_2$ and $\varphi$ are shown in Table~\ref{tab:ProcaSchw}
and are in excellent agreement with the prediction $p=3/2+s$
for intermediate times and spin values $s=0$ for the monopole
and $s=-1$ for dipole and quadrupole. 
A similar analysis for the larger mass parameter $M\mu=0.2$
also leads to good agreement between numerical and frequency-domain
results, albeit with slightly larger discrepancies of $\lesssim7~\%$.

Before we discuss in detail the behaviour of Proca fields in rapidly
rotating Kerr backgrounds, we briefly test our code in the case
of a smaller rotation rate $a/M=0.5$ where the slow-rotation
approximation of \cite{Pani:2012vp,Pani:2012bp} is expected to provide
rather accurate results. For this purpose, we have evolved
a Proca field with a mass parameter $M\mu_V=0.4$
(simulation v2K1\_m040 in Table~\ref{tab:ProcaSetup}) which can be shown
to result in a {\em stable} mode, i.e., it does not satisfy the
superradiance criterion (\ref{eq:MFSRcond}). For this configuration
we observe an extended transient resulting from the wave packet
impinging on the black hole followed by a slowly decaying QN ringdown
signal after $t\sim200~M$. Fitting a damped sinusoid to this
ringdown part, we find a ringdown frequency $M\omega_{11} =
0.389-\imath0.0023$ in excellent agreement with semi-analytic calculations
in the slow-rotation approximation \cite{Pani:2012vp,Pani:2012bp}.

\subsection{Instability of Proca fields in highly spinning Kerr backgrounds}
\label{ssec:ProcaKerrHigh}
\begin{figure}
\begin{center}
\begin{tabular}{ccc}
\includegraphics[width=0.33\textwidth]{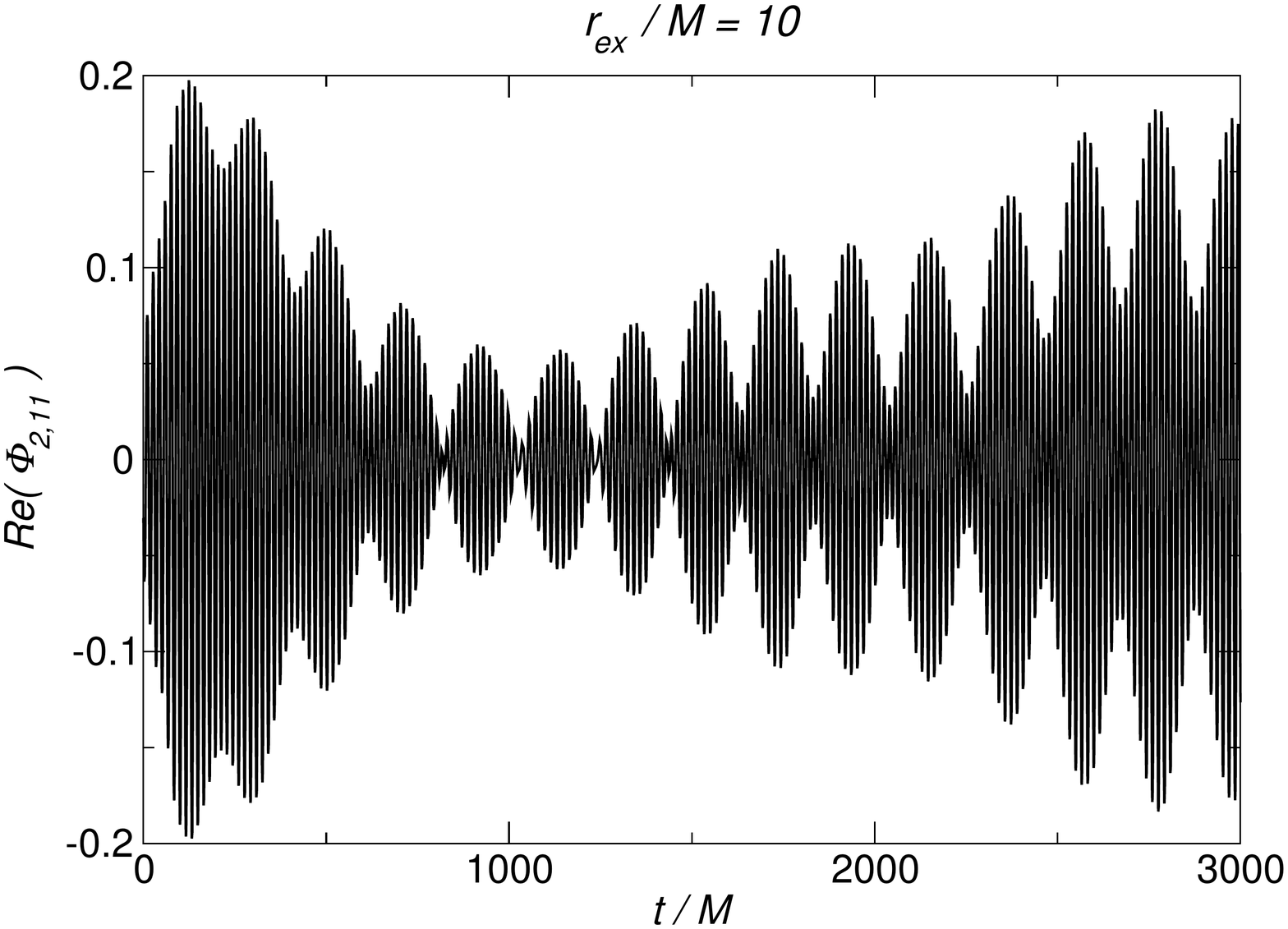} &
\includegraphics[width=0.33\textwidth]{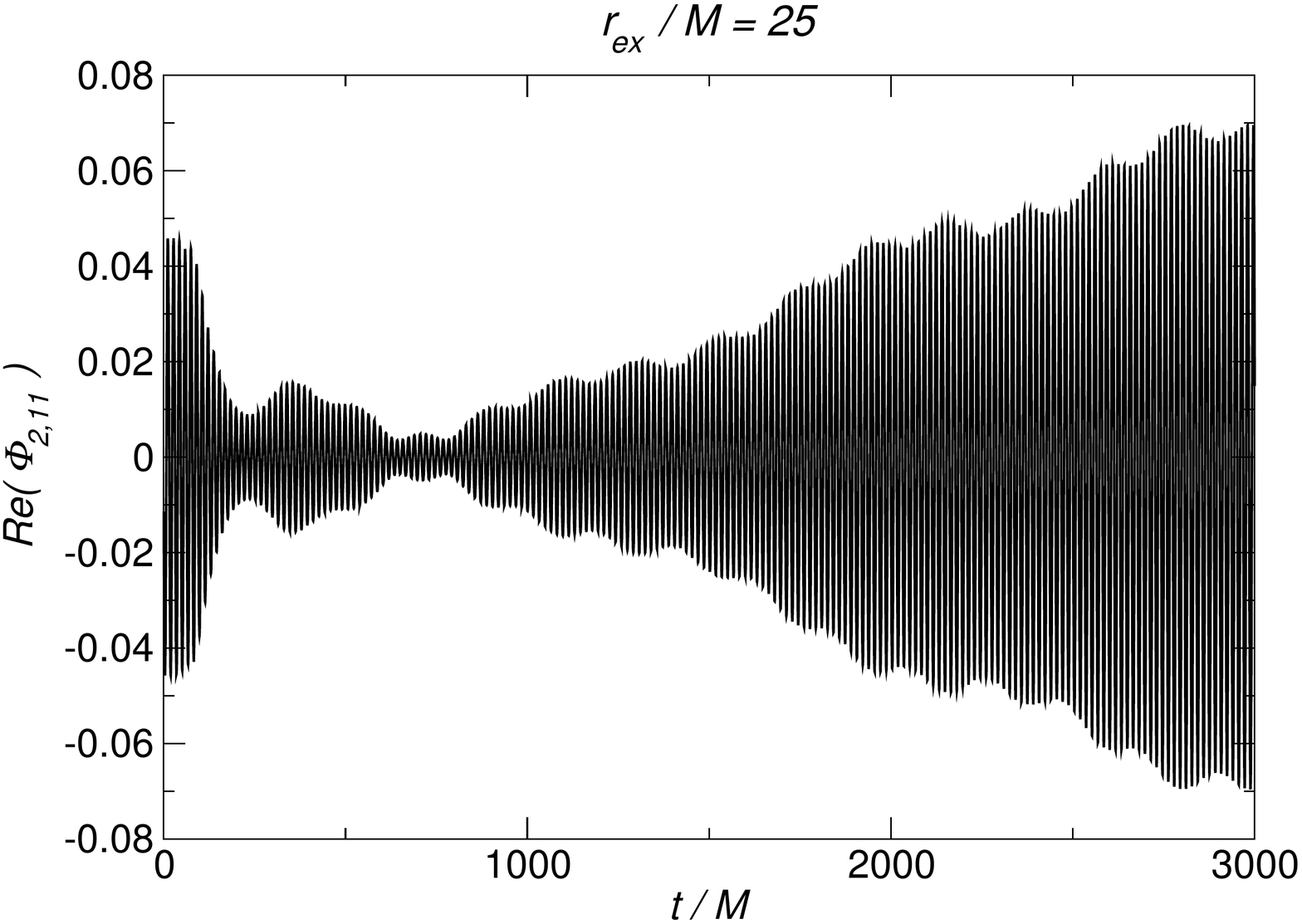} &
\includegraphics[width=0.33\textwidth]{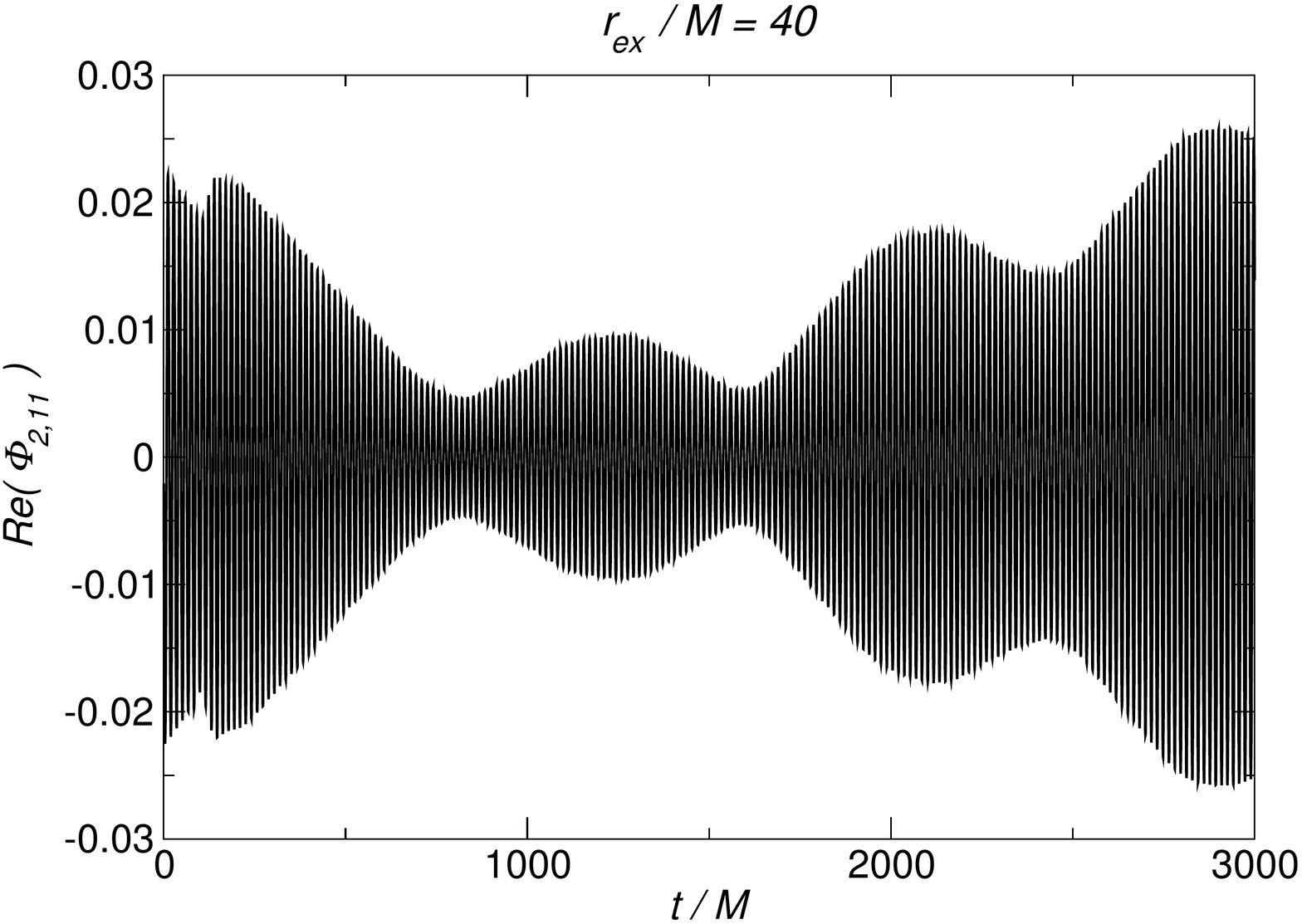} \\
\includegraphics[width=0.33\textwidth]{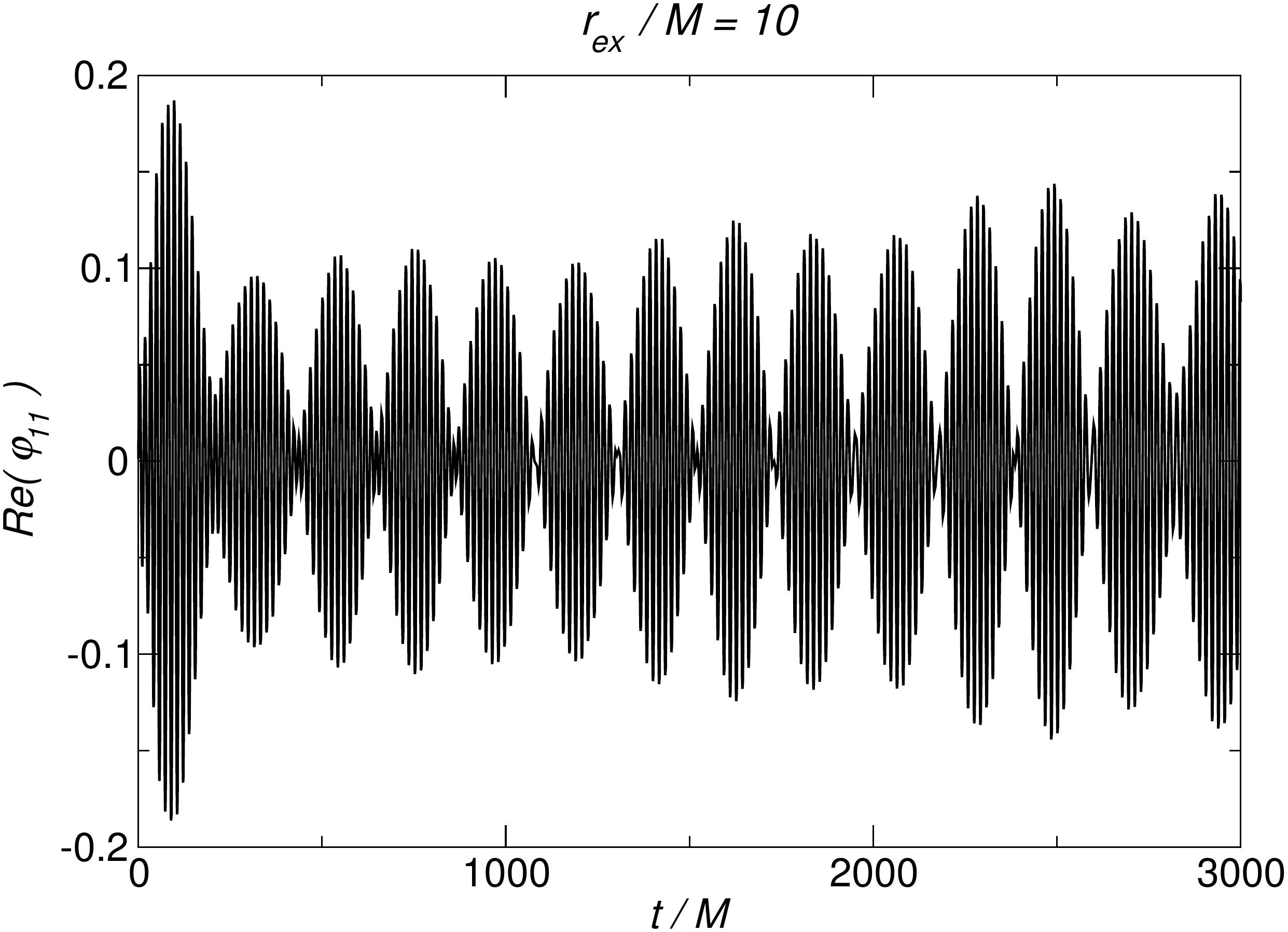} &
\includegraphics[width=0.33\textwidth]{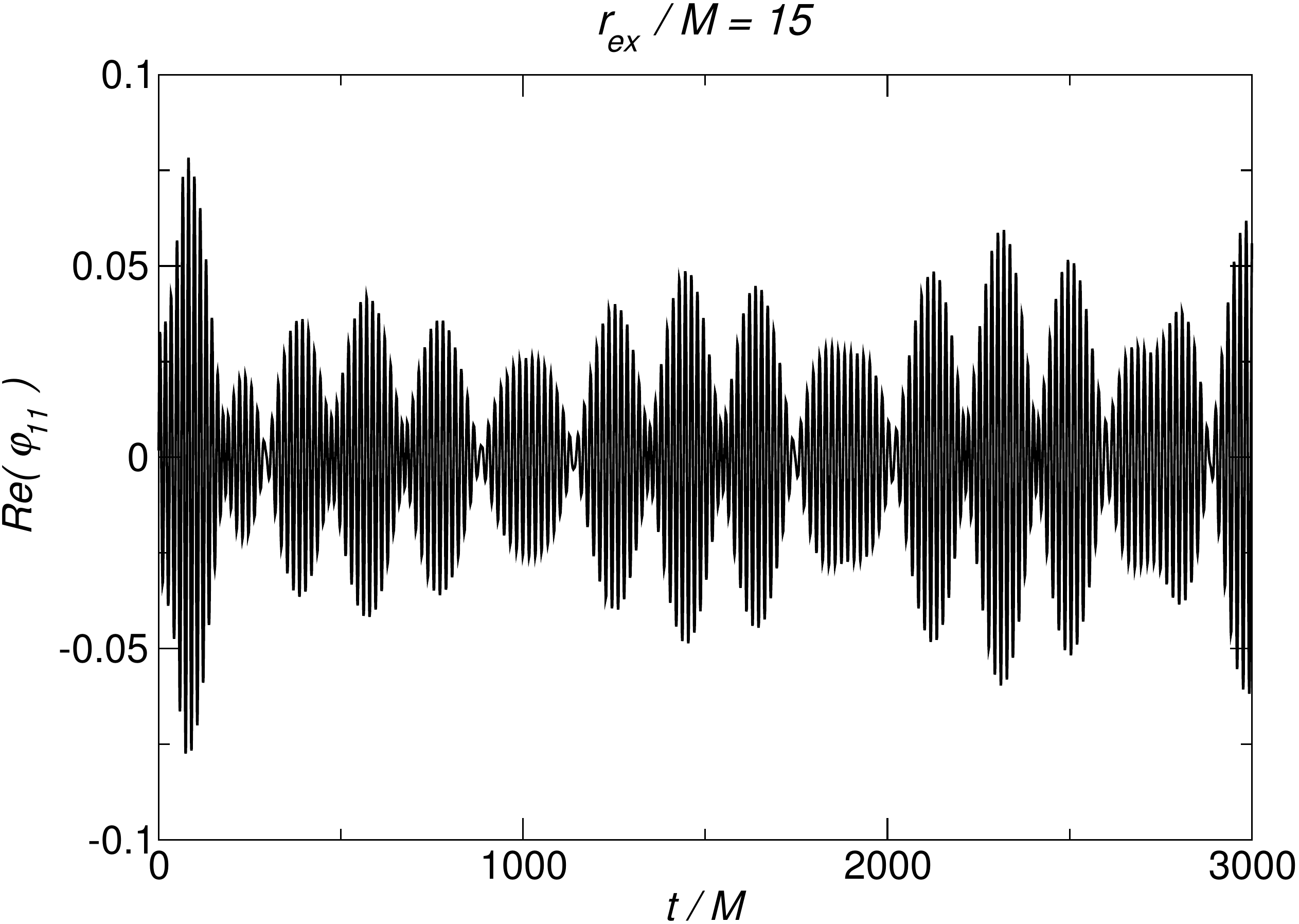} &
\includegraphics[width=0.33\textwidth]{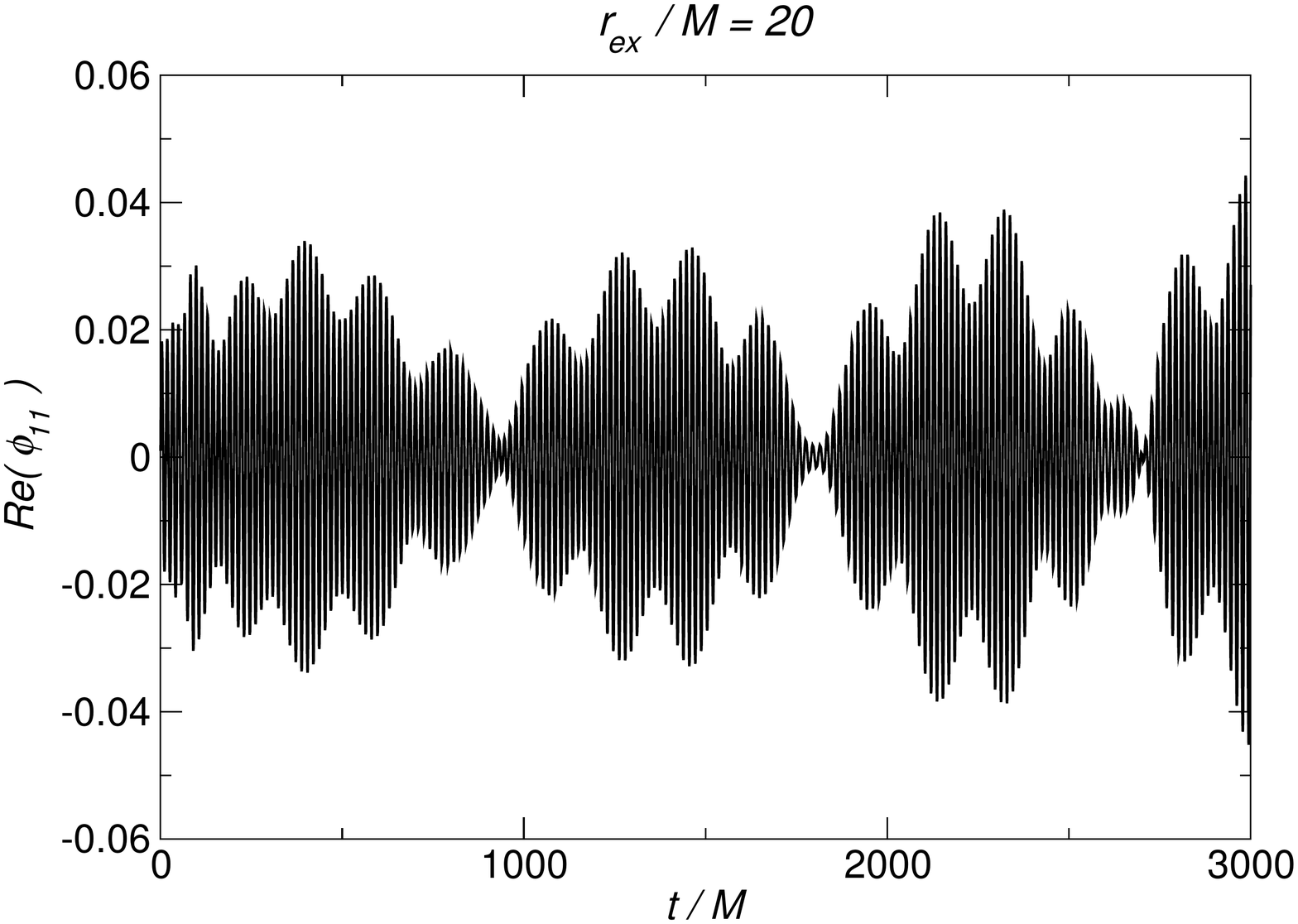} 
\end{tabular}
\end{center}
\caption{\label{fig:Phi2_099_040}
Time evolution of the Proca field with mass coupling $M\mu_V=0.40$
in Kerr background with $a/M=0.99$, at different extraction radii.
We plot the $l=m=1$ mode of the Newman-Penrose scalar $\Phi_2$ (upper panels)
and of the scalar component $\varphi$ (lower panels).
}
\end{figure}

According to the superradiant condition (\ref{eq:MFSRcond}), rapidly
rotating black holes are most likely to induce superradiance phenomena
in ambient vector fields and we therefore discuss the case of Proca
fields in a Kerr background with $a/M=0.99$ in this section.
In fact, recent calculations \cite{Pani:2012vp,Pani:2012bp}
indicate that the maximum instability growth rate is realized in this
background spacetime for a mass parameter of about $M\mu_V=0.4$
and our discussion will focus on this case supplemented
by additional simulations with $M\mu_V=0.42$, $0.44$ and $1.0$.
The time evolution of the $l=m=1$ multipole of the Newman-Penrose
scalar $\Phi_2$ as well as the scalar component $\varphi$ obtained
for the case $M\mu_V=0.4$ is shown for several choices of the
extraction radius $r_{\rm ex}$ in Fig.~\ref{fig:Phi2_099_040}.
Animations of the numerical evolutions are available online
\cite{webpageAnimation}.

\begin{figure}
\begin{center}
\begin{tabular}{cc}
\includegraphics[width=0.50\textwidth]{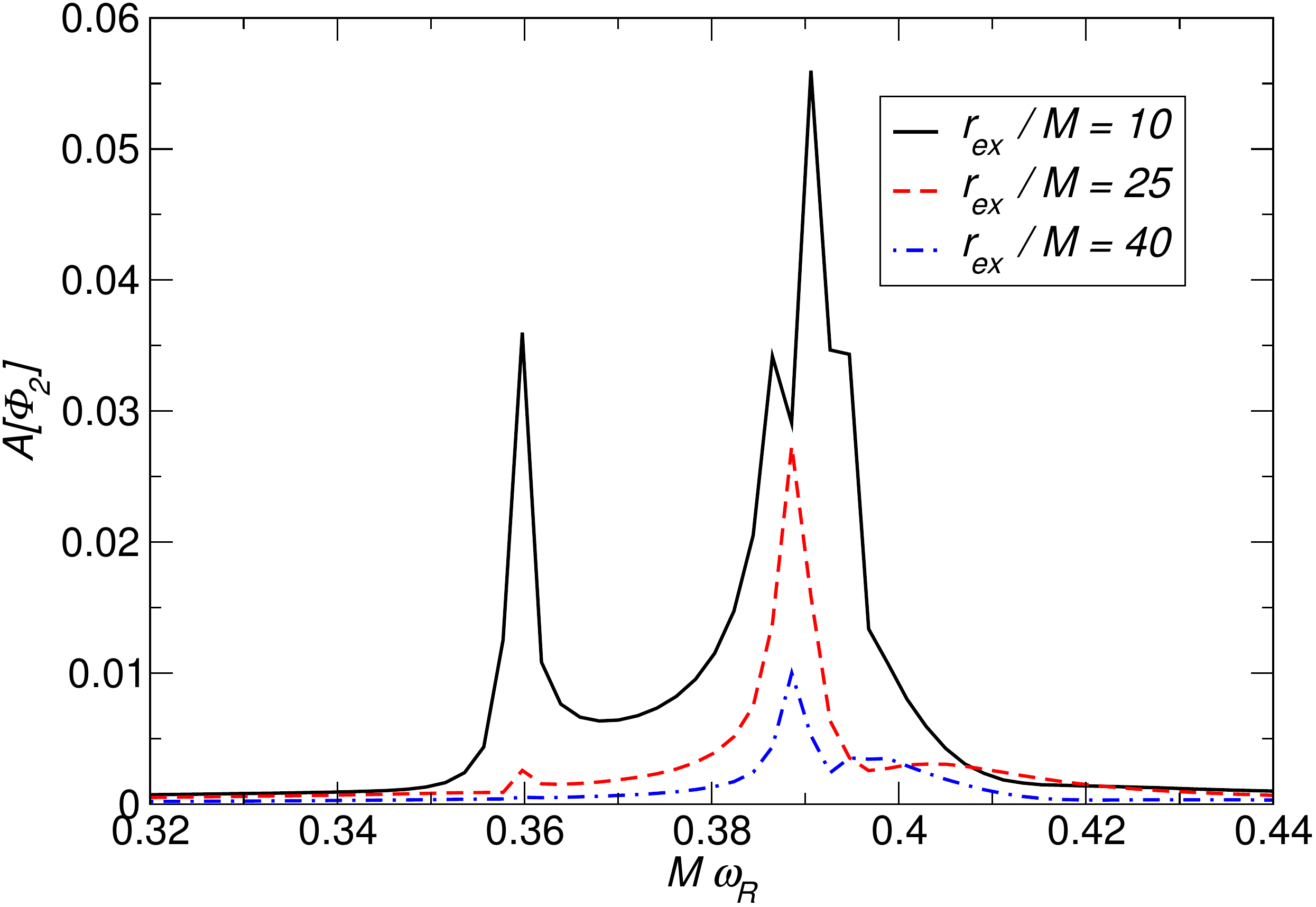} &
\includegraphics[width=0.50\textwidth]{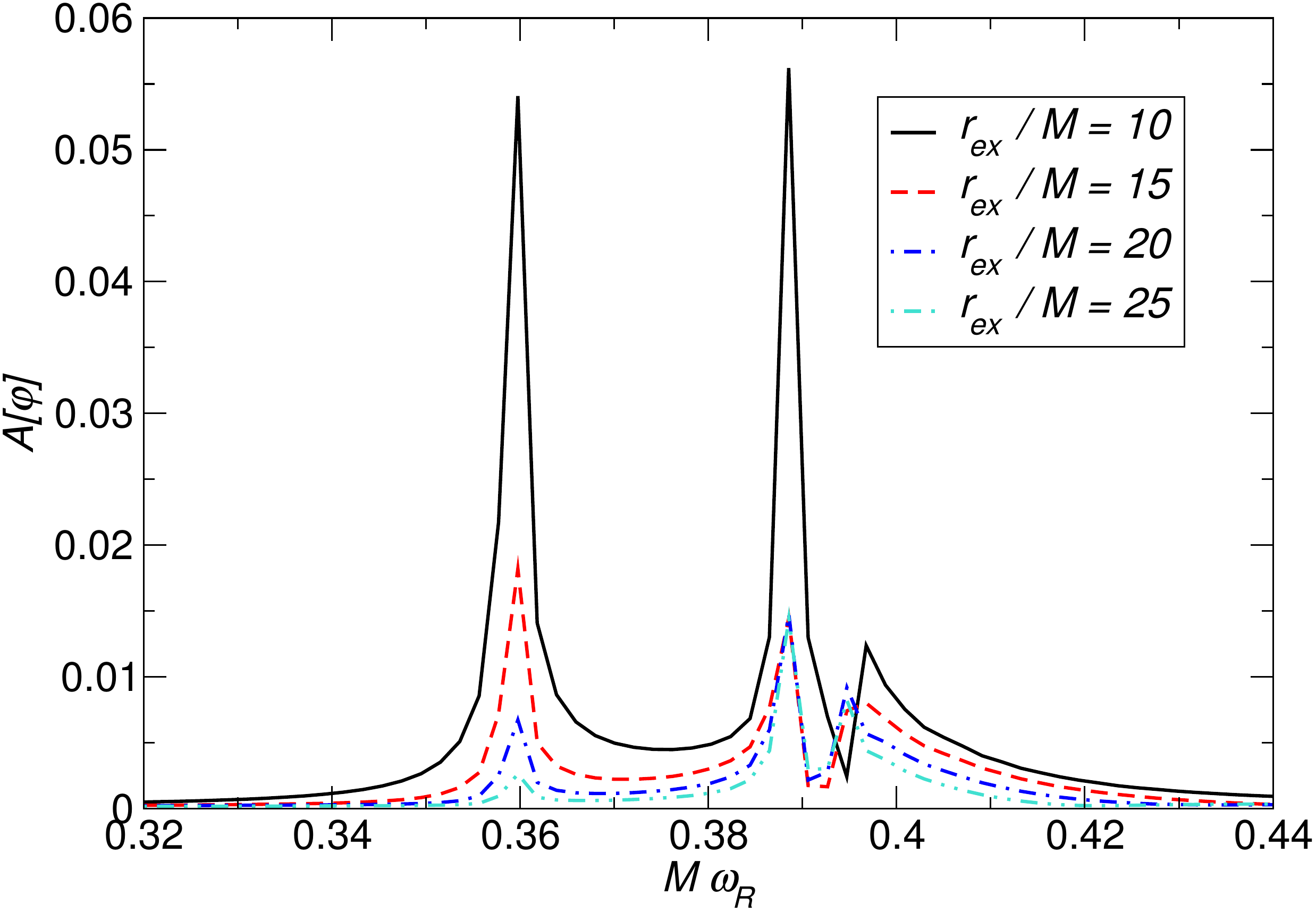} 
\end{tabular}
\end{center}
\caption{\label{fig:Spec_099_040}
Spectra of the $l=m=1$ mode of the Newman-Penrose scalar $\Phi_2$ (left
panel) and of the scalar component $\varphi$ (right panel) excited by
the Proca field with mass coupling $M\mu_V=0.40$ in Kerr background
with $a/M=0.99$.
Different lines correspond to the waveforms measured at different
extraction radii.
}
\end{figure}

{\bf{\noindent{\em{Beating of modes.}}}}
The time evolutions of the Proca field shown in Fig.~\ref{fig:Phi2_099_040}
reveal strong amplitude modulations similar to those
observed for scalar fields in Fig.~\ref{fig:waveformsSSm042_11}.
They also depend sensitively
on the location of the measurement specified by the extraction radius $r_{\rm ex}$
and, again, we interprete this feature as
a beating effect. In order to study this in more detail,
we have computed the Fourier spectra of the waveforms at different
radii and show the results in Fig.~\ref{fig:Spec_099_040}. The
curves clearly reveal several peaks corresponding to separate
mode contributions (fundamental mode or overtones)
of the dipole field and the relative amplitude
of these peaks varies significantly with
extraction radius. Consider, for
example, the spectrum of the dipole of $\Phi_{2}$ in the
left panel of Fig.~\ref{fig:Spec_099_040}:
At $r_{\rm ex}=10~M$ the amplitude of the lowest frequency mode is about
two thirds of the amplitude of the strongest peak near
$M\omega_R=0.39$, but its relative strength rapidly decreases at
larger extraction radii. This is also reflected in the
observed time evolutions of the $\Phi_2$ dipole
in the upper panels of Fig.~\ref{fig:Phi2_099_040}:
the waveform extracted at $r_{\rm ex}=10~M$
exhibits a high frequency modulation
which is weakly present at $r_{\rm ex}=25~M$
and entirely absent in the waveform measured at $r_{\rm ex}=40~M$.
Likewise, the high-frequency modulation of the time component
$\varphi$ weakens at larger radii as the relative
amplitude of the two lowest peaks in the spectrum
(right panel of Fig.~\ref{fig:Spec_099_040}) decreases.

There remains one important issue: which conditions lead to the beating of modes?
While beating is expected to be a generic feature, it is not always excited.
Here, we attempt to determine indicators for the (non-)observability of beating effects.
Because we have performed a multipole expansion of the fields, and beating is present on a fixed multipole,
then clearly a beating pattern requires the excitation of different overtones. 
For this reason, a modulation of the amplitude is only triggered by evolutions of generic initial pulses
\footnote{as opposed to pure bound states. It is important to note that by pure bound states we mean stationary solutions
of the linearized field equations in the Kerr background; therefore effects such as mode coupling are already taken into account:
pure modes do not couple to other multipoles when the expansion basis is taken to be spheroidal harmonics.}.
Furthermore, our numerical
simulations indicate no dependence on the background spacetime;
Fig.~\ref{fig:waveformsSSm042_11} shows beating effects of the
scalar field around Schwarzschild and Kerr BHs. Beating is therefore
{\em not} related to superradiance. On the other hand, the excitation
of different modes is merely a necessary condition but not sufficient
to trigger amplitude modulations. This becomes evident in the evolutions
of Gaussian initial data for the vector field with small mass parameters
in a Schwarzschild background which only exhibit a quasi-normal ringdown
and tail. In order to shed further light on this question, we
therefore consider the entire set of Proca evolutions in BH backgrounds
and summarize our results as follows:
\noindent{(i)} We do not observe beating nor any
long-lived modes for mass parameters
$M\mu_V=0.1,0.2$ in Schwarzschild background;
\noindent{(ii)} we observe
beating in the dipole mode for $M\mu_V=0.4,0.42,0.44$
in rapidly rotating Kerr; 
\noindent{(iii)}
we observe beating in the quadrupole but not in the dipole for $M\mu_V=1.0$ in the background of a highly spinning Kerr BH.
Let us first consider in our interpretation of these observations the
case of small mass parameters $M\mu_{S,V}$. The absence of long-lived
modes in our simulations is most likely a consequence of low-mass
bound states being concentrated far from the black holes. For our
choice of initial Gaussian pulses centered around $r_0=12~M$, these
states are therefore only weakly excited and play no
significant role in the evolution.
Furthermore, beating patterns for small mass parameters have
particularly large periods. We can therefore not rule out that
amplitude modulations might become visible in these evolutions
at late times far larger than the evolution times feasible with
our numerical framework. We
suspect a similar reason is behind our not observing beating modulation
in the dipole mode in case of large mass coupling $M\mu_V=1$.
We therefore tentatively conclude that two conditions need to be met
in addition to the presence of at least two modes in order to observe
beating on time scales of $\sim10^4~M$. (i) the bound states
must not be localized far away from the peak of the initial data and
(ii) beating periods need to be sufficiently short.

{\bf{\noindent{\em{Instability of Proca fields around Kerr.}}}}
The most striking feature of the time evolutions in
Fig.~\ref{fig:Phi2_099_040} is the amplitude growth of the signal
over time (ignoring the early transient stage). We interprete this
growth as a signature of the superradiant instability of massive
vector fields and estimate the growth rate of the $m=1$ dipole
of $\Phi_2$ obtained for $M\mu_V=0.4$ and $a/M=0.99$ as
\begin{align}
  M\omega_{I} \sim(5\pm 1)\cdot 10^{-4}
      &\quad\rightarrow\quad \frac{\tau}{M} \sim (3.3\pm 1.7)\times 10^3\,,
\end{align}
and about half that value for the instability rate of $\varphi$.
We believe the discrepancy of the growth rates of $\Phi_2$ and
$\varphi$ is due to the different growth scales of the axial and
polar sector and the fact that $\varphi$ only sees the axial sector.
We note that our estimates for the growth rates are of the same order
of magnitude as the value $\tau/M \sim 10^3$
derived from extrapolation of slow-rotation calculations;
cf.~Eq. (98) in Ref.~\cite{Pani:2012bp}
and Fig.2 in Ref.~\cite{Pani:2012vp}. The superradiant instability
time scales for vector fields are thus up to four orders of
magnitude larger than those of their scalar counter parts which
renders possible their identification in the numerical evolutions
presented in this work.

%
%

\section{Conclusions}
\label{sec:MFconclusion}

We have performed an extended study of the time evolution of massless
and massive scalar and vector fields in Schwarzschild and Kerr BH
background spacetimes with spin parameters up to $a/M=0.99$. Our
results are consistent with published results obtained in the
frequency domain in so far as these are available.
For evolutions involving exclusively short-lived modes, we observe
the known pattern of an initial transient followed by an
exponentially damped ringdown phase and late-time tails.

For the
case of massive fields, there exists a second class of long-lived modes
called bound states whose evolution exhibits a significantly richer
structure. In particular, we have identified a beating pattern caused
by the interference of different modes (the fundamental mode and
overtones) as manifest in the Fourier spectra calculated from the
time evolutions. The relative amplitude of different modes in the spectra
and, thus, the specific shape of the beating pattern
strongly depends on the observation radius. We believe that these
beating effects provide an explanation for an apparent discrepancy between
time and frequency domain calculations of instability growth rates
of scalar fields around Kerr backgrounds.
Specifically, Ref.~\cite{Strafuss:2004qc} (see their Fig. 3) reported
a time evolution lasting about $3\times 10^3M$ of a massive scalar
field with $M\mu_S=0.25$ in a Kerr background with $a/M=0.9999$
and obtained a growth timescale
approximately two orders of magnitude smaller
than that predicted by frequency-domain calculations.
For this particular configuration, the frequency-domain
estimate, Eq.~\eqref{eq:boundstates}, predicts a beating period
$\tau\sim 5\times 10^3 M$, about the entire duration of their
time evolution, and we believe that their estimate of the growth
rate was correspondingly contaminated by the resulting amplitude
modulation.

Based on our numerical results, we conjecture that two conditions are
required for the presence of beating: \noindent{(i)} generic initial
configurations of massive fields as opposed to single-frequency 
bound state initial data,
and \noindent{(ii)} initial data able to excite these bound states.
For extremely small masses for instance, the bound states are localized
far from the BH. Accordingly, initial data peaked far from the BH would
be required to excite them.

In an accompanying paper Dolan \cite{Dolan:2012yt} has investigated the massive scalar field instability as well as 
a system involving massless scalars but enclosed by a mirror in the time domain. Specifically, the author employs a 
``coupled $1+1$ dimensional'' numerical scheme which allows for significantly longer evolutions times up to $t\sim 10^6M$.
This work finds bound state frequencies and instability time scales in excellent agreement with frequency domain calculations
and, due to the generic initial configurations, confirms/ supports our findings of the beating phenomena.

Finally, we have confirmed the existence of superradiant unstable
massive vector fields around rapidly spinning BHs.
In contrast to massive scalar fields, the instability
of Proca fields is stronger by about four orders of magnitude leading to
growth times as short as $\tau/M\sim 3.3\times 10^3$ in
rapidly spinning Kerr BH backgrounds. It is instructive
to translate this number for the case of realistic astorphysical black
hole candidates.
For a solar mass BH and a supermassive BH of the size of
SagittariusA$^{\ast}$ ($M\sim4.1\cdot10^6M_{\odot}$) at the center of
our galaxy we obtain timescales of $\tau\sim 9~\mathrm{ms}$
and $\tau\sim4\cdot10^4~\mathrm{s}$,
respectively.

It has been argued by Pani {\em et al.}
\cite{Pani:2012vp} that the angular momentum thus extracted from the
black hole provides a mechanism to observationally constrain the mass
of the photon. In view of the short time scales even for a supermassive
black holes, spin measurements of the BH at the center of the Milky Way,
may indeed constrain the photon mass to unprecedented levels.
The exciting prospect of using large, supermassive black holes to
understand the microscopic world raises several questions:

\noindent{(1) Influence of accretion disks.} 
Astrophysical black holes are
not isolated, but typically are surrounded by accretion disks. Can the
interaction with matter kill the instability? It was argued previously \cite{Pani:2012vp}
that the superradiant instability is a global mode, on scales larger
than the BH. On these scales matter is electrically neutral, and the coupling should
be negligible.
Therefore, it is not likely that (thin)
accretion disks can quench the instability; thin disks are expected to lie along the equator
of a Kerr BH and are not expected to interact strongly with boson clouds that extend well off the equator.
The influence of thick disks or other effects
is unknown at the moment, but it is surely important to study these effects in more detail.

\noindent{(2) Self-interacting scalar fields:} One class of interesting
problems involves massive scalar fields whose dynamics are described by
additional non-linear terms, modelling their self-interaction.  This open
issue has first been addressed by Yoshino \& Kodama \cite{Yoshino:2012kn}
who modelled the collapse of a so-called bosenova.

\noindent{(3) Backreaction effects:} As far as we are aware, all studies
exploring massive fields have been performed in the linear regime.
Therefore it is of utmost interest to explore the fully non-linear
regime, which allows for the investigation of the backreaction of
the spacetime, such as the spin-down of the BH due to (subsequent)
superradiant scattering.  This type of studies would enable us to
glance at the end-state of the superradiant instability or, possibly,
equilibrium configurations.

The present study marks crucial first steps to explore an entire
playground of exciting future applications of massive fields in BH
spacetimes.

Animations of the evolutions can be found online \cite{webpageAnimation}.

\section{Acknowledgements}
We thank Sam Dolan, Paolo Pani and Jo\~ao Rosa for useful comments
and discussions and for making their data available to us.  We are
indebted to Hideo Kodama and Hirotaka Yoshino for explanations
regarding time-evolution of bound states.  We thank Gaurav Khanna
for useful correspondence.  We are especially indebted to S\'ergio
Almeida for all his hard work on the ``Baltasar Sete-S\'ois''
cluster.  We thank all participants of the YITP-T-11-08 workshop on
``Recent advances in numerical and analytical methods for black hole
dynamics'' for useful discussions.  We thank the Yukawa Institute
for Theoretical Physics at Kyoto University for their kind hospitality during the early stages of this work.
This work was supported by the {\it DyBHo--256667} ERC Starting
Grant, the NRHEP--295189 FP7--PEOPLE--2011--IRSES Grant,
the {\it CBHEO--293412} FP7-PEOPLE-2011-CIG Grant,
the {\it ERC-2011-StG 279363--HiDGR} ERC Starting Grant,
and by FCT
- Portugal through PTDC projects FIS/098025/2008, FIS/098032/2008,
CTE-ST/098034/2008, CERN/FP/123593/2011.  H.W. is funded by FCT through
grant SFRH/BD/46061/2008. U.S. acknowledges support by the
from
the Ram\'on y Cajal  Programme and Grant FIS2011-30145-C03-03  of the
Ministry of Education and Science of Spain.
A.I. was supported by JSPS Grant-in-Aid for Scientific Research Fund 
(C) 22540299 and (A) 22244030.

Computations were performed on the ``Baltasar Sete-Sois'' cluster at IST,
the cane cluster in Poland through PRACE DECI-7 ``Black hole dynamics
in metric theories of gravity'',  
on MareNostrum in Barcelona through BSC grant AECT-2012-2-0005,
on Altamira in Cantabria through BSC grant AECT-2012-3-0012,
on Caesaraugusta in Zaragoza through BSC grants AECT-2012-2-0014 and AECT-2012-3-0011,
XSEDE clusters SDSC Trestles and NICS Kraken
through NSF Grant~No.~PHY-090003, Finis Terrae through Grant
CESGA-ICTS-234 and the COSMOS supercomputer, part of the DiRAC HPC
Facility which is funded by STFC and BIS.
The authors thankfully acknowledge the computer resources, technical
expertise and assistance provided by the Barcelona Supercomputing
Centre---Centro Nacional de Supercomputaci\'on
and by Andrey Kaliazlin for computational support and technical advice with COSMOS.

\appendix
\section{Flux formula }\label{app:Fluxformula}
From the Lagrangian ${\cal L}$, 
\begin{align}
\frac{1}{\sqrt{ -g}}{\cal L} := &
        - \frac{1}{2}g^{\mu\nu}\Psi^{\ast}_{,\mu}\Psi^{}_{,\nu} - \frac{\mu_S^2}{2}\Psi^{\ast}\Psi-V(\Psi)
        - \frac{1}{4}F^{\mu\nu}F_{\mu\nu} - \frac{\mu_V^2}{2}A_{\nu}A^{\nu}
        - \frac{k_{\rm axion}}{2}\Psi \,^{\ast}F^{\mu\nu}F_{\mu\nu} 
        + J^{(S)}_{\nu} A^{\nu}
\,,
\end{align}
we obtain, under the Lorenz condition \eqref{eq:MFLG},
the equations of motion
\begin{align}
& \left(\nabla^{\mu}\nabla_{\mu} - \mu_S^2 \right) \Psi - \frac{k_{\rm axion}}{2} \,^{\ast}F^{\mu\nu}F_{\mu\nu} - V^{'}(\Psi) = 0
\\ 
& \left(\nabla^{\mu}\nabla_{\mu} - \mu_V^2 \right) A^\nu - R_\mu{}^\nu A^\mu  
  - 2 k_{\rm axion} \,^{\ast} F^{\nu\mu} \nabla_{\mu} \Psi = -J^{(S)\nu}
\end{align}
and the stress-energy tensor
\begin{align} 
T_{\mu \nu} = & T^{A}_{\mu \nu} + T^{\Psi}_{\mu \nu} \,, 
\\ 
& T^{A}_{\mu \nu} := 
        F_{\mu \alpha}F_\nu{}^{\alpha} 
      - \frac{1}{4}g_{\mu \nu} F_{\alpha \beta}F^{\alpha \beta} 
      + \mu_V^2 A_\mu A_\nu 
      - \frac{1}{2}\mu_V^2 A^\alpha A_\alpha g_{\mu \nu} \,, 
\label{T:A}
\\ 
& T^{\Psi}_{\mu \nu} := 
        \frac{1}{2} \left( \nabla_\mu \Psi^{\ast} \nabla_\nu \Psi + \nabla_{\mu} \Psi \nabla_{\nu} \Psi^{\ast} \right) 
      - \frac{1}{2} g_{\mu \nu} 
        \left( \nabla_{\mu} \Psi^{\ast} \nabla^{\mu} \Psi
               + \mu_S^2 \Psi^{\ast} \Psi + 2 V(\Psi) 
        \right)  \,. 
\label{T:phi}
\end{align}
Note that the Chern-Simons term does not alter the stress-energy tensor. 

\medskip
We are concerned with stationary axisymmetric spacetimes.  Let us
denote by $t^\mu$ and $\phi^\mu$, respectively, the stationary and the
axisymmetric Killing field.  The rigidity theorem states that there exists
a Killing vector field $\chi^\mu$, which is normal to, hence null on,
the horizon ${\cal H}^+$.  Such a Killing vector field is given by the
combination of $t^\mu$ and $\phi^\mu$,
\begin{equation}
  \chi^\mu = t^\mu + \Omega_H \phi^\mu \,,
\end{equation}
where $\Omega_H$ denotes the angular velocity of the Killing horizon 
of the black hole, and for the Kerr metric
\begin{equation}
 \Omega_H = \frac{a}{r_{+}^2+a^2} \,, 
\end{equation}
with $r_{+}$ being the horizon radius $r_{+} = M+ \sqrt{M^2-a^2}$ in 
the Boyer-Lindquist coordinates. 

\medskip
Let us introduce the energy current $J_\mu$ by
\begin{equation}
  J_\mu = - T_{\mu \nu}t^\nu 
        = - T_{\mu \nu}^\Psi t^\nu - T^A_{\mu \nu} t^\nu \,,   
\end{equation}
and consider two time-slices $\Sigma_1$ and $\Sigma_2 
\subset \{ J^+(\Sigma_1)\setminus \Sigma_1 \}$, 
which intersect the event horizon ${\cal H}^+$ at $t=t_1$ and $t=t_2$, 
respectively.  
(One may view $t$ as the Killing parameter of $t^\mu$). 
Then the energy flux ${\cal F}$ that flows into the black hole in 
the time-interval ${\cal I}=\{{\cal H}^+: t_1<t< t_2 \}$ is given by
\begin{eqnarray}
 {\cal F} &=& \int_{\Sigma_2} {\rm d} \Sigma^\mu J_\mu 
                  -\int_{\Sigma_1} {\rm d} \Sigma^\mu J_\mu   
     = \int_{\cal I} {\rm d} N n^\mu J_\mu    
\nonumber \\
     &=& - \int_{\cal I} {\rm d} N \chi^\mu J_\mu = \int_{H} 
         {\rm d} S \left< \chi^\mu t^\nu T_{\mu \nu}\right> \,.  
\label{Flux}
\end{eqnarray}
Here $n^\mu$ denotes the normal to ${\cal H}^+$, 
${\rm d} N$ the volume element of the horizon ${\cal H}^+$, 
and ${\rm d} S$ the area element of the cross-section 
$H= {\cal H}^+\cap \Sigma$, and $\left< \cdot \right> $ 
expresses the time average along the horizon. 
Superradiant scattering occurs when the energy flux 
going into the horizon becomes negative:
\begin{equation}
  {\cal F} < 0 \,. 
\end{equation}

\medskip 
We consider below the flux with respect to the scalar field 
and the vector field separately. 
For the scalar field, the mode decomposition is defined by  
\begin{equation}
 \pounds_t \Psi = - i\omega \Psi \,, \quad 
 \pounds_\phi \Psi = im\Psi \,, 
\end{equation}
where $\omega$ denotes the frequency and $m$ the angular quantum number. 
It implies, in particular, 
\begin{equation}
 \pounds_\chi \Psi = - i(\omega - m\Omega_H) \Psi \,.  
\end{equation}
Then, substituting (\ref{T:phi}) into (\ref{Flux}) and using the fact that 
$g_{\mu \nu}t^\nu \chi^\mu = 0$ on the horizon, 
we can immediately find the flux formula for $\Psi$: 
\begin{equation}
  {\cal F}^\Psi 
  = \omega (\omega - m\Omega_H)\int_H {\rm d} S \left<| \Psi |^2 \right> \,,  
\label{flux:phi}
\end{equation}
and read off the superradiance condition 
\begin{equation}
0<\omega<m\Omega_H \,. 
\label{condi:suprad}
\end{equation} 

\medskip 
For the vector field, substituting (\ref{T:A}) into (\ref{Flux}), 
we have 
\begin{eqnarray}
 {\cal F}^A 
  &=& \int_H {\rm d} S 
      \left< \chi^\mu F_{\mu \alpha} t^\nu F_\nu{}^\alpha \right> 
   + \mu_V^2 \int_H {\rm d} \left< \chi^\mu A_\mu t^\nu A_\nu \right> \,, 
\end{eqnarray}
where we have again used $g_{\mu \nu}\chi^\mu t^\nu = 0$ on the horizon. 

\medskip 
For the vector field, the mode decomposition is defined by 
\begin{equation}
 (\pounds_t A)_\mu = -i \omega A_\mu \,, \quad 
 (\pounds_\chi A)_\mu = -i (\omega - m\Omega) A_\mu \,. 
\label{mode-decomp}
\end{equation}
Now, noting 
\begin{equation} 
 \chi^\mu F_{\mu \alpha} 
  = \pounds_\chi A_\alpha - \nabla_\alpha (A_\mu \chi^\mu) \,, \quad 
 t^\nu F_{\nu \beta} = \pounds_t A_\beta - \nabla_\beta (A_\nu t^\nu) \,, 
\end{equation}
and using Eq.~\eqref{mode-decomp} and the Lorenz condition~\eqref{eq:MFLG}
we find 
\begin{eqnarray}
 \chi^\mu F_{\mu \alpha} t^\nu F_{\nu}{}^\alpha 
 &=& g^{\alpha \beta} \pounds_\chi A_\alpha \pounds_t A_\beta 
     + \chi^\mu A_\mu \nabla^\alpha \nabla_\alpha (t^\nu A_\nu) 
     + \mbox{divergence terms} \,. 
\end{eqnarray}
On the horizon integral, we ignore the divergence terms and then obtain 
\begin{eqnarray}
 {\cal F}^A 
  &=& \int_H {\rm d} S 
      \left< 
             {\rm Re}(\chi^\mu F_{\mu \alpha}t^\nu F_{\nu}{}^{\alpha *}) 
      \right>   
    + \mu_V^2 \int_H {\rm d} S 
              \left< {\rm Re}(\chi^\mu A_\mu t^\nu A_\nu^*)\right> 
\nonumber \\  
   &=& \omega(\omega- m\Omega)\int_H {\rm d} S \left< |A|^2 \right>  
      + \mu_V^2 \int_H {\rm d} S \left< |\chi^\mu A_\mu t^\nu A_\nu| \right> 
      -  \int_H {\rm d} S 
         \left< {\rm Re} 
              \left(
                  \chi^\mu A_\mu  \nabla^\alpha \nabla_\alpha (t^\nu A_\nu^*) 
              \right)
           \right> \,.
\end{eqnarray}
If we impose 
\begin{equation}
  \chi^\mu A_\mu = 0 \,\, \mbox{on ${\cal H}^+$} \,, 
\end{equation}
we get the simple formula, similar to the scalar field case (\ref{flux:phi}), 
\begin{equation} 
 {\cal F}^A = \omega(\omega- m\Omega)
              \int_H {\rm d} S \left< |A|^2 \right>  \,,   
\end{equation}
and the superradiance condition for the vector field is again given by 
(\ref{condi:suprad}).

\newpage
\section{Spin-weighted spherical harmonics}\label{app:Ylmm}
Here,  we list the spin-weighted spherical harmonics up to $l=2$ in spherical
coordinates $\{\theta,\phi\}$ and Cartesian coordinates $\{x,y,z\}$
given by Eq.~\eqref{eq:CartCoords}. 
In case of spin-weight $s=0$ we obtain:
\begin{description}
\item [$l=0$] 
\begin{align}
\label{eq:s0Y00}
Y^R_{00} = \frac{1}{\sqrt{4\pi}}
\,, &\quad
Y^I_{00} = 0
\end{align}
\item [$l=1$]
\begin{subequations}
\begin{align}
Y^R_{10} = & \sqrt{\frac{3}{4\pi}} \cos\theta
         =   \sqrt{\frac{3}{4\pi}} \frac{z}{r}
\,, \quad
Y^I_{10} = 0
\,,\\
Y^R_{11} = & -\sqrt{\frac{3}{8\pi}} \sin\theta \cos\phi
         =   -\sqrt{\frac{3}{8\pi}} \frac{x}{r}
\,, \quad
Y^I_{11} = -\sqrt{\frac{3}{8\pi}} \sin\theta \sin\phi
         = -\sqrt{\frac{3}{8\pi}} \frac{y}{r}
\,,\\
Y^R_{1-1} = & -Y^R_{11}
\,, \quad
Y^I_{1-1} =  Y^I_{11}
\end{align}
\end{subequations}
\item [$l=2$]
\begin{subequations}
\begin{align}
Y^R_{20} = & \sqrt{\frac{5}{16\pi}} (3\cos^2\theta - 1) 
         =   \sqrt{\frac{5}{16\pi}} \left(3\frac{z^2}{r^2} - 1\right)
\,, \quad
Y^I_{20} = 0
\,,\\
Y^R_{21} = & -\sqrt{\frac{5}{8\pi}} \cos\theta \sin\theta \cos\phi
         =   -\sqrt{\frac{5}{8\pi}} \frac{xz}{r^2}
\,, \quad
Y^I_{21} =   -\sqrt{\frac{5}{8\pi}} \cos\theta \sin\theta \sin\phi
         =   -\sqrt{\frac{5}{8\pi}} \frac{yz}{r^2}
\,,\\
Y^R_{22} = & \sqrt{\frac{15}{32\pi}} \sin^2\theta  \cos(2\phi)
         =   \sqrt{\frac{15}{32\pi}} \frac{x^2-y^2}{r^2}
\,, \quad
Y^I_{22} = \sqrt{\frac{15}{32\pi}} \sin^2\theta \sin(2\phi)
         = \sqrt{\frac{15}{8\pi}} \frac{xy}{r^2}
\,\\
\label{eq:s0Y2m2}
Y^R_{2-1} = & - Y^R_{21}\,, \quad
Y^I_{2-1} =     Y^I_{21}\,, \quad 
Y^R_{2-2} =     Y^R_{22}\,, \quad
Y^I_{2-2} =   - Y^I_{22}\,
\end{align}
\end{subequations}
\end{description}
where $\cos(2\phi) = \cos^2\phi - \sin^2\phi$,
$\sin(2\phi) = 2\cos\phi \sin\phi$,
$\cos(3\phi) = 4 \cos^3\phi - 3 \cos\phi$ and
$\sin(3\phi) = 4 \cos^2\phi \sin\phi - \sin\phi$.
We further summarize the $s=-1$ spin-weighted spherical harmonics up to
$l=2$, where we have also defined $\rho^2 = x^2 + y^2 $.
\begin{description}
\item [$l=0$] 
\begin{align}
\label{eq:sm1Y00}
_{-1}Y^R_{00} = & 0
\,,\quad
_{-1}Y^I_{00} =   0
\end{align}
\item [$l=1$]
\begin{subequations}
\begin{align}
_{-1}Y^R_{1-1} = &  \sqrt{\frac{3}{16\pi}}\cos\phi (\cos\theta-1)
               =    \sqrt{\frac{3}{16\pi}}\frac{x(z-r)}{r\rho}
\,,\\ 
_{-1}Y^I_{1-1} = & - \sqrt{\frac{3}{16\pi}}\sin\phi (\cos\theta-1)
               =   - \sqrt{\frac{3}{16\pi}}\frac{y(z-r)}{r\rho}
\,\\
_{-1}Y^R_{10} = & -\sqrt{\frac{3}{8\pi}}\sin\theta
             =    -\sqrt{\frac{3}{8\pi}}\frac{\rho}{r}
\,,\quad
_{-1}Y^I_{10} =   0
\,\\
_{-1}Y^R_{11} = & - \sqrt{\frac{3}{16\pi}}\cos\phi(1+\cos\theta)
               =  - \sqrt{\frac{3}{16\pi}}\frac{x(z+r)}{r\rho}
\,,\\ 
_{-1}Y^I_{11} = & - \sqrt{\frac{3}{16\pi}}\sin\phi(1+\cos\theta)
               =  - \sqrt{\frac{3}{16\pi}}\frac{y(z+r)}{r\rho}
\end{align}
\end{subequations}

\item [$l=2$]
\begin{subequations}
\label{eq:sm1Y}
\begin{align}
_{-1}Y^R_{2-2} = &  \sqrt{\frac{5}{16\pi}} \sin\theta (\cos\theta-1) (2\cos^2\phi - 1) 
               =    \sqrt{\frac{5}{16\pi}} \frac{(z-r)(x^2 - y^2)}{r^2\rho} 
\,,\\ 
_{-1}Y^I_{2-2} =& - \sqrt{\frac{5}{ 4\pi}} \cos\phi \sin\phi \sin\theta (\cos\theta-1)
               =  - \sqrt{\frac{5}{ 4\pi}} \frac{xy(z-r)}{r^2\rho} 
\,\\
_{-1}Y^R_{2-1} = &  \sqrt{\frac{5}{16\pi}} \cos\phi (2\cos^2\theta - \cos\theta - 1 )
               =    \sqrt{\frac{5}{16\pi}} \frac{x(2z^2-zr-r^2)}{r^2\rho} 
\,,\\ 
_{-1}Y^I_{2-1} =& - \sqrt{\frac{5}{16\pi}} \sin\phi (2\cos^2\theta - \cos\theta - 1 )
               =  - \sqrt{\frac{5}{16\pi}} \frac{y(2z^2-zr-r^2)}{r^2\rho} 
\,\\
_{-1}Y^R_{20} = & - \sqrt{\frac{15}{8\pi}} \cos\theta\sin\theta
              =   - \sqrt{\frac{15}{8\pi}} \frac{z\rho}{r^2} 
\,,\quad
_{-1}Y^I_{20} =   0
\,\\
_{-1}Y^R_{21} = & - \sqrt{\frac{5}{16\pi}} \cos\phi ( 2\cos^2\theta + \cos\theta - 1) 
              =   - \sqrt{\frac{5}{16\pi}} \frac{x(2z^2+zr-r^2)}{r^2\rho} 
\,,\\ 
_{-1}Y^I_{21} = & - \sqrt{\frac{5}{16\pi}} \sin\phi ( 2\cos^2\theta + \cos\theta - 1)
              =   - \sqrt{\frac{5}{16\pi}} \frac{y(2z^2+zr-r^2)}{r^2\rho} 
\,\\
\label{eq:sm1Y22}
_{-1}Y^R_{22} = & \sqrt{\frac{5}{16\pi}} \sin\theta ( \cos\theta+1) (2\cos^2\phi - 1)
              =   \sqrt{\frac{5}{16\pi}} \frac{(z+r)(x^2 - y^2)}{r^2\rho} 
\,,\\ 
_{-1}Y^I_{22} = & \sqrt{\frac{5}{ 4\pi}} \cos\phi \sin\phi \sin\theta ( \cos\theta+1)
              =   \sqrt{\frac{5}{ 4\pi}} \frac{xy(z+r)}{r^2\rho} 
\end{align}
\end{subequations}

\end{description}


\bibliographystyle{h-physrev4}
\bibliography{Proca_Bib}

\end{document}